\documentclass[american,english,aps]{revtex4}
\usepackage[T1]{fontenc}
\usepackage[latin9]{inputenc}
\usepackage{amsmath}
\usepackage{amssymb}
\usepackage{graphicx}
\usepackage{esint}

\makeatletter

\providecommand{\tabularnewline}{\\}

\@ifundefined{textcolor}{}
{%
 \definecolor{BLACK}{gray}{0}
 \definecolor{WHITE}{gray}{1}
 \definecolor{RED}{rgb}{1,0,0}
 \definecolor{GREEN}{rgb}{0,1,0}
 \definecolor{BLUE}{rgb}{0,0,1}
 \definecolor{CYAN}{cmyk}{1,0,0,0}
 \definecolor{MAGENTA}{cmyk}{0,1,0,0}
 \definecolor{YELLOW}{cmyk}{0,0,1,0}
}

\usepackage{babel}\usepackage{amsfonts}\usepackage{babel}
\providecommand{\U}[1]{\protect\rule{.1in}{.1in}}
\makeatletter\providecommand{\tabularnewline}{\\}\@ifundefined{textcolor}{}{\definecolor{BLACK}{gray}{0}\definecolor{WHITE}{gray}{1}\definecolor{RED}{rgb}{1,0,0}\definecolor{GREEN}{rgb}{0,1,0}\definecolor{BLUE}{rgb}{0,0,1}\definecolor{CYAN}{cmyk}{1,0,0,0}\definecolor{MAGENTA}{cmyk}{0,1,0,0}\definecolor{YELLOW}{cmyk}{0,0,1,0}}\makeatletter\providecommand{\tabularnewline}{\\}\@ifundefined{textcolor}{}{\definecolor{BLACK}{gray}{0}\definecolor{WHITE}{gray}{1}\definecolor{RED}{rgb}{1,0,0}\definecolor{GREEN}{rgb}{0,1,0}\definecolor{BLUE}{rgb}{0,0,1}\definecolor{CYAN}{cmyk}{1,0,0,0}\definecolor{MAGENTA}{cmyk}{0,1,0,0}\definecolor{YELLOW}{cmyk}{0,0,1,0}}\makeatother\makeatother\makeatother

\makeatother

\usepackage{babel}
\begin{document}

\title{Relaxation of excited spin, orbital, and valley qubit states in single
electron silicon quantum dots}

\author{Charles Tahan}

\affiliation{Laboratory for Physical Sciences, 8050 Greenmead Dr, College Park,
MD 20740}

\author{Robert Joynt}

\affiliation{Physics Department, University of Wisconsin-Madison, 1150 University
Ave., Madison, WI 53706}
\begin{abstract}
We expand on previous work that treats relaxation physics
of low-lying excited states in ideal, single electron, silicon quantum
dots in the context of quantum computing. These states are of three
types: orbital, valley, and spin. The relaxation times depend sensitively
on system parameters such as the dot size and the external magnetic
field. Generally, however, orbital relaxation times are short in strained
silicon ($10^{-7}$ to $10^{-12}$ s), spin relaxation times are long,
($10^{-6}$ to $\gg$ 1 s), while valley relaxation times are expected
to lie in between. The focus is on relaxation due to emission or absorption
of phonons, but for spin relaxation we also consider competing mechanisms
such as charge noise. Where appropriate, comparison is made to reference
systems such as quantum dots in III-V materials and silicon donor
states. The phonon bottleneck effect is shown to be rather small in
the silicon dots of interest. We compare the theoretical predictions
to some recent spin relaxation experiments and comment on the possible
effects of non-ideal dots. 
\end{abstract}
\maketitle

\section{Introduction}

The spin of an electron in silicon may act as an information carrier
in future information technologies, from quantum computers to spintronics.
For quantum information applications, the spin of cold, localized
electrons in silicon can make a qubit with a low memory-error-rate
due to the purifiability of the spin environment (a spin-0 nuclear
isotope is available) and silicon's inherently weak spin-orbit interaction,
which isolates information stored in the electron spin from charge
movement and other noise. Quantum dots, in addition, allow for ready
tunability and alignment of the confined electron (for physical transport,
computation via qubit-qubit coupling, readout, initialization) \cite{ART-Friesen-PracticalQDQCinSi-2002},
the potential for a fabrication route with present-day lithographic
techniques, and the enabling of a fast, DC-controlled two-qubit gate
based on Heisenberg exchange \cite{ART-Loss/DiVincenzo-QCinQDs-1998}.
The goal of designing and constructing quantum computers based on
quantum dots requires characterization of all their physically relevant
properties. For this, it is necessary to have a full toolbox of experimental
diagnostics - in this case, one electron excited-state lifetime measurements
as a function of external parameters such as temperature and magnetic
field, and to be able to interpret these measurements in the light
of theory.

In this paper we present calculations of the dominant spin relaxation
processes in ideal silicon quantum dot spin qubits \cite{THESIS-Tahan}
along with calculations for orbital and valley relaxation. Our particular
emphasis is decay due to phonon coupling, since this mechanism is
believed to be responsible for $T_{1}$ across wide parameter regimes
relevant to quantum information applications. We will indicate in
detail how it can be distinguished experimentally from other mechanisms.

A full, experimentally verified theory of the energy relaxation processes
of the excited electronic states in silicon quantum dots is important
for several reasons. First, it accomplishes a major step on the experimental
path to determining the quantum coherence times of isolated spins
in silicon (a preeminent goal in the verification of a qubit). Second,
it helps corroborate our understanding of the material system and
better characterizes the device under scrutiny; an example would be
transport spectroscopy of nearby energy levels and their line widths.
Third, it provides necessary parameters needed for the design of future
experiments, systems, and architectures. Indeed, if the dominant spin
qubit relaxation mechanisms are as we predict, we can not only validate
the $T_{1}$ lifetimes of silicon qubits, but also retrieve the relative
magnitudes of the dominant spin-orbit coupling contributions inherent
in the device (important for both silicon quantum computing and silicon
spintronics applications), as well as the nature and energies of the
various states above the two spin-qubit states. Finally, the lifetimes
of excited orbital states are relevant to optical pumping schemes,
many-phonon decoherence calculations, transport spectroscopy, beyond
single-spin qubit implementations \cite{ART-Smelyanskiy-2003,ART-Shi-Coppersmith-2011},
and other areas of quantum control.

It has long been known that localized spins in silicon can have exceedingly
long lifetimes, even at relatively high temperatures ($>$ 2K) \cite{ART-FeherGere-1959}.
Theoretical predictions of spin decoherence times are notoriously
difficult as many mechanisms can relax a spin, even in the more robust
case of direct energy relaxation, or $T_{1}$ processes, where a quantum
of energy is lost to the environment. In silicon, for example, energy
relaxation processes may depend on the many-valley nature of the conduction
band electrons; neglecting this effect leads to incorrect predictions
(to many orders of magnitude) \cite{ART-Pines-SpinRelaxationInSilicon-1957,ART-Abrahams-DonorElectronSpinRelaxationInSilicon-1957}.
The key realization came in 1960 from Roth \cite{ART-Roth-gFactorandSpinLatticeRelaxationinGeandSi-1960,ART-Roth-SpinLatticeInteractionElectrons-}
and, independently, Hasegawa \cite{ART-Hasegawa-SpinLatticeRelaxationinSiGe-1960}
- that spin mixing to the (1s-like) valley manifold states explains
the {}``fast\textquotedblright{} relaxation observed for donor electrons.
Soon after, \textit{\emph{Feher, Gere, Wilson }}\textit{et al. }\cite{ART-FeherGere-1959,ART-Wilson/Feher-ESR3-1961}
and Castner \cite{ART-Castner-RamanSpinLatticeinSi-1963,ART-Castner-OrbachSpinLatticeinSi-1967}
thoroughly fleshed out the experiments and theories of donor state
lifetimes in silicon. Castner was the first to calculate relaxation
across different valley donor states. This body of work was the basis
of some of the first proposals for electron spin qubits in silicon
as a basis for a quantum computer \cite{ART-Kane-QC}. This reinvigoration
of the field has led to the extension of these original relaxation
theories to new nanostructures like donors and quantum dots in strained
silicon \cite{ART-Tahan-DecoherenceinSiQCs-2002,THESIS-Tahan,ART-Glavin/Kim}
and in III-V quantum dots \cite{ART-Khaetskii-SpinRelaxationInDots-1999,ART-Khaetskii-ZeemanFlipInDots-2001}.

The region of qubit interest in our case refers to a spin qubit with
finite magnetic field, well below any degeneracy with higher orbital
or valley states. We can summarize the key results of this paper and
prior work on qubit relaxation relevant to qubit and quantum computer
design in silicon quantum dots as follows:
\begin{enumerate}
\item We are concerned with the lifetimes of excited states of a 0-dimensional
(0D) localized electron in silicon. The relaxation of 1D and 2D mobile
electron spins has been investigated for spintronics applications.
This has led to some misconceptions about electron spin relaxation
in 1D and 2D vs. 0D. They are in fact very different. For mobile electrons,
scattering plays the key role \cite{ART-Glazov-2DEGspin,ART-Tahan-2deg},
and spin rotation between or during scattering events is the driver
of loss of spin memory\ (D'yakonov-Perel and Elliott-Yafet effects
\cite{BOOK-Optical-Orientation}). This normally leads to spin lifetimes
on the order of microseconds in silicon \cite{ART-Tryshkin/Lyon-ESRof2DEGs-2003,ART-Tahan-2deg}.
Since scattering is not an issue in 0D, this is an unjustified worry
for spin qubits for silicon quantum computers. However, spin lifetime
measurements in silicon quantum wells can help determine relevant
spin-orbit coupling parameters needed for quantum spin relaxation
calculations.
\item Spontaneous emission of a phonon is the dominant mechanism determining
the spin-flip time, $T_{1}$ ($T_{1}$ being the characteristic time
for true energy relaxation to the environment via a phonon or photon),
at temperatures around 100 mK. Photon emission is negligible because
of the much lower density of final photon states. Phonon emission
accompanied by a spin-flip occurs due to spin-orbit mixing within
the crystal. Of the ``bulk'' Roth \cite{ART-Roth-gFactorandSpinLatticeRelaxationinGeandSi-1960}
and Hasagawa \cite{ART-Hasegawa-SpinLatticeRelaxationinSiGe-1960}
mechanisms that are relevant for donors in unstrained silicon, the
``valley repopulation\textquotedblright{} mechanism disappears with
increasing {[}001{]} strain as is common in silicon quantum wells
\cite{ART-Tahan-DecoherenceinSiQCs-2002}. The ``one-valley\textquotedblright{}
mechanism goes to zero if the magnetic field is parallel to one of
the three crystallographic axes, and goes as $B^{5}$ for other directions
\cite{ART-Glavin/Kim}. 
\item The effect of germanium in a SiGe QD heterostructure is not significantly
detrimental to relaxation times for growth-typical Ge concentrations
\cite{ART-Tahan-DecoherenceinSiQCs-2002} (in the virtual crystal
approximation).
\item The spin-orbit coupling (SOC) in lateral Si quantum dots comes predominantly
from structural inversion asymmetry (SIA) and symmetry-breaking due
to interface effects leading to both Rashba-like and Dresselhaus-like
SOC terms \cite{BOOK-Optical-Orientation,ART-Ivchenko-2006,ART-Ivchenko-2008,ART-Prada-spin-orbitSiGe}.
(The relative magnitude of Dresselhaus-like and Rashba-like SOC in
silicon quantum wells or dots has yet to be verified experimentally,
let alone systematically across samples.) Each term gives a characteristic
magnetic field anisotropy in $T_{1}.$ Overall, $1/T_{1}$ is proportional
to the seventh power of the magnetic field $B$ \cite{THESIS-Tahan}.
This contrasts sharply with GaAs quantum dots and Si donor states.
For these two cases $1/T_{1}\propto B^{5}$ (though for different
reasons). The ratio of the Rashba-like and Dresselhaus-like terms
are expected to be sample dependent since in silicon they are due
solely to interface effects (silicon, unlike III-Vs, has no bulk-inversion
asymmetry (BIA)).
\item Direct coherent rotations due to nearby spins in the semiconductor
are possible. \ 

\begin{enumerate}
\item At zero and low magnetic fields, direct dipole-dipole magnetic coupling
with the central electron qubit and other electrons in the environment
can occur. \ These rotations are technically coherent processes,
but they result in spin flips that change the longitudinal component
of the qubit magnetization and thus appear like $T_{1}$ processes.
These processes do not depend on $B$ and can set upper limits on
observed $T_{1}$ times. \ The strength of this interaction is reduced
as the inhomogeneity of the electron line widths increases, though
even one electron spin exactly at resonance 200 nm away can cause
200 ms effective $T_{1}$ lifetimes. A full theory is beyond the scope
of this paper.
\item At zero fields, direct electron qubit - nuclei flip-flops are possible
leading to $T_{1}$-like rotation. At finite fields this mechanism
is suppressed due to the mismatch of the electrons and nuclei respective
g-factors.
\item These mechanisms may in some cases be corrected via spin-echo techniques
or suppressed by freezing out the background spins 
\end{enumerate}
\item Other mechanisms for longitudinal spin relaxation, $T_{1}$, such
as hyperfine coupling to Si$^{29}$nuclei in natural Si, 1/f noise,
and Johnson noise, are estimated to be small in the parameter ranges considered
here, though further work is required to verify this. The magnetic field dependence of these effects allows them to
be distinguished from spin-phonon coupling.
\item The phonon bottleneck effect operates rather weakly in the parameter
regime of interest for quantum dot applications. It can be calculated
in a theory that goes beyond the electric dipole approximation; the
result is only a slight enhancement of $T_{1}.$
\item Orbital relaxation in strained silicon is much faster than in bulk
silicon in some important cases. Surprisingly, the rate of spontaneous
decay from the first excited orbital state in silicon quantum dots
can be comparable to that of GaAs quantum dots, which are commonly
expected to relax more efficiently due to that crystal's piezoelectric
nature. This has important implications for optically-induced motional
spin-charge transduction (readout) and optical pumping (initialization)
of spin qubits \cite{ART-Friesen-ReadoutLetter-2003}, making the
former harder and the latter easier, as well as for excited state
spectroscopy.
\item Excited valley state relaxation can be slow in silicon due to a small
matrix element connecting valley states of different symmetry; this
leads to the hope of valley qubits \cite{talk-valley-march,talk-amsterdam-03,ART-Smelyanskiy-2003}.
The phonon emission rate has a maximum at the Umklapp phonon energy
(11 meV and 23 meV for transverse and longitudinal phonons in silicon)
that connects valley minima from one Brillouin zone to its nearest
neighbor \cite{ART-Castner-RamanSpinLatticeinSi-1963}. This leads
to---at the longest---nanosecond relaxation times \cite{ART-Soykal-PRL-2012}
in P donors with their large valley splitting (\textasciitilde{}10
meV), but in quantum dots is suppressed due to a large energy mismatch.
However, the valley index in general cannot be considered a good quantum number
in quantum dots due to large valley-orbit mixing \cite{ART-friesen-valley-orbit}.
For the same reason, ``valley relaxation\textquotedblright{} in realistic
devices can be dominated by orbital relaxation due to mixing with
nearby orbital levels. The situation is different and more favorable
in some cases such as for Li donors 
\cite{ART-Smelyanskiy-2003,ART-Soykal-PRL-2012}.
\item Non-ideal interfaces in silicon quantum dots may effect the relaxation
processes; these effects are microscopic in origin \cite{ART-Ando-Propertiesof2D-1982,ART-Chutia-Valley,ART-friesen-valley-orbit}
and are not considered quantitatively in this paper.
\end{enumerate}
This content is arranged as follows. We begin with an introduction
to the single electron states in silicon quantum dots typical of heterostructures
used for qubits today. We follow with a discussion of the electron-phonon
interaction, the dominant relaxation mechanism in these devices. We
then use that theory to calculate the orbital relaxation (no spin
flip) of low-lying excited states to states having the same valley
index. \ We do this first using the electric-dipole (ED) approximation
and then including all multipoles. \ This gives useful numbers for
excited orbital state lifetimes as well as a quantitative idea of
the extent of validity of the ED approximation in further calculations.
Then the main subject of this paper is tackled, namely the spin flip
mechanisms relevant to quantum dots. In this section, we review and
adapt the previously known {}``bulk\textquotedblright{} spin-flip
mechanisms to the quantum dot case. We then consider new mechanisms
due to structural inversion asymmetry which give the dominant spin
relaxation mechanisms for most of the magnetic field range. \ This
is followed by comparisons with spin relaxation mechanisms (noise,
nuclei) and valley relaxation, which completes the narrative of excited
lifetimes in single electron quantum dots relevant to quantum computing.
Where possible we compare the results for quantum dots with those
for P donors in silicon and GaAs quantum dots, both of which are relevant
reference systems.

The final section summarizes the relation of theory to experiment.
It is difficult to make sharp predictions for the absolute magnitude
of $T_{1}$ because of strong dependences on quantities that are usually
somewhat uncertain, particularly the dot size, as well as a reliance
on bulk material constants which may vary in nanostructures. We show
how to overcome this problem by combining measurements of qubit spin
lifetimes with measurements of excited orbital state energies and
lifetimes. It is precisely for this reason that we deal in such detail
with the excited states.

\section{Silicon Quantum Dot states\label{sec:Silicon-Quantum-Dot}}

This work is concerned mainly with lateral quantum dots formed by
heterostructure confinement in the growth ($z$) direction and lateral
confinement by metallic top gates. Figure 1 shows some of the heterostructure
choices possible in constructing these dots, from modulation doped
two-dimensional electron gas (2DEG) structures depleted by top gates
to accumulation mode inversion layer devices in MOSFET-like structures.
We will concern ourselves here predominately with the SiGe QW QD case,
although our considerations should carry over to Si MOSFET dot structures
as well.

In semiconductors, the electron wave functions are superpositions
of Bloch states at the bottom of the conduction band (CB), so the
indirect band-gap, ``many-valley'' nature of silicon (as opposed
to a single $\Gamma$-valley in GaAs) takes on great importance. In
a biaxially-strained silicon QW, the number of states is doubled relative
to GaAs, but reduced from the 6-fold degeneracy of electrons in the
bulk. The conduction band (CB) minima located at $\boldsymbol{k}=\left(0,0,\pm k_{0}\right)$
with $k_{0}=0.85k_{max}$ have band energies lower than the minima
at $\boldsymbol{k}=\left(\pm k_{0},0,0\right)$ and $\boldsymbol{k}=\left(0,\pm k_{0},0\right)$
by about 0.1 to 0.15 eV at typical strain values (20-30\% Ge in the
barriers) \cite{ART-schaffler-SIGE}. Thus the $\boldsymbol{k}=\left(0,0,\pm k_{0}\right)$
valleys are populated but not the $\boldsymbol{k}=\left(0,\pm k_{0},0\right)$
and $\boldsymbol{k}=\left(\pm k_{0},0,0\right)$ valleys \cite{BOOK-YuCardona}.
A similar splitting is at work in Si MOSFET-type structures, though
here the physical origin of the lifting of the degeneracy is due to
anisotropy of the silicon effective mass (and in some cases local
strain). In the absence of magnetic fields and valley-splitting effects,
the electronic ground state in this system is four-fold degenerate
(spin and valley). When potentials or boundary conditions that mix
the two valleys are present, as they always will be to some extent,
there will be excited valley states corresponding to different linear
combinations of the valley minima. So each valley state has its own
identical set of orbital and spin states and an additional quantum
number is needed to specify which valley state the electron occupies.

A magnetic field splits the degeneracy of the spin states. The valley
degeneracy is split by the hard confinement of the QW interfaces (or
the impurity potential in that case) and is influenced by a number
of parameters, especially the magnitude of the electric field in the
growth direction and the sharpness of the confining potential. For
the moment, we assume for simplicity that any static magnetic field
is small or directed in the plane of the QW (so that the orbital functions
are unperturbed) and that the well walls, located at $z=\{0,d\},$
are smooth. In these circumstances, we may write 
\[
\psi_{m}^{(i)}=F_{m}^{(i)}(x,y)F_{m}^{(i)}(z)\left[\alpha_{+z}^{(i)}u_{+z}(\mathbf{r})e^{ik_{z}z}+\alpha_{-z}^{(i)}u_{-z}(\mathbf{r})e^{-ik_{z}z}\right].
\]
 $F_{m}^{(i)}(z)$ is the envelope function obtained by solving the
confinement problem in the effective mass approximation for the $m$-th
orbital, $\alpha_{\pm z}^{(i)}$ are the coefficients weighting the
two valleys for the $i$-th valley state ($i=1,2$ for strained silicon,
$i=1,2,3,4,5,6$ for bulk silicon), and $u_{\pm z}(\mathbf{r})$ are
the lattice-periodic parts of the Bloch functions, $u_{j}(\mathbf{r})\exp(i\mathbf{k}_{j}\cdot\mathbf{r})$,
at the conduction band minimum $\mathbf{k}_{j}$. $\alpha_{z}^{(i=+)}=\alpha_{-z}^{(+)}=1/\sqrt{2}$
for the symmetric valley (``sin-like'') state and $\alpha_{z}^{(i=-)}=-\alpha_{-z}^{(-)}=1/\sqrt{2}$
for the antisymmetric valley (``cosine-like'') state. It is often
convenient to expand the Bloch functions into a sum, 
\[
u(\mathbf{r})=\sum_{G}C_{\mathbf{G}}\exp[i\mathbf{r}\cdot\mathbf{G}],
\]
where $C_{\mathbf{G}}$ weight the Fourier components of expansion
(independent of $\mathbf{r}$) and $\mathbf{G}$ are the reciprocal
lattice vectors. The wave functions for a more realistic device, calculated
in the tight-binding theory of Ref.\cite{ART-Boykin-ValleySplitting-2003},
are shown in Figure \ref{fig:Results-from-1D} (where the ``Kohn-Luttinger''
oscillations are evident but the lattice periodic oscillations are
not included). A donor vs. dot energy level comparison is shown in
Figure \ref{fig:The-circled-numbers}.

Until now we have only concerned ourselves with ``perfect\textquotedblright{}(completely
flat) interfaces. In these cases the valley and orbital states are
well defined - valley index being a generally good quantum number
- much like the isolated donor case. In reality most interfaces are
imperfect; they have alloy disorder (Si vs. Ge atoms), steps due to
miscut, steps due to growth layer formation, and even interface states
and traps (especially with respect to Si/SiO$_{2}$ interfaces). This
leads to mixing of the valley and orbital wave functions as well as
diminishing of the valley splitting (due to interference) \cite{ART-Chutia-Valley,ART-friesen-valley-orbit},
and as such, wave functions and splittings that vary from device to
device and dot to dot. These microscopic variations are not considered
in this paper. Fortunately, these effects are often not large corrections
to the calculations below, as the experimentally accessible energy
splittings come into the equations at a much higher power than the
relevant (experimentally inaccessible) matrix elements. As we go we
will point out differences from our theory for the ideal and likely
situations, focusing on experimentally accessible signatures.

\begin{figure}[ptb]

\begin{centering}
\includegraphics[scale=0.4]{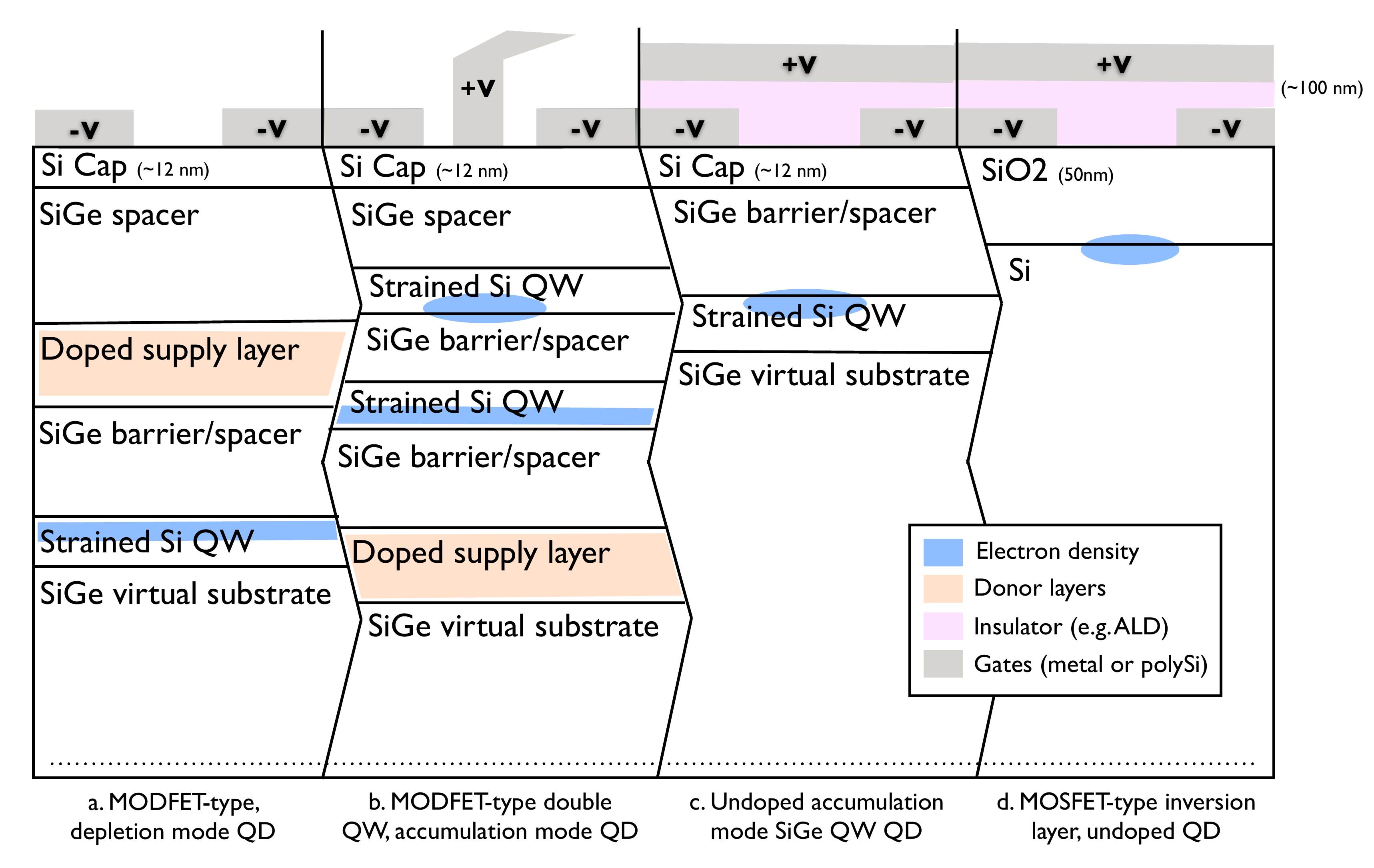} 
\par\end{centering}

\caption{\selectlanguage{american}%
Different types of quantum dot heterostructures relevant to the theory
in this paper.\foreignlanguage{english}{ }\selectlanguage{english}%
}
\end{figure}

\section{Electron-Phonon Interaction}

Phonons are quanta of lattice vibrations, that is, mechanical motion
that gives time-dependent stress. This alters the band structure by
shifting band energies and lifting degeneracies \cite{BOOK-ridley-quantumsemi,BOOK-YuCardona}.
It is typically assumed that this effect does not change the band
curvature (effective masses) but does shift the energy states of interest.
The shift in energy of the band edge per unit elastic strain is called
the \emph{deformation potential} and is common to all semiconductors
and solids. In polar crystals, distortion of the lattice can also
create large internal electric fields which affect the electron. This
is the piezoelectric interaction. Ionic crystals like GaAs suffer
from piezo-phonons, which are often very efficient at electron scattering;
silicon, being non-polar and centrosymmetric, has none. While optical
strain also exists in materials with two atoms per unit cell, such
as silicon, optical phonons in silicon have a narrow bandwidth centered
at a much higher energy than a quantum computer would operate. We
briefly review the theory of the electron-phonon deformation potential
interaction outlined by Herring and Vogt \cite{ART-Herring/Vogt-Deformation-Theory-1955}.
\begin{table}[ptb]

\begin{centering}
\begin{tabular}{cc}
\hline 
Constant  & Value\tabularnewline
\hline 
\hline 
$e$  & $1.6\times10^{-19}$ C\tabularnewline
\hline 
$\hbar$  & $1.05\times10^{-34}$ J s\tabularnewline
\hline 
$c$  & $3\times10^{8}$ m/s\tabularnewline
\hline 
$\epsilon_{0}$  & $8.85\times10^{-12}$ C$^{2}$/Nm$^{2}$ \tabularnewline
\hline 
$\epsilon_{Si}$  & 11.8\tabularnewline
\hline 
$g_{\Vert}$(Si)  & 1.999\tabularnewline
\hline 
$g_{\perp}$(Si)  & 1.998\tabularnewline
\hline 
$\Xi_{u}$(Si)  & 8.77 eV=1.4$\times10^{-18}$ J\tabularnewline
\hline 
$\Xi_{d}$(Si)  & 5 eV=8$\times10^{-19}$ J\tabularnewline
\hline 
$\rho$(Si)  & 2330 kg/m$^{3}$\tabularnewline
\hline 
$v_{l}$(Si)  & 9330 m/s\tabularnewline
\hline 
$v_{t}$(Si)  & 5420 m/s\tabularnewline
\hline 
$a_{0}$(Si) & 0.543 nm\tabularnewline
\hline 
$k_{max}$(Si) & $2\pi/a_{0}$\tabularnewline
\hline 
$k_{B}$  & 1.38$\times10^{-23}$ J/K\tabularnewline
\hline 
\end{tabular}
\par\end{centering}

\caption{\selectlanguage{american}%
Physical constants and materials parameters for bulk silicon.  \label{tab:Physical-constants-and}\foreignlanguage{english}{ }\selectlanguage{english}%
}

\label{tab: physical}
\end{table}

The energy shift of a non-degenerate band edge due to strain is given
by 
\begin{equation}
H_{eL}=\sum_{\alpha,\beta}U_{\alpha\beta}\Xi_{\alpha\beta}^{(i)},\label{eq:eL-Hamiltonion}
\end{equation}
 where 
\[
U_{\alpha\beta}=\frac{1}{2}\left(\frac{\partial u_{i}}{\partial x_{i}}+\frac{\partial u_{j}}{\partial x_{i}}\right)
\]
 is the strain tensor and $\mathbf{\Xi}^{(i)}$ is the deformation
potential tensor for the $i$th silicon CB valley. Since our electron
is confined to massively strained {[}001{]} silicon, we need only
include the $\left(0,0,\pm k_{0}\right)$ valleys which have deformation
potential tensors given by 
\begin{equation}
\mathbf{\Xi}^{\pm z}=\Xi_{d}\delta_{\alpha\beta}+\Xi_{u}K_{\alpha}^{(i)}K_{\beta}^{(i)}=\left(\begin{array}{ccc}
\Xi_{d} & 0 & 0\\
0 & \Xi_{d} & 0\\
0 & 0 & \Xi_{d}+\Xi_{u}
\end{array}\right),\label{eq:deformationMatrix}
\end{equation}
 where $\Xi_{d}$ relates to pure dilatation and $\Xi_{u}$ is associated
with shear strains. $\widehat{\mathbf{K}}^{(i)}$ is a unit vector
in the direction of the $i$th valley. For silicon under compressive
stress along {[}001{]}, opposing valleys move in energy identically.
In order to associate phonon modes with strain, we can expand the
unit cell displacement $\mathbf{u(r)}$ in plane waves, 
\[
\mathbf{u}(\mathbf{r})=\sum_{\boldsymbol{q}\lambda}\left[\mathbf{e}\left(\mathbf{q},\lambda\right)a_{\boldsymbol{q}\lambda}e^{i\mathbf{q}\cdot\mathbf{r}}+\mathbf{e}^{\ast}\left(\mathbf{q},\lambda\right)a_{\boldsymbol{q}\lambda}^{\ast}e^{-i\mathbf{q}\cdot\mathbf{r}}\right].
\]
 $a_{\boldsymbol{q}\lambda}\mathbf{\ }$destroys a phonon with wavevector
q and polarization $\lambda$ (2 transverse and 1 longitudinal; see
details in Table 2) of a phonon; $\mathbf{e}\left(\mathbf{q},\lambda\right)$
is its unit displacement vector. \ This results in a strain tensor
due to a phonon of 
\begin{equation}
U(\mathbf{q},\lambda)_{\alpha\beta}=\frac{i}{2}\left[(\mathbf{e}_{\alpha}\left(\mathbf{q},\lambda\right)q_{\beta}+\mathbf{e}_{\beta}\left(\mathbf{q},\lambda\right)q_{\alpha})a_{\boldsymbol{q}\lambda}^{\ast}e^{-i\mathbf{q}\cdot\mathbf{r}}+(\mathbf{e}_{\alpha}\left(\mathbf{q},\lambda\right)q_{\beta}+\mathbf{e}_{\beta}\left(\mathbf{q},\lambda\right)q_{\alpha})a_{\boldsymbol{q}\lambda}e^{i\mathbf{q}\cdot\mathbf{r}}\right].\label{eq:phononStrainTensor}
\end{equation}
 The operators $a_{\boldsymbol{q}\lambda}$ and $a_{\boldsymbol{q}\lambda}^{\ast}$
have matrix elements
\begin{align*}
\left\langle n_{\boldsymbol{q}\lambda}-1\right\vert a_{\boldsymbol{q}\lambda}\left\vert n_{\boldsymbol{q}\lambda}\right\rangle  & =\sqrt{\hbar n_{\boldsymbol{q}\lambda}/2M_{c}\omega_{\boldsymbol{q}\lambda}},\\
\left\langle n_{\boldsymbol{q}\lambda}+1\right\vert a_{\boldsymbol{q}\lambda}^{\ast}\left\vert n_{\boldsymbol{q}\lambda}\right\rangle  & =\sqrt{\hbar\left(n_{\boldsymbol{q}\lambda}+1\right)/2M_{c}\omega_{\boldsymbol{q}\lambda}},
\end{align*}
 $M_{c}$ is the mass of the crystal and $n_{\boldsymbol{q}\lambda}=1/\left(e^{\hbar\omega_{\mathbf{q}\lambda}/kT}-1\right)$
is the phonon occupation number of the mode with wave number $\mathbf{q}$
and polarization $\lambda$. The complete electron-phonon Hamiltonian
must be summed over phonon modes and polarizations. For a {[}001{]}
strained-silicon quantum well, it can be written succinctly as 
\begin{equation}
H_{ep}=\sum_{\lambda=1}^{3}\sum_{\mathbf{q}}iq~\left[a_{\boldsymbol{q}\lambda}^{\ast}e^{-i\mathbf{q}\cdot\mathbf{r}}+a_{\boldsymbol{q}\lambda}e^{i\mathbf{q}\cdot\mathbf{r}}\right]\times\left[\Xi_{d}\mathbf{e}_{x}\left(\mathbf{q},\lambda\right)\hat{q}_{x}+\Xi_{d}\mathbf{e}_{y}\left(\mathbf{q},\lambda\right)\hat{q}_{y}+\left(\Xi_{d}+\Xi_{u}\right)\mathbf{e}_{z}\left(\mathbf{q},\lambda\right)\hat{q}_{z}\right].\label{eq:Helectronphonon}
\end{equation}

We are especially concerned with the anisotropic effects due to the
massive strain of the system in question. As can be seen from Eq.
\ref{eq:deformationMatrix} and the deformation constant values in
Table \ref{tab:Physical-constants-and}, the shift in energy of a
specific valley due to an acoustic phonon is very anisotropic. In
the case of bulk Si, the six conduction band minima are equidistant
from the $\Gamma$-point and thus form an isotropic response to phonon
deformations. This means essentially that transverse phonons will
not contribute to the relaxation times for intervalley transitions
of the same symmetry (that is, $\alpha_{n}^{(i)}=\alpha_{n}^{(j)}$
for initial state $i$ and final state $j$). Another way to see this
is to consider the electron-lattice matrix element between different
plane wave states at the same minimum (Equation 3.29 of Ref. \cite{ART-Hasegawa-SpinLatticeRelaxationinSiGe-1960}),
\[
\left\langle \psi_{m}^{(i)}\right\vert H_{eL}\left\vert \psi_{n}^{(i)}\right\rangle _{\boldsymbol{q}t}=a_{\mathbf{q}t}\left[i\mathbf{e}(\mathbf{q},t)\cdot\Xi^{(i)}\cdot\mathbf{q}\right]f_{mn}^{(i)}(\mathbf{q})+c.c.
\]
 where $f_{mn}^{(i)}(\mathbf{q})=\int F_{m}^{(i)}(\mathbf{r})e^{i\mathbf{q}\cdot\mathbf{r}}F_{n}^{(i)}d\mathbf{r}$.
\ We have used the polarization index $t$ to indicate a transverse
phonon. \ The matrix element between two dot wave functions is then
\[
\left\langle \sum\alpha_{m}^{(i)}\psi_{m}^{(i)}\right\vert H_{eL}\left\vert \sum\alpha_{n}^{(i)}\psi_{n}^{(i)}\right\rangle _{\boldsymbol{q}t}=a_{\boldsymbol{q}t}\left[i\mathbf{e}\left(\mathbf{q},t\right)\cdot\sum\alpha_{m}^{(i)}\alpha_{n}^{(i)}\Xi^{(i)}\cdot\mathbf{q}\right]f_{mn}^{(i)}(\mathbf{q})+c.c.
\]
 It's easy to see from the above equation that if $\sum\alpha_{m}^{(i)}\alpha_{n}^{(i)}\Xi^{(i)}$
is proportional to the identity matrix (assuming $\alpha_{m}=\alpha_{n}=1$),
then the transverse phonon matrix elements must be zero since $\mathbf{e}_{t1}(\mathbf{q},t1)\perp\mathbf{e}_{t2}(\mathbf{q},t2)\perp\mathbf{q}$.
The point is that in strained silicon, only the $\pm z$ minima are
occupied so unlike the bulk silicon case, transverse phonons will
contribute. This turns out to be very important in relaxation calculations,
as will be seen below. 
\begin{table}[ptb]

\begin{centering}
\begin{tabular}{|c|c|c|c|}
\hline 
 & Longitudinal $(s=l)$  & Transverse $(s=t_{1})$  & Transverse $(s=t_{2})$\tabularnewline
\hline 
\hline 
$e_{x}$  & $\sin\theta\cos\phi$  & $\sin\phi$  & $-\cos\theta\cos\phi$\tabularnewline
\hline 
$e_{y}$  & $\sin\theta\sin\phi$  & $-\cos\phi$  & $-\cos\theta\sin\phi$\tabularnewline
\hline 
$e_{z}$  & $\cos\theta$  & 0  & $\sin\theta$\tabularnewline
\hline 
\end{tabular}
\par\end{centering}

\caption{\selectlanguage{american}%
Polarization components.  \foreignlanguage{english}{ }\selectlanguage{english}%
}

\label{cap:Polarization-components.}
\end{table}

\begin{figure}[ptb]
\begin{centering}
\includegraphics{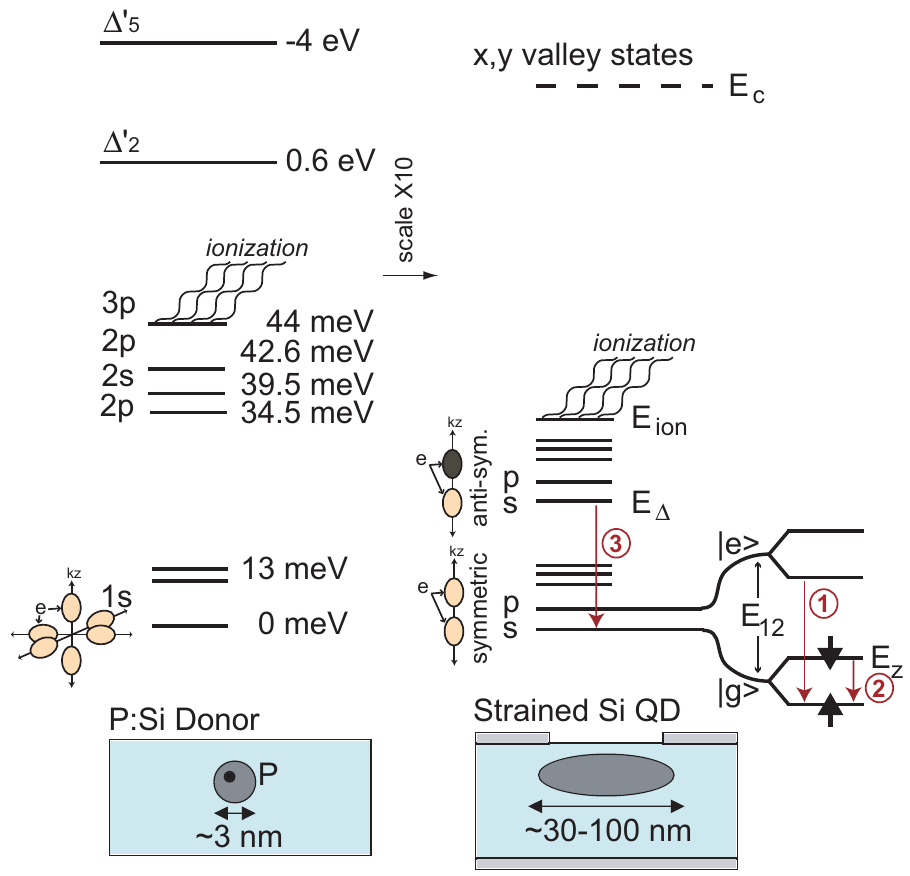} 
\par\end{centering}

\caption{\selectlanguage{american}%
Energy level diagram for P donors in Si (left) and Si quantum dots
(right). Note the reduction in energy scale from left to right. In
both cases, the electron exists as a superposition of different conduction
valley minima within the crystal. Different low-lying dot levels (all
$s$-like) result from the sharp donor potential and different amplitudes
in the six valleys shown. Low-lying dot states arise from different
orbital wave functions with a further valley splitting (far right)
between symmetric and anti-symmetric combinations of the two valleys,
$\pm z$. The circled numbers represent relevant relaxation processes:
(1) orbital relaxation across the first energy gap; (2) spin relaxation
of the electron qubit; and (3) valley relaxation. Note that the valley
splitting can vary from 0 to meV and may fall within the energy levels
of the low-lying Zeeman and orbital states.\foreignlanguage{english}{
\label{fig:The-circled-numbers}}\selectlanguage{english}%
}
\end{figure}

\section{Orbital Relaxation in Strained Si Quantum Dots}

The line widths and characteristic behavior (dependence on magnetic
field, etc.) of the lowest lying excited states in quantum dot systems
are usually very relevant for characterizing a spin-based qubit. We
begin by considering the relaxation of an excited state that involves
no spin flip and takes place within the same valley state (assuming
for now no valley-orbit mixing). Spontaneous emission of a single
acoustic phonon is the dominant relaxation mechanism for excited electronic
states in Si at QC temperatures ($<$100 mK) (an ansatz guided by
empirical evidence for silicon donors \cite{ART-Feher-1959,ART-FeherGere-1959}
and GaAs quantum dots \cite{ART-Hanson-RevModQDs}). The deformation
potential approach \cite{ART-Herring/Vogt-Deformation-Theory-1955}
is easily modified to include strain \cite{ART-Tahan-DecoherenceinSiQCs-2002}.
The phonon-induced energy shift is very anisotropic for a silicon
conduction band valley. Because of this the results for strained silicon
are quite different from those of bulk silicon \cite{ART-Abrahams-DonorElectronSpinRelaxationInSilicon-1957}.

In the bulk, the six conduction band minima are equidistant from the
$\Gamma$-point and thus form an isotropic response to phonon deformations
(specifically the case where $\alpha_{i}=1$ for all $i$ across both
states). So the angular integral over transverse phonons averages
to zero. For example, the transition from the ground, 1s-like symmetric
state to the 2p-like, symmetric state in bulk silicon has no transverse
phonon contribution \cite{ART-Abrahams-DonorElectronSpinRelaxationInSilicon-1957}.
The same transition in strained silicon does have such a contribution,
because the cubic symmetry of the six minima has been broken by strain,
and indeed it is the largest term. This greatly increases the relaxation
rate since the inverse sound velocity comes into the rate equations
with a very high power as we will now show.

Fermi's Golden Rule, between states of arbitrary spin, 
\begin{equation}
\Gamma=\frac{2\pi}{\hbar}~\left\vert \left\langle nsv\right\vert H_{ep}\left\vert ms^{\prime}v^{\prime}\right\rangle \right\vert ^{2}~\delta\left(E_{ph}-E_{ns,ms^{\prime}}\right),\label{eq:GoldenRuleOrbital}
\end{equation}
 is the basis for our phonon relaxation rate calculations. Here, $\left\vert nsv\right\rangle $
is the state $\psi_{ns}^{(v)}$ of the electron in the dot on level
$n$ with spin state $s$ and on valley manifold $v$, including the
effect of a magnetic field\emph{.} We assume an isotropic phonon spectrum
such that the energy of the phonon is $E_{ph}=\hbar\omega_{\mathbf{q}\lambda}$,
where $\omega_{\mathbf{q}\lambda}=v_{\lambda}\left\vert \mathbf{q}\right\vert $
and $v_{\lambda}$ is the velocity of the mode $\lambda.$ Setting
$s=s^{\prime}$ for orbital relaxation without a spin-flip and $v=v^{\prime}$,
$E_{ns,ms^{\prime}}=E_{mn}$ is the energy splitting between states
$m$ and $n$, $H_{ep}$ is the electron-phonon interaction of Eq.
\ref{eq:Helectronphonon}. Summing over phonon modes, using $\hat{q}=(\sin\theta\cos\phi,\,\sin\theta\sin\phi,\,\cos\theta)$
and $\hat{e}_{l}=\hat{q}\perp\hat{e}_{t1}\perp\hat{e}_{t2}$ (see
Table \ref{cap:Polarization-components.}) for the wave vector and
polarization vectors and using the electric dipole (ED) approximation,
$e^{i\mathbf{q}\cdot\mathbf{r}}\approx1+i\mathbf{q}\cdot\mathbf{r}$,
we find for the phonon-induced relaxation rate, 
\begin{equation}
\Gamma_{mn}^{ED}=\frac{\left\vert E_{mn}\right\vert ^{5}}{\hbar^{6}\pi\rho_{Si}}\left\{ \left(\left\vert M_{x}^{\left(mn\right)}\right\vert ^{2}+\left\vert M_{y}^{\left(mn\right)}\right\vert ^{2}\right)\Upsilon_{xy}+\left\vert M_{z}^{\left(mn\right)}\right\vert ^{2}\Upsilon_{z}\right\} \left(n_{B}\left(E_{mn}\right)+1\right),\label{eq:orbital}
\end{equation}
 where 
\begin{align}
\Upsilon_{xy} & =\frac{35\Xi_{d}^{2}+14\Xi_{d}\Xi_{u}+3\Xi_{u}^{2}}{210v_{l}^{7}}+\frac{2\Xi_{u}^{2}}{105v_{t}^{7}},\label{eq:upsilon}\\
\Upsilon_{z} & =\frac{35\Xi_{d}^{2}+42\Xi_{d}\Xi_{u}+15\Xi_{u}^{2}}{210v_{l}^{7}}+\frac{\Xi_{u}^{2}}{35v_{t}^{7}},\nonumber 
\end{align}
the matrix elements are $\vec{M}^{\left(mn\right)}=\langle F_{m}|\vec{r}|F_{n}\rangle$,
$\rho_{Si}$ is the mass density of Si, $E_{mn}$ is the energy gap
between orbital states, $\ v_{\ell}$ and $v_{t}$ are the longitudinal
and transverse sound velocities. The single electron envelope functions
can be calculated by solving the Poisson and Schroedinger equations
directly as in Ref. \cite{ART-Friesen-PracticalQDQCinSi-2002} or,
as is normally done, by approximating the potential as a parabola,
giving harmonic oscillator states defined by the fundamental energies
$E_{10}=\hbar\omega_{x,y}$. With the envelope functions $F_{0}=\left(2/\pi\right)^{1/4}x_{0}^{-1/2}\exp\left(-x^{2}/x_{0}^{2}\right)$
and $F_{1}=\left(2/\pi\right)^{1/4}(2/\sqrt{x_{0}^{3}})x\exp\left(-x^{2}/x_{0}^{2}\right)$,
the matrix element is given by 
\begin{equation}
\left\vert M_{x}^{\left(10\right)}\right\vert ^{2}=\frac{\hslash^{2}}{m_{t}E_{10}}=\left(\frac{x_{0}}{2}\right)^{2},\label{eq:mxy}
\end{equation}
where $2\sqrt{\left\langle x^{2}\right\rangle }=\max\{x_{0},y_{0}\}=\sqrt{2}\hbar/\sqrt{m_{t}\Delta}$
is the lateral size of the dot, $L$, and $x_{0}$ and $y_{0}$ are
the dot sizes in the $x$ and $y$ directions. We define $\Delta\equiv E_{1}-E_{0}$.
Thus, the orbital relaxation rate from the lowest orbital state for
a slightly asymmetric (non-degenerate excited state), parabolic dot
is 
\begin{equation}
\Gamma_{\Delta}^{ED}=\frac{2\Xi_{u}^{2}}{105v_{t}^{7}}\frac{\Delta^{4}}{\hbar^{2}\pi\rho_{Si}m_{t}}\left(n_{B}\left(\Delta\right)+1\right)\label{eq:Orbital-Parabolic}
\end{equation}
(we have used the fact that $v_{l}\sim2v_{t}$ to eliminate terms
due to longitudinal phonons). Because $E_{mn}$ appears in the fifth
power (general case) or fourth power (parabolic dots) in Equations
\ref{eq:orbital} and \ref{eq:Orbital-Parabolic} for orbital relaxation,
an accurate value for $E_{mn}$ is much more important than equivalent
accuracy in the wave function matrix elements. The dipole approximation
is valid until roughly $qL>>1$ ($L$ is the maximal linear size of
the dot) when the relaxation rate starts to decrease due to phonon
bottleneck effects. We treat this effect explicitly in the next section.

\section{Phonon Bottleneck Effect\label{sec:Phonon-Bottleneck-Effect}}

When the dot becomes very small ($L$ comparable to a few interatomic
spacings), the relaxation rate is reduced. This is due to the impossibility
of satisfying simultaneously energy and momentum conservation during
an electron-acoustic-phonon scattering event \cite{ART-Benisty-PhononBottle}.
Mathematically, the phonon bottleneck effect is due to the fast oscillating
exponential factor: emission of a phonon with wave vector $\boldsymbol{q}$
is unlikely when $q>2\pi/L$ \cite{ART-Benisty-Weisbuch-PhononBottle-1991}.
Table \ref{cap:Phonon-bottleneck-numbers} charts this transition
for parabolic Si quantum dots. Note that in a lateral quantum dot,
unlike excitonic quantum dots, Auger processes and electron-hole scattering
do not play a role in negating phonon bottleneck. Assuming parabolic
dots, only dots with fundamental energies $\Delta\approx1-2$ meV
are small enough (and virtually impossible to construct with laterally-gated
devices) for phonon-bottleneck effects to make a significant impact
on increasing the orbital relaxation times. 
\begin{table}[ptb]
\begin{centering}
\begin{tabular}{c|c|ccc}
\hline 
$\Delta$ (meV)  & $L$ (nm) - Si  & $\lambda$ (nm) - Si($v_{t}$) & $1/\Gamma_{ED}$ (Eq. \ref{eq:orbital}) & $1/\Gamma_{exact}$ (Eq. \ref{eq:exact_orbital})\tabularnewline
\hline 
0.05  & 127  & 447 & $2.5\times10^{-7}$s  & $4.0\times10^{-7}$ s \tabularnewline
0.1  & 90  & 223 & $1.6\times10^{-8}$ s  & $3.8\times10^{-8}$ s \tabularnewline
0.2 & 63 & 112 & $9.7\times10^{-10}$ s  & $5\times10^{-9}$ s \tabularnewline
0.3 & 52 & 75 & $1.9\times10^{-10}$ s  & $1.8\times10^{-9}$ s \tabularnewline
0.4 & 45 & 56 & $6.0\times10^{-11}$ s & $9.5\times10^{-10}$ s\tabularnewline
0.5  & 40  & 47 & $2.5\times10^{-11}$ s  & $6\times10^{-10}$ s \tabularnewline
1  & 28  & 22 & $1.6\times10^{-12}$ s  & $1.4\times10^{-10}$s \tabularnewline
2  & 20  & 11 & $9.7\times10^{-14}$ s  & $3.8\times10^{-11}$ s \tabularnewline
3  & 16  & 7.4 & $1.9\times10^{-14}$ s  & $1.9\times10^{-11}$ s \tabularnewline
8 & 10 & 2.8 & $3.8\times10^{-16}$ s & $9.6\times10^{-12}$ s\tabularnewline
10  & 9  & 2.2 & $1.6\times10^{-16}$ s & $1.4\times10^{-11}$ s\tabularnewline
\hline 
\end{tabular}
\par\end{centering}

\caption{\selectlanguage{american}%
Characteristic numbers for the phonon bottleneck effect in Si and
orbital relaxation rates. $\Delta$ is the first excitation energy,
$L$ is the lateral dimension, and $\lambda$ is the wavelength of
the transverse phonon with the resonant energy.\label{tab:Characteristic-numbers-for}\selectlanguage{english}%
}
\end{table}
\begin{figure}[ptb]
\begin{centering}
\includegraphics[scale=0.45]{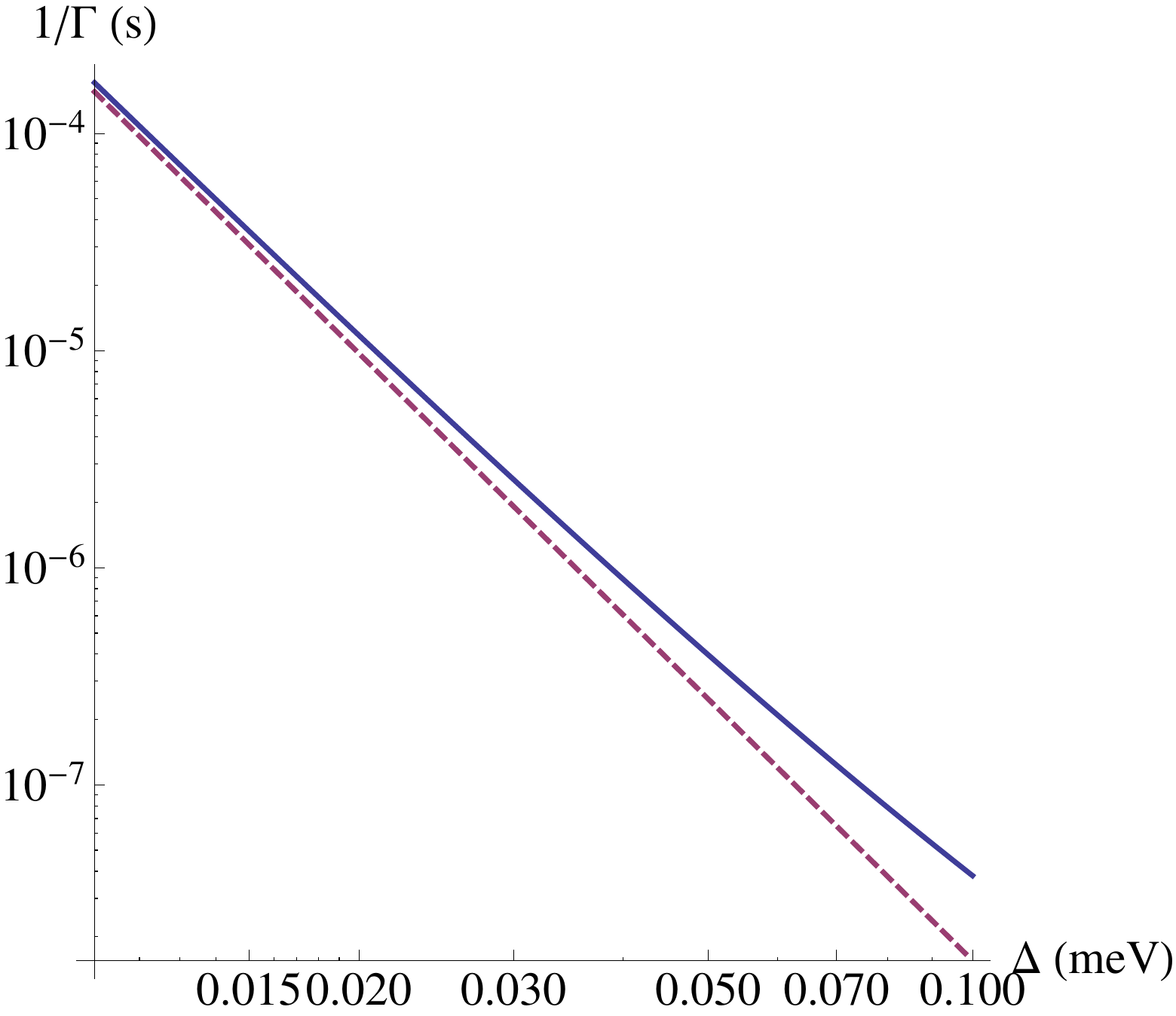}\includegraphics[scale=0.45]{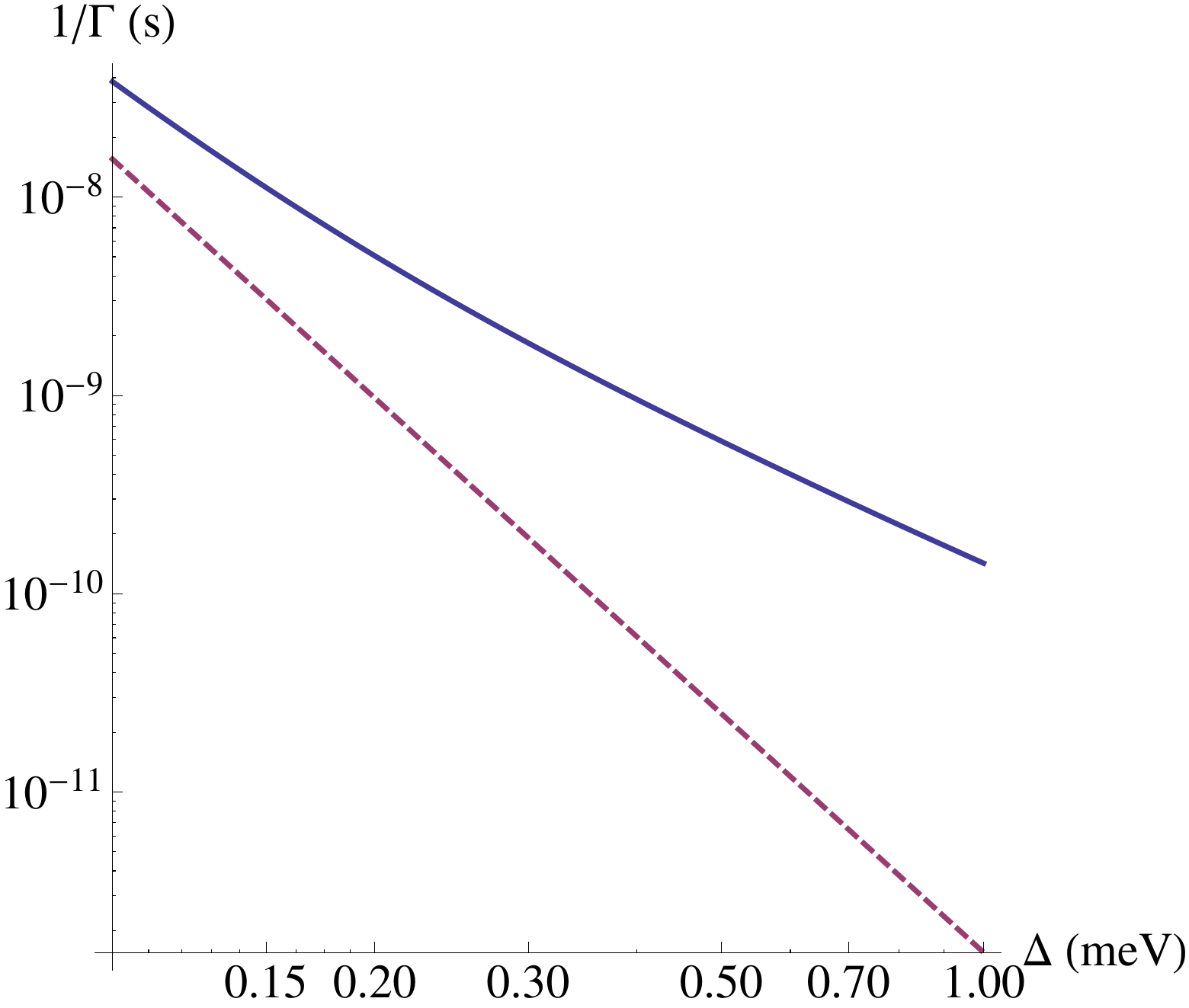} 
\par\end{centering}

\begin{centering}
\includegraphics[scale=0.45]{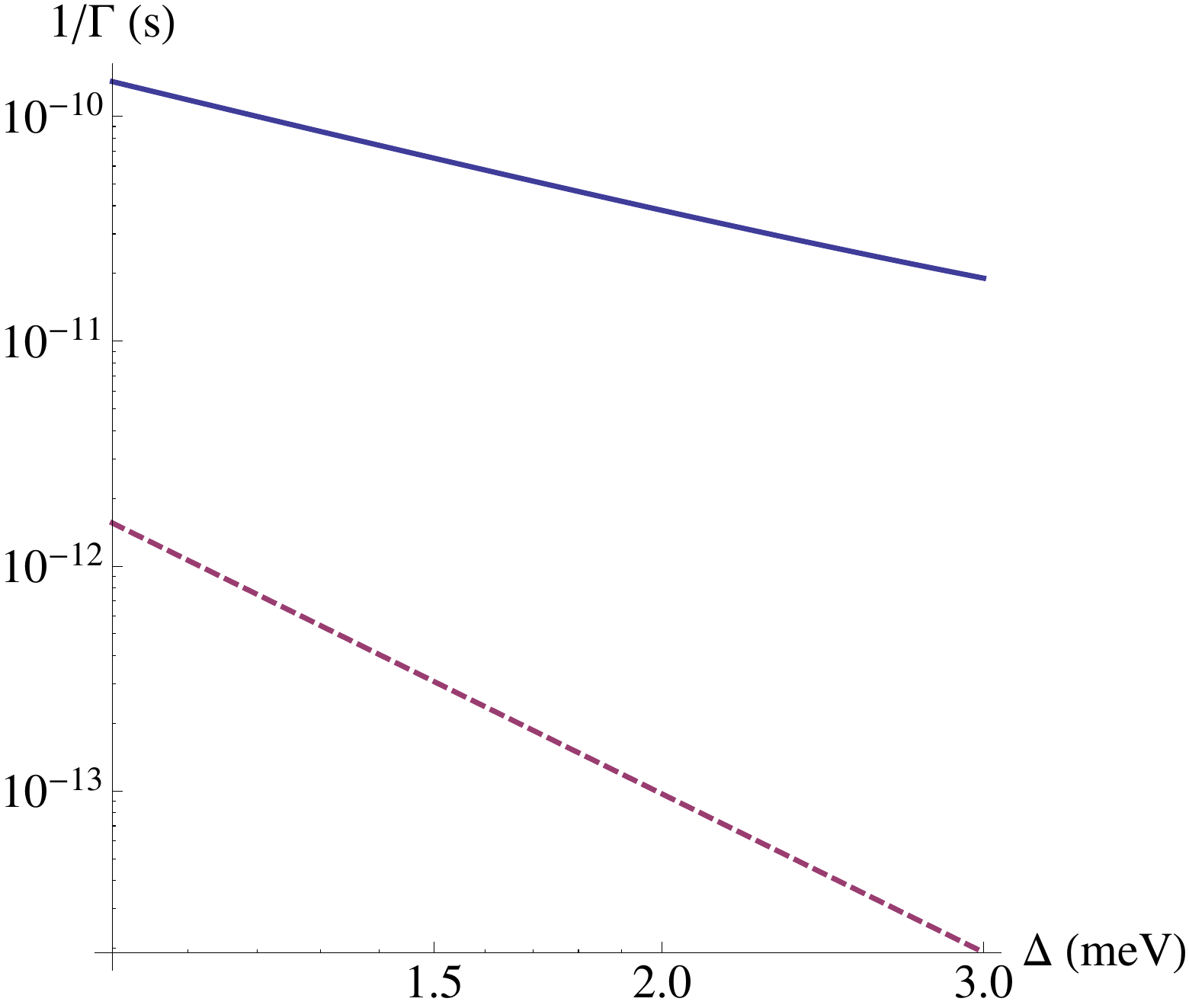}\includegraphics[scale=0.45]{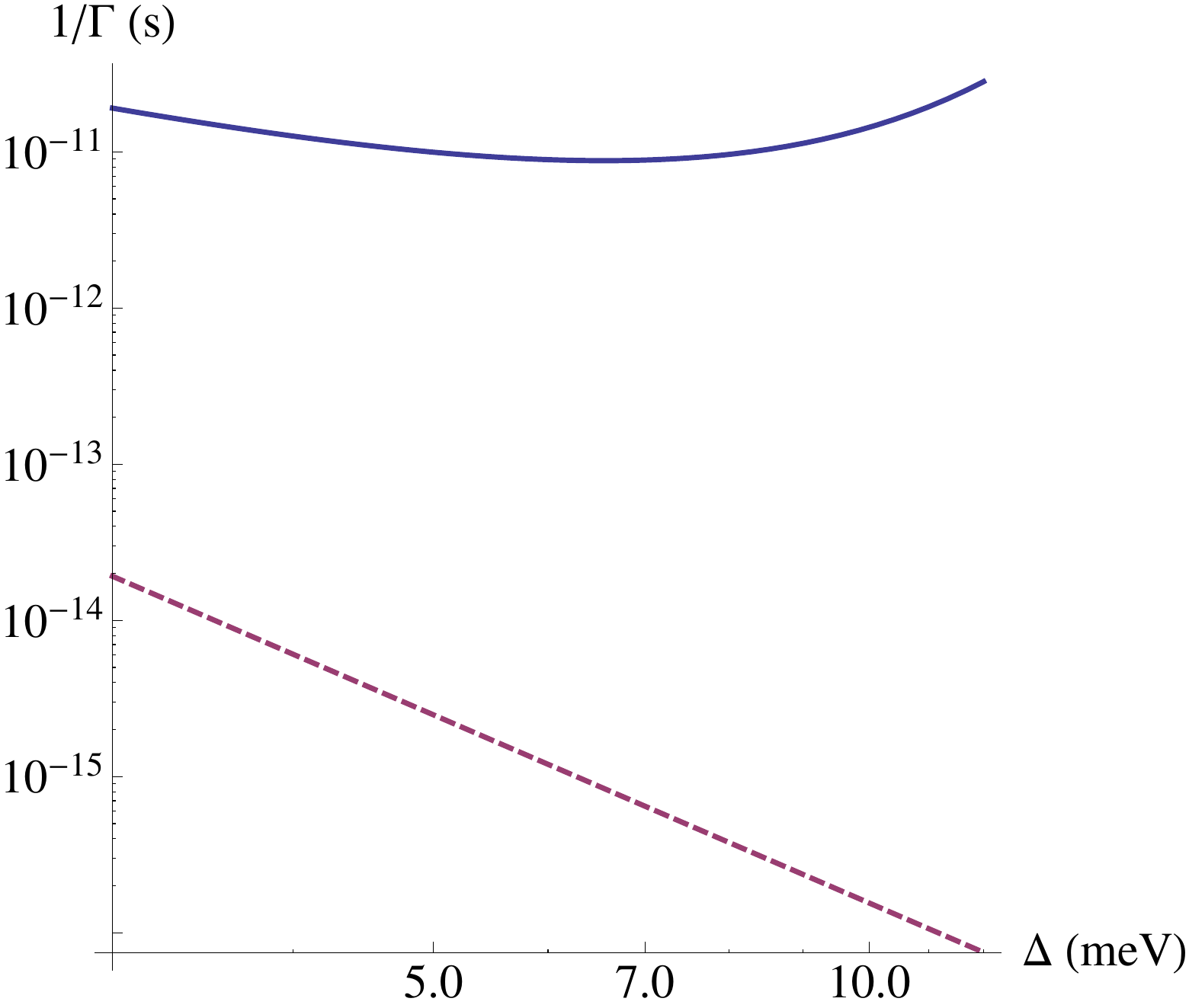} 
\par\end{centering}

\caption{\selectlanguage{american}%
Relaxation rates, $1/\Gamma$, for the transition from the first excited
orbital state to the ground state by emission of a phonon versus orbital
energy splitting, $\Delta$. The exact results including all multipole
contributions are given by the blue solid lines while the electric
dipole approximation results are given by the purple dashed lines.\foreignlanguage{english}{\label{fig:Orbital-relaxation-times}}\selectlanguage{english}%
}
\end{figure}

We outline the calculation for Si, including the valley effects, in
Appendix \ref{sub:Orbital-relaxation-Appendix}. The chief difficulty
is to include all multipole moments. The result is that the orbital
relaxation rate for a parabolic dot (in all three dimensions) from
its first orbital, excited state to its ground state is given by 
\begin{align}
\Gamma_{\Delta}^{exact}= & \left(n_{B}\left(\Delta\right)+1\right)\frac{\hbar\Delta^{4}}{16\pi\rho_{Si}m_{t}}\times\nonumber \\
 & \left\{ \exp\left(-\frac{1}{4}x_{0}^{2}q_{\Delta l}^{2}\right)\frac{1}{\hbar^{5}v_{l}^{7}}\left[\Xi_{d}^{2}(A_{l}^{\left(0\right)}-A_{l}^{\left(2\right)})+2\Xi_{d}\Xi_{u}(A_{l}^{\left(2\right)}-A_{l}^{\left(4\right)})+\Xi_{u}^{2}(A_{l}^{\left(4\right)}-A_{l}^{\left(6\right)})\right]\right.\nonumber \\
 & \left.+\exp\left(-\frac{1}{4}x_{0}^{2}q_{\Delta t}^{2}\right)\frac{1}{\hbar^{5}v_{l}^{7}}\Xi_{u}^{2}\left[A_{t}^{\left(2\right)}-2A_{t}^{\left(4\right)}+A_{t}^{\left(6\right)}\right]\right\} ,\label{eq:exact_orbital}
\end{align}
where $z_{0}$ is the size of wave function in the growth direction
and $q_{\Delta l}=\Delta/\hbar v_{l}$, $q_{\Delta t}=\Delta/\hbar v_{t}$.
The $A$ coefficients are defined by 
\[
A_{s}^{\left(n\right)}\left(q_{\Delta s}\right)=\int_{1}^{-1}x^{n}\exp\left(\frac{1}{4}(x_{0}^{2}-z_{0}^{2})q_{\Delta s}^{2}x^{2}\right).
\]
The results are plotted in Figure \ref{fig:Orbital-relaxation-times}.
Note that the exact expression, Eq. \ref{eq:exact_orbital}, reduces
to the dipole approximate expression, Eq. \ref{eq:orbital}, when
$q_{\Delta}^{2}=0$ as expected. For large $\Delta,$ the exact solution
for the orbital relaxation rate $\Gamma_{12}$ begins to diverge from
the electric-dipole approximation early on and never falls below a
picosecond or so. Despite this, the electric-dipole approximation
holds well for small energy gaps, $0.1-2$ meV, where a quantum computer
in silicon will most likely operate. This figure shows that there
is no significant benefit in going beyond a few meV. Only around 10
meV does the relaxation time start to increase, but this a relatively
small effect. It does demonstrate, however, that the phonon bottleneck
effect may be experimentally observable in these systems and, more
importantly, that our use of the electric-dipole approximation gives
results below for the spin-flip times that may be considered a lower
bound on the maximum possible time $T_{1}$.

Table \ref{tab:Characteristic-numbers-for} compares the output of
Eq. \ref{eq:orbital} and Eq. \textbf{\ref{eq:exact_orbital}} for
pure orbital relaxation in lateral silicon quantum dots. One can compare
these results to those for GaAs quantum dots \cite{ART-Fujisawa/Tarucha-ArtHydrogenGaAs-2002},
where typical values would be $1/\Gamma=10^{-8}$ s for $\Delta=1$
meV and piezo-phonons dominate. Excited orbital states in strained
silicon typically relax in nanoseconds or faster, corresponding to
a level broadening of a micro-eV or wider. It is possible that a low-lying
excited valley state (of the same spin direction) may be closer in
energy than the orbital level. For bulk silicon, theory and experiment
have found characteristic relaxation times for the $2p$-$1s$ transition
of \textasciitilde{}200 ps \cite{ART-Greenland}.

\section{Spin-flip ($T_{1}$) times}

Our expressions for orbital relaxation in strained silicon can be
extended to the case of a spin-flip transition due to spin-orbit coupling
(SOC) in a QW. We expect that relaxation via a phonon is the dominant
cooling mechanism, in this case mediated by SOC which mixes pure spin
states. This is known to be the case for donor-bound spins in bulk
\cite{ART-FeherGere-1959}. Structural inversion asymmetry has traditionally
been thought to be the dominant source of SOC in silicon quantum wells
due to the large electric field common to modulation-doped or top-gated
SiGe heterostructures. The nature of this SOC has been well described
elsewhere \cite{BOOK-Winkler,ART-Tahan-2deg,ART-Ivchenko-2006,ART-Ivchenko-2008,ART-Prada-spin-orbitSiGe};
it leads to a Rashba term in the Hamiltonian of the electrons. However,
interface effects that break the inversion asymmetry can also lead
to a generalized Dresselhaus-like term \cite{ART-Ivchenko-2008}.
Surprisingly, this can lead to effects of similar or even greater
magnitude than the Rashba term \cite{ART-Ivchenko-2006,ART-Ivchenko-2008,ART-Prada-spin-orbitSiGe}.
We discuss this further below. Here we only note that the zero-field
energy level splittings caused by SOC are small, of the order of $\mu$eV,
which validates our use of a perturbation theory that uses zero-order
electron wave functions and energy levels taken from the SOC-free
Hamiltonians.

Until the appearance of Refs. \cite{ART-Friesen-ReadoutLetter-2003,THESIS-Tahan},
there was no finite spin-flip time prediction for lateral silicon
quantum dots when the external field $\vec{B}$ is parallel to $\widehat{z}$.
Previous theories for $T_{1}$ in silicon had been based on the two
dominant mechanisms relevant to P:Si donors: the ``valley-repopulation''
mechanism (bulk SOC mixing with the six nearby 1s-like states) and
the ``one-valley mechanism'' (bulk SOC mixing with continuum states)
\cite{ART-Hasegawa-SpinLatticeRelaxationinSiGe-1960,ART-Roth-gFactorandSpinLatticeRelaxationinGeandSi-1960,ART-Roth-SpinLatticeInteractionElectrons-}.
Both mechanisms are independent of the size and shape of the localized
electron wave function. We showed rigorously in Ref. \cite{ART-Tahan-DecoherenceinSiQCs-2002}
that the former becomes negligible with {[}001{]} strain. The latter
is slightly modified with strain and goes to zero for certain directions
of the static magnetic field, particularly the {[}001{]} direction
(the most relevant to QC), for both one and two-phonon processes \cite{ART-Glavin/Kim}.
We review these bulk mechanisms here as they are relevant for donor
qubits and in some cases may be seen as residual $T_{1}$ mechanisms
at low magnetic fields in dots. Then we will derive the spin relaxation
times for dots in strained structures due to inversion asymmetry-based
SOC leading to a $T_{1}$ that is finite for $\vec{B}$ parallel to
$\widehat{z}.$

\subsection{Bulk spin-flip mechanisms}

For donor states in bulk silicon, Roth and Hasegawa \cite{ART-Hasegawa-SpinLatticeRelaxationinSiGe-1960,ART-Roth-gFactorandSpinLatticeRelaxationinGeandSi-1960,ART-Roth-SpinLatticeInteractionElectrons-}
identified two mechanisms that have been confirmed experimentally
up to $T=$2 K in P donor spins \cite{ART-Wilson/Feher-ESR3-1961}.
The spin relaxation in the Roth-Hasegawa picture is due to a modulation
of the system's g-factor by acoustic phonons. Both mechanisms are
direct single-phonon processes. The g-tensors for a given conduction
band state can be written as a sum over the g-tensors at each conduction
band minimum, 
\[
\mathbf{g}=\sum_{i}\alpha_{i}\mathbf{g}_{i},
\]
 where $\left|\alpha_{i}\right|^{2}$ is the squared amplitude (\textquotedbl{}population\textquotedbl{}
in the early literature) of the single electron wave function at the
$i$th valley and $\mathbf{g}^{(i)}=g_{\perp}\delta_{\alpha\beta}+(g_{\parallel}-g_{\perp})K_{\alpha}^{(i)}K_{\beta}^{(i)}$.

There are two mechanisms leading to spin flip. \ The {}``valley-repopulation\textquotedblright{}\ mechanism
is due to mixing between the symmetric ground state of the donor electron
and the split-off doublet state where a phonon changes the $\alpha_{i}^{\prime}s$.
The {}``one-valley\textquotedblright{}\ mechanism is due to phonon-induced
modulation of the $\mathbf{g}_{i}$ themselves and subsequent mixing
with nearby conduction bands which are coupled through an inter-band
deformation potential. The two mechanisms are of the same order of
magnitude in the bulk case and complementarily explain the angular
magnetic field dependence of $T_{1}$ in those systems \cite{ART-Wilson/Feher-ESR3-1961}.

Ref. \cite{ART-Tahan-DecoherenceinSiQCs-2002} showed how the valley-repopulation
contribution to the spin-flip becomes negligible with increasing {[}001{]}
compressive strain as is inherent in a silicon quantum dot. This can
be seen easily qualitatively. Consider first a potential with spherical
symmetry. The population amplitudes describing the lowest six conduction
states are given by \cite{ART-Wilson/Feher-ESR3-1961}:
\[
\begin{array}{rl}
Singlet: & \alpha_{11}=\frac{1}{\sqrt{6}}(1,\,1,\,1,\,1,\,1,\,1)\\
Doublet: & \alpha_{21}=\frac{1}{\sqrt{12}}(-1,-1,-1,-1,2,2)\\
 & \alpha_{22}=\frac{1}{\sqrt{4}}(1,\,1,-1,-1,\,0,\,0)\\
Triplet: & \alpha_{31}=\frac{1}{\sqrt{2}}(1,-1,\,0,\,0,\,0,\,0)\\
 & \alpha_{32}=\frac{1}{\sqrt{4}}(0,\,0,\,1,-1,\,0,\,0)\\
 & \alpha_{33}=\frac{1}{\sqrt{4}}(0,\,0,\,0,\,0,\,1,-1),
\end{array}
\]
 in the valley basis $(x,-x,\, y,-y,\, z,-z)$. The six states are
split even at zero strain by non-spherical central-cell corrections
in the donor case (the sharp potential of the donor) and by the interfaces
in the quantum dot case. SOC represented by the anisotropic g-factor
mixes the singlet ground state only with one of the doublet states
\cite{ART-Tahan-DecoherenceinSiQCs-2002}. In QDs the strong compressive
strain in the $z$ direction causes a large relative splitting between
the six valley states; the result is that only the $\alpha_{11}$
and $\alpha_{33}$ states will be populated. These symmetric and antisymmetric
valley states are not mixed by the SOC which results in a vanishing
matrix element. So, in the quantum dot limit ($\pm z$ valleys populated),
the valley-repopulation contribution to the spin-flip rate becomes
negligible. Note that for this mechanism we only consider mixing to
the six lowest states of the donor, all of which have orbital s-like
character. The $2p$ states in a donor are typically $30$ meV (bulk)
to $3$ meV ({[}001{]} strain) away. Mixing with these states will
be considered separately below. 

The one-valley mechanism, however, is relevant to QDs. Roth showed
that in bulk silicon, the contribution from mixing with nearby bands
is described by a Hamiltonian 
\[
H_{one-valley}^{bulk}=A\beta(U_{xy}(\sigma_{x}H_{y}+\sigma_{y}H_{x})+c.p.),
\]
\begin{figure}[ptb]
\begin{centering}
\includegraphics[scale=0.4]{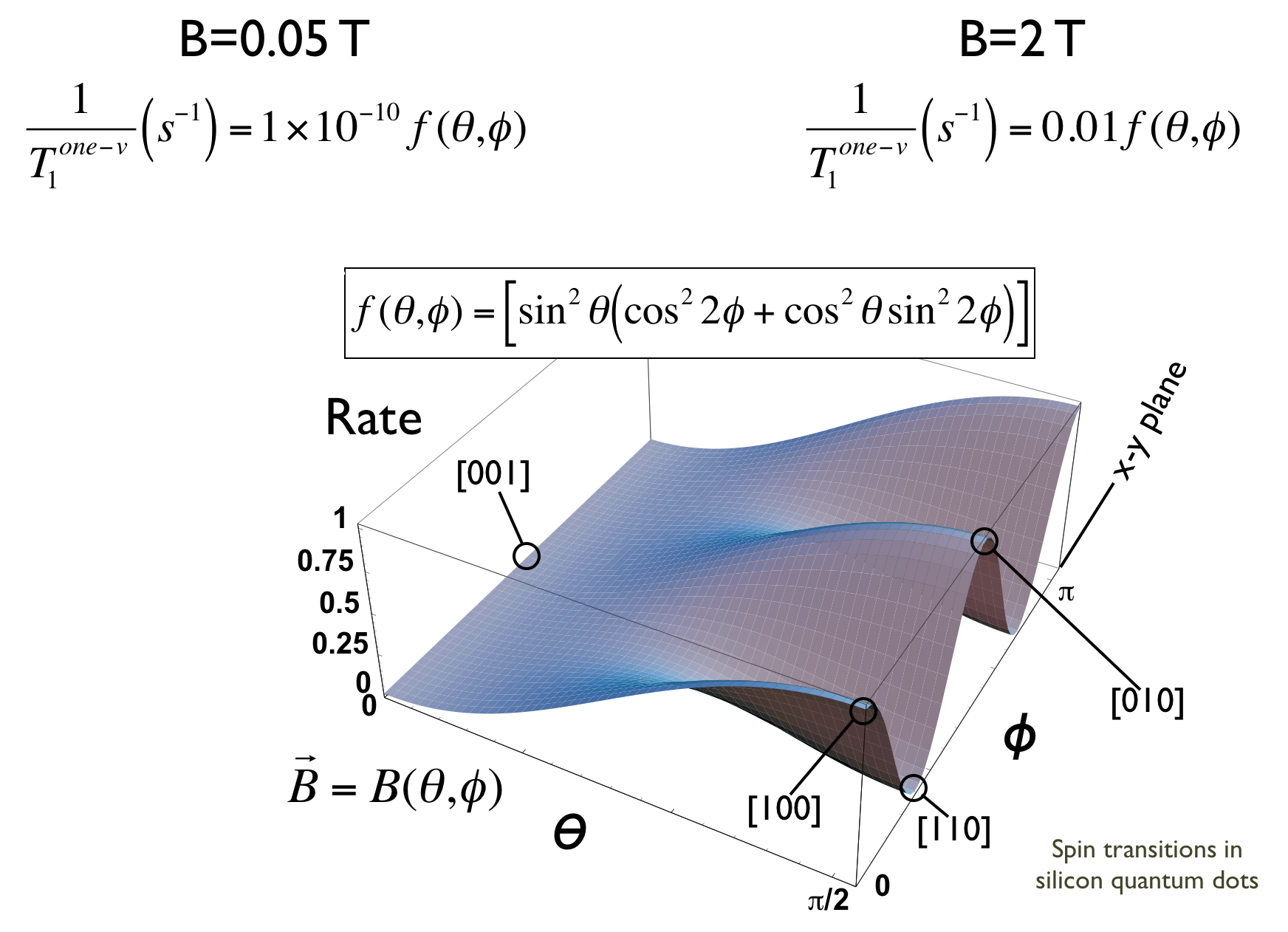} 
\par\end{centering}

\caption{\selectlanguage{american}%
Dependence of the spin relaxation rate on the angle of the external
magnetic field for the one-valley mechanism in strained silicon quantum
dots (or donors in strained silicon).\selectlanguage{english}%
}

\label{fig:Spin-relaxation-one-valley}
\end{figure}
 where c.p. stands for cyclic permutations. Group theoretical considerations
and perturbation theory lead to the conclusion that the dominant contribution
to $A$ comes from mixing with the nearby $\Delta_{2}^{\prime}$ and
$\Delta_{5}$ bands and is given by 
\begin{align*}
A & =\frac{2i\beta}{3m}\frac{\langle\Delta_{2^{\prime}}|p_{z}|\Delta_{2^{\prime}}\rangle\langle\Delta_{2^{\prime}}|D_{xy}|\Delta_{1}\rangle}{E_{12^{\prime}}^{2}E_{15}}\\
 & \times\left\{ \langle\Delta_{1}|p_{x}|\Delta_{5^{\prime}}^{x}\rangle\langle\Delta_{5}^{x}|h_{x}|\Delta_{2^{\prime}}\rangle+\langle\Delta_{1}|h_{x}|\Delta_{5}^{y}\rangle\langle\Delta_{5}^{y}|p_{x}|\Delta_{2^{\prime}}\rangle\right\} \\
 & \times\Delta g_{\perp}\frac{E_{15}}{E_{12}^{2}}\langle\Delta_{2^{\prime}}|D_{xy}|\Delta_{1}\rangle,
\end{align*}
 where $\mathbf{h}=\nabla V\times\mathbf{P}$ is the usual crystal
spin-orbit vector, $E_{ij}$ are the energy gaps to the relevant bands,
and $D$ is the inter-band deformation potential. $H_{one-valley}^{bulk}$
represents a sum over the six minima in the bulk case, but in a lateral
quantum dot the dominant contribution comes only from the $\pm z$
minima and was determined by Glavin and Kim \cite{ART-Glavin/Kim}
to be
\[
H_{one-valley}^{QD}=A\beta U_{xy}(\sigma_{x}H_{y}+\sigma_{y}H_{x}).
\]
 The constant $A$ was experimentally determined by Wilson and Feher
\cite{ART-Wilson/Feher-ESR3-1961} as $A=0.44$. The $T_{1}$ time
due to $H_{one-valley}^{QD}$ can be readily calculated and is given
by \cite{ART-Glavin/Kim}
\[
\frac{1}{T_{1}^{one-valley}}=\frac{2\pi^{4}A^{2}\hbar}{5g^{2}\rho v_{t}^{5}}\left(\frac{g\mu B}{2\pi\hbar}\right)^{5}(1+2n_{B}(g\mu B))\sin^{2}\theta(\cos^{2}2\phi+\cos^{2}\theta\sin^{2}2\phi),
\]
 where $(\theta,\phi)$ define the angle of the magnetic field relative
to {[}001{]}. $n_{B}$ is the Bose function. \ It is evident that
this equation produces infinite relaxation times if the magnetic field
points along the {[}001{]} or {[}011{]} axes. Figure \ref{fig:Spin-relaxation-one-valley}
plots the one-valley relaxation rate as a function of magnetic field
direction. We particularly stress that the $B^{5}$ dependence of
$1/T_{1}$ is the characteristic signature of this mechanism and that
this rate is very small in general.

\subsection{Dot-specific mechanisms\label{sub:Dot-specific-mechanisms}}

Now we will consider SOC that comes from the dot structure itself.
The spin-orbit Hamiltonian for a two dimensional electron is conventionally
written as
\[
H_{so}=\alpha\left(\sigma_{x}k_{y}-\sigma_{y}k_{x}\right)+\beta\left(\sigma_{x}k_{x}-\sigma_{y}k_{y}\right),
\]
 where $k_{x},k_{y}$ are the in-plane wave vector components; $\alpha$
and $\beta$ are the strengths of the so-called Rashba and Dresselhaus
spin-orbit terms. This, combined with electron-phonon coupling, can
also produce spin relaxation, and the effect (with $\alpha=0)$ has
been computed for GaAs QDs \cite{ART-Khaetskii-SpinRelaxationInDots-1999}.
In Si QDs, there is SIA SOC that comes from the fact that the mirror
symmetry $z\leftrightarrow-z$ is broken. This occurs in SiGe/Si/SiGe
heterostructures and in MOSFET-type QDs either by modulation doping
or by a top-gate induced electric field \cite{ART-Tahan-2deg}. In
contrast, Dresselhaus SOC has traditionally been assumed to be absent
in Si structures because bulk Si has inversion symmetry. It was recently
shown \cite{ART-Ivchenko-2008,ART-Prada-spin-orbitSiGe} that this
is not the case. The breaking of inversion symmetry by the interfaces
gives a non-zero Dresselhaus-like term that can be surprisingly large
\cite{ART-Prada-spin-orbitSiGe}. This has yet to be verified experimentally.
Often, experimental measures of spin relaxation involve terms proportional
to $\alpha^{2}+\beta^{2}$ so the terms are hard to verify independently.
Therefore we will keep both terms in $H_{SO}$ and compute the spin
relaxation that comes from these asymmetry-induced effects. The results
differ from those of the GaAs quantum dot analog due to the many-valley
nature of silicon and the dominance of acoustic over piezo-phonons
in silicon.

The orbital energy level splittings are much reduced in quantum dots
relative to donors (see Figure 1) because of the more shallow potential.
It is thus relatively easy to make the Zeeman splitting larger than
the orbital splitting. However, for quantum computing the likely situation
is for the Zeeman splitting to be less than the orbital splitting
to maintain a good qubit manifold. Here we will consider only this
case where the magnetic field splitting is small compared to the orbital
splitting. We comment on this approximation further below. In Si,
the Hamiltonian for the electron-phonon matrix element is, for $v=v'$:
\begin{align*}
\left\langle ms\left\vert H_{ep}\right\vert ns^{\prime}\right\rangle _{\mathbf{q}\lambda} & =i\delta_{s,s^{\prime}}\left\langle m\left\vert \left[a_{\mathbf{q}\lambda}^{\ast}e^{-i\mathbf{q}\cdot\mathbf{r}}+a_{\mathbf{q}\lambda}e^{i\mathbf{q}\cdot\mathbf{r}}\right]\right\vert n\right\rangle \\
 & \ q\ \left[\left(\Xi_{d}\mathbf{e}_{x}\left(\mathbf{q},\lambda\right)\widehat{q}_{x}+\Xi_{d}\mathbf{e}_{y}\left(\mathbf{q},\lambda\right)\widehat{q}_{y}+\left(\Xi_{d}+\Xi_{u}\right)\mathbf{e}_{z}\left(\mathbf{q},\lambda\right)\widehat{q}_{z}\right)\right],
\end{align*}
 where $\left\vert ns\right\rangle $ denotes a state with an electron
in the $n$ th level of the dot with spin $s$. Again, we consider
relaxation processes within the same valley state. A phonon with wave
vector $\vec{q}$ and polarization $\lambda$ is absorbed or emitted
depending on whether $E_{n}>E_{m}$ or $E_{n}<E_{m}.$ $\ s=\uparrow,\downarrow$
is the spin projection on the $z$-axis, defined to be along the external
applied field $\vec{B}=B\left(\sin\Theta\cos\Phi,\sin\Theta\sin\Phi,\cos\Theta\right).$
Hence $\left\vert \uparrow\right\rangle =\begin{pmatrix}e^{-i\Phi/2}\cos\Theta/2\\
e^{i\Phi/2}\sin\Theta/2
\end{pmatrix},$ etc.

The matrix elements of $H_{so}$ are 
\begin{align*}
\left\langle ms\left\vert H_{so}\right\vert ns^{\prime}\right\rangle  & =\left\langle ms\left\vert \left[\alpha\left(\sigma_{x}k_{y}-\sigma_{y}k_{x}\right)+\beta\left(\sigma_{x}k_{x}-\sigma_{y}k_{y}\right)\right]\right\vert ns^{\prime}\right\rangle \\
 & =im_{t}E_{mn}\left[\left(\alpha y_{mn}+\beta x_{mn}\right)\sigma_{x}^{ss^{\prime}}-\left(\alpha x_{mn}+\beta y_{mn}\right)\sigma_{y}^{ss^{\prime}}\right],
\end{align*}
 where $x_{mn}=\left\langle m\left\vert x\right\vert n\right\rangle $
is the dipole matrix element for the dot states, \foreignlanguage{american}{$\hat{\sigma}_{x}^{\uparrow\downarrow}=-\cos\varphi\cos\vartheta-i\sin\varphi$
and $\hat{\sigma}_{y}^{\uparrow\downarrow}=-\sin\varphi\cos\vartheta+i\cos\varphi$,}
and where we have used the trick \foreignlanguage{american}{$(p_{x})_{kn}=imE_{kn}x_{kn}/\hbar$};
we use units with $\hbar=1.$ $H_{so}$ causes the eigenstates to
be mixtures of up and down spin states. For example, if the unperturbed
orbital ground states $\left\vert 0\uparrow\right\rangle ^{\left(0\right)}$
and $\left\vert 0\downarrow\right\rangle ^{\left(0\right)}$ are perturbed
by $H_{so}$ , the new eigenstates $\left\vert 0\uparrow\right\rangle ^{\left(1\right)}$
and $\left\vert 0\downarrow\right\rangle ^{\left(1\right)}$ are 
\begin{align*}
\left\vert 0\uparrow\right\rangle ^{\left(1\right)} & \approx\left\vert 0\uparrow\right\rangle ^{\left(0\right)}+im_{t}\sum_{m\neq0}\left(1+g\mu_{B}B/E_{m}\right)\left(\left(\alpha y_{mn}+\beta x_{mn}\right)\sigma_{x}^{\downarrow\uparrow}-\left(\alpha x_{mn}+\beta y_{mn}\right)\sigma_{y}^{\downarrow\uparrow}\right)\left\vert m\downarrow\right\rangle ,\\
\left\vert 0\downarrow\right\rangle ^{\left(1\right)} & \approx\left\vert 0\downarrow\right\rangle ^{\left(0\right)}+im_{t}\sum_{m\neq0}\left(1-g\mu_{B}B/E_{m}\right)\left(\left(\alpha y_{mn}+\beta x_{mn}\right)\sigma_{x}^{\uparrow\downarrow}-\left(\alpha x_{mn}+\beta y_{mn}\right)\sigma_{y}^{\uparrow\downarrow}\right)\left\vert m\uparrow\right\rangle ,
\end{align*}
 where $m_{t}$ is the transverse mass \foreignlanguage{american}{and
we have expanded around $g\mu B$ with $\frac{1}{E_{nk}\pm g\mu B}=\frac{1}{E_{nk}}\left(1\mp\frac{g\mu B}{E_{nk}}+...\right)$}.
The $\left(1\right)$ superscript indicates first order in $\left\vert H_{so}\right\vert /E_{m}$
- the spin-orbit splitting compared to the orbital excitation energies.
It is important to compute the $g\mu_{B}B/E_{m}$ correction for reasons
that will soon become apparent.

Our interest is in the matrix element
\[
\left\langle 0\uparrow\left\vert H_{ep}\right\vert 0\downarrow\right\rangle _{\mathbf{q}\lambda},
\]
 where the $\vec{q}\lambda$ subscript indicates that there is a phonon
in the final state. We find
\begin{align*}
\left\langle 0\uparrow\left\vert H_{ep}\right\vert 0\downarrow\right\rangle _{\mathbf{q}\lambda} & =i\delta_{s,s^{\prime}}\left\langle (0)\left\vert a_{\mathbf{q}\lambda}\right\vert (1)\right\rangle q\\
 & \left[\left(\Xi_{d}\widehat{e}_{x}\left(\lambda\right)\widehat{q}_{x}+\Xi_{d}\widehat{e}_{y}\left(\lambda\right)\widehat{q}_{y}+\left(\Xi_{d}+\Xi_{u}\right)\widehat{e}_{z}\left(\lambda\right)\widehat{q}_{z}\right)\right]\times\\
 & \left[\left\langle 0\uparrow\right\vert ^{\left(0\right)}-im_{t}\sum_{m\neq0}\left(1+g\mu_{B}B/E_{m}\right)\left(\left(\alpha y_{0m}+\beta x_{0m}\right)\sigma_{x}^{\downarrow\uparrow}-\left(\alpha x_{0m}+\beta y_{0m}\right)\sigma_{y}^{\downarrow\uparrow}\right)^{\ast}\left\langle m\downarrow\right\vert \right]\\
 & \times\delta_{s,s^{\prime}}e^{i\vec{q}\cdot\vec{r}}\\
 & \times\left[\left\vert 0\downarrow\right\rangle ^{\left(0\right)}+im_{t}\sum_{m\neq0}\left(1-g\mu_{B}B/E_{m}\right)\left(\left(\alpha y_{0m}+\beta x_{0m}\right)\sigma_{x}^{\uparrow\downarrow}-\left(\alpha x_{0m}+\beta y_{0m}\right)\sigma_{y}^{\uparrow\downarrow}\right)\left\vert m\uparrow\right\rangle \right]
\end{align*}
 and we make the electric dipole approximation $e^{i\mathbf{q}\cdot\mathbf{r}}\approx1+i\mathbf{q}\cdot\mathbf{r},$
which gives
\begin{align*}
\left\langle 0\uparrow\left\vert H_{ep}\right\vert 0\downarrow\right\rangle _{\vec{q}\lambda} & =-2im_{t}\sum_{i=x,y,z}\sum_{m\neq0}\frac{g\mu_{B}B}{E_{m}}r_{m0}^{\left(i\right)}~q~\left(\left(\alpha y_{0m}+\beta x_{0m}\right)\sigma_{x}^{\uparrow\downarrow}-\left(\alpha x_{0m}+\beta y_{0m}\right)\sigma_{y}^{\uparrow\downarrow}\right)\\
 & \times\sqrt{\frac{1}{2M_{c}\omega_{\vec{q}\lambda}}}\left[\left(\Xi_{d}\widehat{e}_{x}\left(\lambda\right)\widehat{q}_{x}+\Xi_{d}\widehat{e}_{y}\left(\lambda\right)\widehat{q}_{y}+\left(\Xi_{d}+\Xi_{u}\right)\widehat{e}_{z}\left(\lambda\right)\widehat{q}_{z}\right)\right]q_{i}.
\end{align*}
\foreignlanguage{american}{Because $\left(H_{SO}\right)_{kn}^{\downarrow\uparrow}=-\left(H_{SO}\right)_{kn}^{\downarrow\uparrow}$,
the overall matrix element is reduced by roughly $g\mu B/E_{m}$.
This is the manifestation of the so-called Van Vleck cancellation.}

We do the thermal average over phonon states and apply Fermi's Golden
Rule. \ This yields
\begin{align*}
\frac{1}{T_{1}} & =2\pi\sum_{\vec{q}\lambda}\left[1+2n_{B}\left(\omega_{\vec{q}\lambda}\right)\right]\left\vert \left\langle 0\uparrow\left\vert H_{ep}\right\vert 0\downarrow\right\rangle _{\vec{q}\lambda}\right\vert ^{2}\delta\left(g\mu_{B}B-v_{\lambda}q\right)\\
 & =\frac{m_{t}^{2}}{2\pi^{2}\rho}\left(g\mu_{B}B\right)^{7}\sum_{\lambda}\frac{1}{v_{\lambda}^{7}}\sum_{m,n\neq0}S_{m}S_{n}^{\ast}\frac{1}{E_{m}E_{n}}\times\\
 & \int d\Omega_{q}\left(\widehat{q}\cdot\vec{r}_{m0}\right)\left(\widehat{q}\cdot\vec{r}_{n0}\right)\left[\left(\Xi_{d}\widehat{e}_{x}\left(\lambda\right)\widehat{q}_{x}+\Xi_{d}\widehat{e}_{y}\left(\lambda\right)\widehat{q}_{y}+\left(\Xi_{d}+\Xi_{u}\right)\widehat{e}_{z}\left(\lambda\right)\widehat{q}_{z}\right)\right]^{2},
\end{align*}
 where
\[
S_{m}=\left(\left(\alpha y_{0m}+\beta x_{0m}\right)\sigma_{x}^{\uparrow\downarrow}-\left(\alpha x_{0m}+\beta y_{0m}\right)\sigma_{y}^{\uparrow\downarrow}\right).
\]
 The integral is over the directions of $\vec{q}=q\left(\sin\theta\cos\phi,\sin\theta\sin\phi,\cos\theta\right).$
($\theta$ and $\phi$ are not the same as $\Theta$ and $\Phi,$
which give the directions of the magnetic field.) 

Following Ref. \cite{ART-Khaetskii-ZeemanFlipInDots-2001} we now
define the dot polarization tensor 
\[
\xi_{ij}=-2e^{2}\sum_{m}\frac{\left(x_{i}\right)_{m0}\left(x_{j}\right)_{0m}}{E_{m}},
\]
 where the sum is over all the orbital states. The opposite valley
states are not included in the sum since the intervalley electron-phonon
coupling is assumed to be small (this could be different in non-ideal
interfaces). It is also reasonable to neglect $z_{m0},$ since the
spatial extent of the wave function in the growth direction is small
compared to $x_{0},y_{0}$ (at least by a factor of 10). The result
(with $\hbar$ restored), is 
\begin{align}
\frac{1}{T_{1}} & =\Upsilon_{xy}\frac{m_{t}^{2}}{\pi\hbar^{10}\rho_{Si}}\left(g\mu_{B}B\right)^{7}\frac{1}{e^{4}}\left[1+2n_{B}\left(g\mu_{B}B\right)\right]\times\nonumber \\
 & \{\left[\left(\alpha^{2}+\beta^{2}\right)\left(\xi_{xx}\xi_{xx}+\xi_{yx}\xi_{yx}+\xi_{xy}\xi_{xy}+\xi_{yy}\xi_{yy}\right)+2\alpha\beta\left(\xi_{xx}\xi_{xy}+\xi_{yx}\xi_{yy}+\xi_{xy}\xi_{xx}+\xi_{yy}\xi_{yx}\right)\right]\nonumber \\
 & \times\left(\frac{3}{4}+\frac{1}{4}\cos2\Theta\right)\nonumber \\
 & +\left(\alpha^{2}-\beta^{2}\right)\left(\xi_{xx}\xi_{xx}+\xi_{yx}\xi_{yx}-\xi_{xy}\xi_{xy}-\xi_{yy}\xi_{yy}\right)\nonumber \\
 & \times\sin^{2}\Theta\cos2\Phi\nonumber \\
 & +\frac{1}{2}\left[\left(\alpha^{2}+\beta^{2}\right)\left(\xi_{xx}\xi_{xy}+\xi_{yx}\xi_{yy}+\xi_{xy}\xi_{xx}+\xi_{yy}\xi_{yx}\right)+2\alpha\beta\left(\xi_{xx}\xi_{xx}+\xi_{yx}\xi_{yx}+\xi_{xy}\xi_{xy}+\xi_{yy}\xi_{yy}\right)\right]\nonumber \\
 & \times\sin^{2}\Theta\sin2\Phi\}.\label{eq:T1BeforeParabolic}
\end{align}
Note that we include a factor of two in the phonon population multiplier,
$1+2n_{b}$, so to satisfy the traditional definition of $T_{1}$
where both relaxation and excitation are possible (although at very
low temperature this term goes to one). The most striking qualitative
feature of this expression is the $B^{7}$ dependence \cite{THESIS-Tahan};
this can be considered as the characteristic feature of dot-specific
SOC and contrasts with the $B^{5}$ dependence of bulk SOC, as well
as GaAs quantum dots (which are dominated by piezo-phonon relaxation
in energy regimes of interest \cite{ART-Khaetskii-ZeemanFlipInDots-2001}).
In addition, there is field anisotropy. To understand this anisotropy
note that the diagonal elements $\xi_{xx}\approx\xi_{yy}$ are likely
to dominate the off-diagonal elements $\xi_{xy}$ and $\xi_{yx}.$
Examination of the expression then shows that the largest term in
$1/T_{1}$ is proportional to $\left(\alpha^{2}+\beta^{2}\right)$
$\left(3+\cos2\Theta\right)/4.$ This does not vanish when $\vec{B}$
is along the $z$-axis $\left(\Theta=0\right),$ again in contrast
to the Roth-Hasegawa contributions. If one wishes to determine $\alpha$
and $\beta$ individually, then the smaller contribution proportional
to $\left(\alpha^{2}-\beta^{2}\right)\sin^{2}\Theta\cos2\Phi$ must
be measured. It could be enhanced if $\xi_{xx}$ is very different
from $\xi_{yy}$ which would be the case for a very elliptical dot.

\begin{figure}
\includegraphics[scale=0.4]{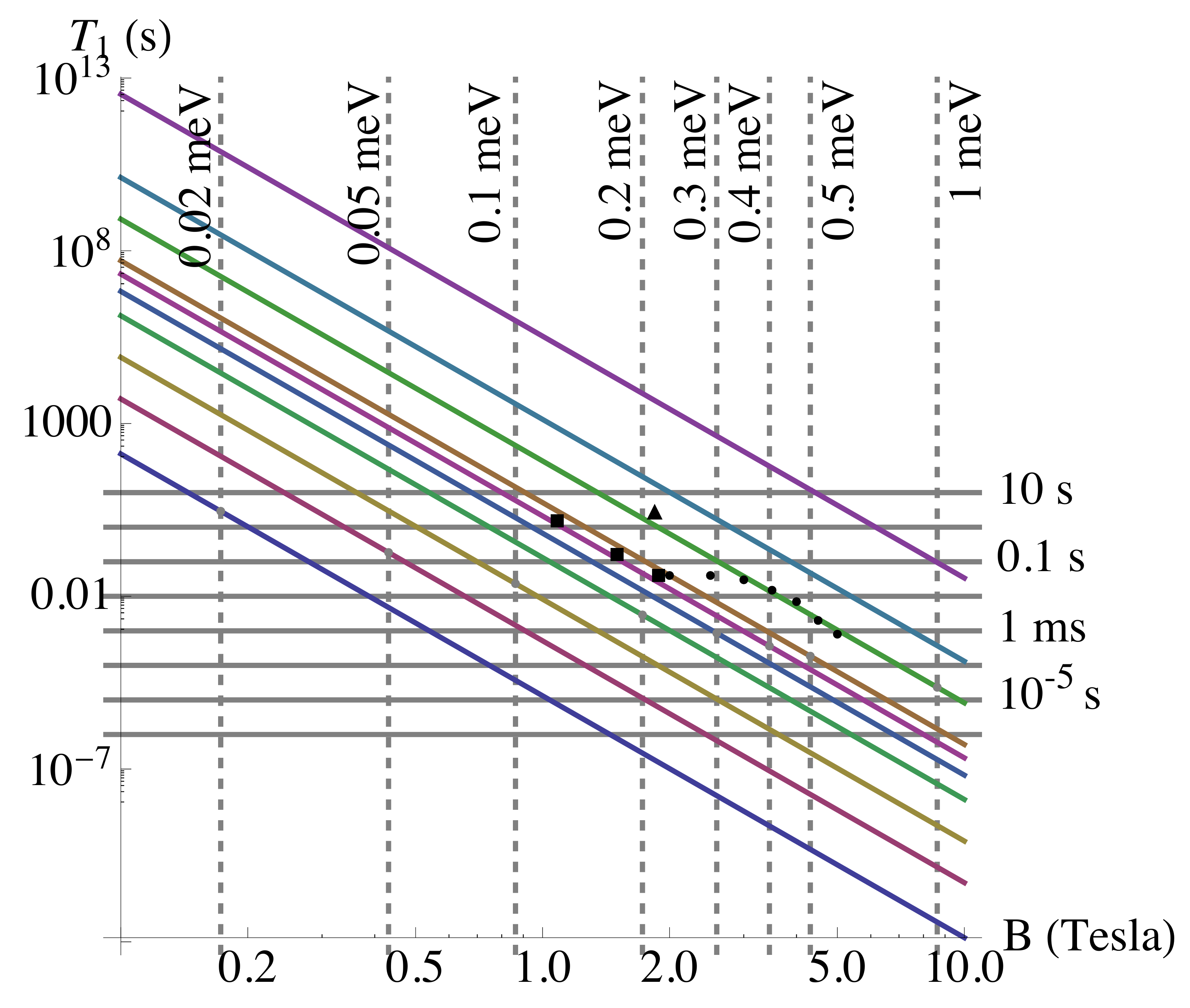} \caption{\selectlanguage{american}%
Spin relaxation time, $T_{1}$, for an ideal, circular quantum dot
as a function of magnetic field as calculated from Eq. \ref{eq:anisotropy}
for SOC constant value (which can vary greatly from device to device)
of $\sqrt{\alpha^{2}+\beta^{2}}=4\mu eV\cdot$nm (taken from a SiGe
QW experiment \cite{ART-Wilamowsky/Rossler-2002} - see text) and
$B||[110]$ ($T_{1}$ for $B||\hat{z}$ is a factor of 2 longer).
Diagonal lines for spin relaxation times from bottom to top are for
different orbital energy splittings (dot size gets smaller going up)
of $\Delta=0.02$ meV, $0.05$ meV, 0.1 meV, 0.2 meV, 0.3 meV, 0.4
meV, 0.5 meV, 1 meV, 2 meV, and 8 meV, respectively.\foreignlanguage{english}{
Theory is only appropriately compared to experiment below degeneracy
points (given by intersection of vertical and diagonal lines). Any
possible spin ``hot spots'' would occur at degeneracy between Zeeman
and first excited state splitting (which is approximately 0.3-0.4
meV in these size dots). Points represent presently published experimental
data for a SiGe quantum dot from HRL (squares) \cite{ART-Hayes-T1Si},
a SiO$_{2}$ quantum dot (circles) from UCLA \cite{ART-House-Jiang-T1},
and a SiGe quantum dot (triangle) from Wisconsin \cite{ART-Simmons-T1}.\label{fig:Spin-relaxation-time-1}}\selectlanguage{english}%
}
\end{figure}
The polarization tensors (matrix elements) can be calculated numerically,
or in the small magnetic field limit (where the first excitation energy
is much less than the Landau energy), the zero B-field parabolic matrix
elements can be used as a decent approximation. We can include the
magnetic field in a circular dot explicitly if $B||z$ by utilizing
the Fock-Darwin states \cite{Book-Jacak}. In this case $\left\langle 00\right\vert x\left\vert 01\right\rangle =\left\langle 00\right\vert y\left\vert 10\right\rangle =\left\vert \sqrt{2}L_{B}/2\right\vert ,$
where $\hbar\omega_{0}$ is the fundamental energy of the dot, $L_{B}=2\sqrt{\hbar/(m^{*}\Omega)}$,
$\Omega=\sqrt{\omega_{0}^{2}+\omega_{c}^{2}/4}$, $\omega_{c}=eB/m^{*}$,
$E_{00,01}=\hbar\omega_{-}$, $E_{00,10}=\hbar\omega_{+}$, and $\omega_{\pm}=\sqrt{\omega_{0}^{2}+\omega_{c}^{2}/4}\pm\omega_{c}/2$.
Then, taking into account dipole selection rules (only transitions
$n_{\pm}^{^{\prime}}=n_{\pm}\pm1$ are allowed), 
\[
\xi_{xx}=\xi_{yy}=-e^{2}\left(\frac{L_{B}}{2}\right)^{2}\left[\frac{1}{\hbar\omega_{-}}+\frac{1}{\hbar\omega_{+}}\right]\,\text{and }\,\xi_{xy}=\xi_{yx}=0,
\]
In the $B=0$ limit, with $L=L_{B}(B=0)$, reduces to $\xi_{xx}=\frac{e^{2}}{m^{*}\omega_{x}^{2}}$
and $\xi_{yy}=\frac{e^{2}}{m^{*}\omega_{y}^{2}}$ for an elliptical
dot.
\begin{align*}
\end{align*}
To estimate the overall quantitative magnitude we shall assume a circular
dot with a parabolic potential (note also that for a circular dot
$\xi_{xx}(B_{z})=\xi_{yy}(B_{z})=\xi_{xx}(0)$$ $. \ Then with $\xi_{xx}=\xi_{yy}=\xi$
and $\xi_{xy}\approx\xi_{yx}=0$ and 
\begin{equation}
\frac{1}{T_{1}^{QD}}=\frac{1}{105}\Xi_{u}^{2}\frac{\left(\alpha^{2}+\beta^{2}\right)}{\pi\hbar^{6}\rho_{Si}v_{t}^{7}}\frac{\left(g\mu_{B}B\right)^{7}}{\Delta^{4}}\left(3+\cos2\Theta\right)\left[1+2n_{B}\left(g\mu_{B}B\right)\right]\label{eq:anisotropy}
\end{equation}
 where we have used 
\begin{equation}
\left\vert \frac{\xi}{e^{2}}\right\vert ^{2}=\frac{\hbar^{4}}{m_{t}^{2}\Delta^{4}}=\frac{m_{t}^{2}L^{8}}{2^{4}\hbar^{4}}\label{eq:xi}
\end{equation}
(for reference we find $\Delta=\frac{2\hbar^{2}}{m_{t}L^{2}}$). The
contribution of longitudinal $(\ell)$ phonons is suppressed by roughly
a fifth in $\Upsilon_{xy}$ and is neglected. For reference, the spin
relaxation rate can also be written in terms of the dipole matrix
elements between $1s$ and $2p$, $M^{(10)}$, as (assuming mixing
to one excited state): 
\begin{equation}
\frac{1}{T_{1}^{QD}}=\frac{4}{105}\Xi_{u}^{2}\frac{\left(\alpha^{2}+\beta^{2}\right)}{\pi\hbar^{6}\rho_{Si}v_{t}^{7}}\frac{\left(g\mu_{B}B\right)^{7}\left|M^{10}\right|^{4}}{\Delta^{2}}\left(3+\cos2\Theta\right)\left[1+2n_{B}\left(g\mu_{B}B\right)\right].\label{eq:anisotropy-wrt-ME}
\end{equation}

The magnitudes of $\alpha$ and $\beta$ are material system and device
specific. Wilamowski et al. \cite{ART-Wilamowsky/Rossler-2002} have
measured $\alpha_{W}=0.55/\sqrt{2}\times10^{-12}$ eV$\cdot$cm =
$4$ $\mu eV$$\cdot$nm via 2DEG spin relaxation in Si/SiGe quantum
wells. On the other hand, for a SiGe/Si/SiGe well and a field of $10^{7}V/m$
(roughly a factor of 2 larger E-field than typical SiGe QW QDs but
about right for SiO$_{2}$ dots), Prada et al. \cite{ART-Prada-spin-orbitSiGe}
theoretically find that $\beta>\alpha$ and $\beta=\mbox{5.77 }\mu$eV$\cdot$nm
$=9.2\times10^{-34}J-m$; note that they mention that the $\beta$
term could decrease in (typical) heterostructure quantum wells with
miscut, and that it will very from device to device. Calculations
of the spin relaxation in 2DEGs \cite{ART-Glazov-2DEGspin,ART-Tahan-2deg}
give $1/T_{1}^{2DEG}\propto\left(\alpha^{2}+\beta^{2}\right)$ assuming
minimal cyclotron effects or $B\parallel(x,y)$; so the two results
are consistent if we attribute the Wilamowski result as due to $\beta$.
Nestoklen et al. \cite{ART-Nestoklon-Ivchenko}, also theoretically,
find a value for $\beta$ roughly 6 times smaller than Wilamowski
et al. \cite{ART-Wilamowsky/Rossler-2002} ($\beta_{N}=15.6$, $\alpha_{N}=5.2$
$\mu eV$$\cdot$nm). Note that the measured line width in Ref. \cite{ART-Wilamowsky/Rossler-2002}
does not depend on the in-plane orientation of the magnetic field,
implying (at least for that device) that one SOC term dominates over
the other \cite{ART-Glazov-2DEGspin,ART-Tahan-2deg}.

Figure \ref{fig:Spin-relaxation-time-1} plots Equation \ref{eq:anisotropy}
for $B\parallel[110]$ as a function of orbital energy splitting using
the result of Wilamowski et al., $\alpha_{W}=\sqrt{\left(\alpha^{2}+\beta^{2}\right)}$,
and substituting values from Table \ref{tab:Physical-constants-and}.
Also shown are some recent experimental results \cite{ART-Hayes-T1Si,ART-House-Jiang-T1,ART-Simmons-T1}
which follow the $B^{7}$ trend but generally show longer lifetimes.
We defer our comparison to experiment to the Discussion section below.
Note that our results are for the electric dipole approximation, shown
in the previous section to be a good approximation below \textasciitilde{}1
meV. Calculations that attempt to calculate the spin relaxation beyond
the dipole approximation \cite{ART-Bulaev-Loss,ART-Raith-Fabian}
have shown the possibility of fast spin relaxation (``hot spot'')
when the orbital and Zeeman energies are degenerate: a result of orbital-spin
level mixing. This mixing is contingent on the nature of the SOC mixing
--- Dresselhaus (no mixing) or Rashba (mixing). Unfortunately we do
not know the nature of the SOC in these devices and indeed it may
depend on microscopic details and very from device to device (even
on the same chip). An additional complication arises when the first
excited state is a valley state or valley-like - a situation we will
discuss later in the text. So a hot spot is not assured. Therefore,
we mark these crossovers in energy in Figure \ref{fig:Spin-relaxation-time}
as a possible position of interesting physics (fast relaxation) which
may tell us more about the nature of these states. Our results are
relevant for the quantum computing situation where $g\mu B<\Delta$
and likely a good approximation beyond the crossing point (with the
cyclotron modified wave function incorporated). 
\begin{figure}[ptb]
\begin{centering}
\includegraphics[scale=0.65]{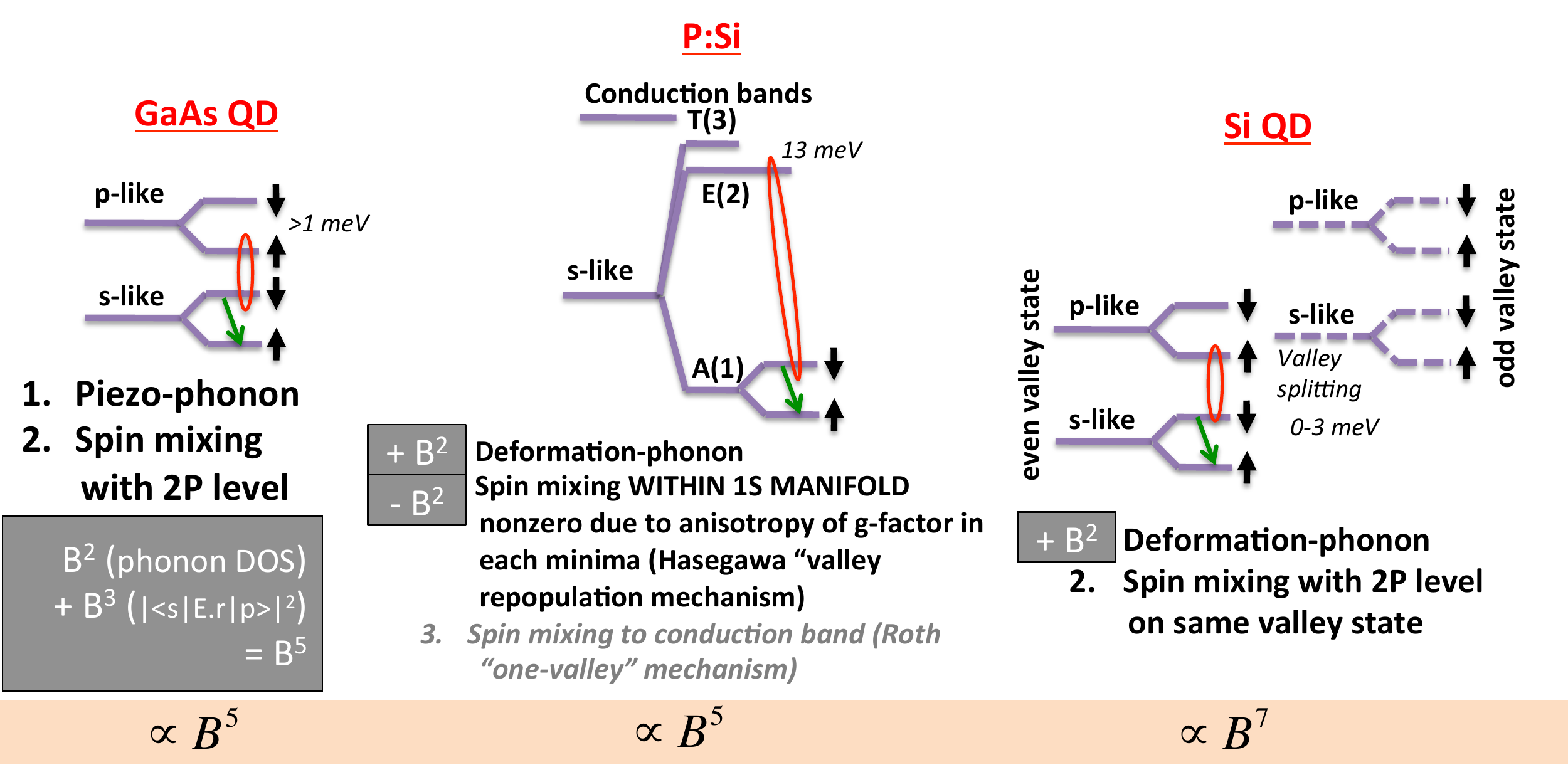}
\par\end{centering}

\caption{\selectlanguage{american}%
Schematic summary of the mechanisms behind the different field dependancies
of the spin relaxation rate for GaAs dots, for which $1/T_{1}\sim B^{5}$
, for Si donor states dots, for which $1/T_{1}\sim B^{5}$, and for
ideal Si dots, for which $1/T_{1}\sim B^{7}$.\selectlanguage{english}%
}
\end{figure}

\subsection{Electrical and Magnetic Noise}

Another possible mechanism that could limit $T_{1}$ in Si/SiGe quantum
dots is the electric and magnetic noise coming from trapped charges
and other two-level systems, noise in the circuitry, thermal and quantum
current fluctuations in nearby conductors, etc. In this section we
will point out how the presence of this sort of noise could be indicated
by lifetime measurements, and how it can be distinguished from the
other sources of noise we have been considering in this paper.

Electrical noise from a random field $\vec{E}\left(t\right)$ can
produce spin relaxation if there is spin-orbit coupling present. Relaxation
can occur by two distinct mechanisms: (1) spin-orbit-mediated virtual
excitation to higher orbital states with spin flip and (2) modulation
of the Rashba field. These correspond roughly to the Elliot-Yafet
and D'yakonov-Per'el mechanisms in bulk.

For the first mechanism we have a Hamiltonian 
\[
H=-\frac{1}{2}g\mu_{B}\vec{B}\cdot\vec{\sigma}-e\vec{E}\left(t\right)\cdot\vec{r}+\alpha\left(\sigma_{x}\frac{\partial}{\partial y}-\sigma_{y}\frac{\partial}{\partial x}\right)+\beta\left(\sigma_{x}\frac{\partial}{\partial x}-\sigma_{y}\frac{\partial}{\partial y}\right).
\]
This produces relaxation that is physically analogous to the spin-phonon
mechanism, and the derivation is parallel, so we omit it. \ We obtain

\[
\frac{1}{T_{1}}=\eta\frac{\left(em_{t}g\mu_{B}B\right)^{2}}{\hbar^{2}}S_{E_{x}}\left(g\mu_{B}B/\hbar\right)\left(\frac{4^{2}\hbar^{4}}{m_{t}^{2}\Delta^{4}}\right)\left(\alpha+\beta\right)^{2}\left(1+\cos^{2}\Theta\right)\left(1-\sin2\Phi\right)\propto S_{E_{x}}\left(g\mu_{B}B/\hbar\right)B^{2}L^{8},
\]
 where $S_{E_{x}}\left(g\mu_{B}B/\hbar\right)$ is the spectral density
of the $E_{x}$ autocorrelation function, evaluated at the qubit operating
frequency, and $L$ is a measure of the diameter of the dot. $\eta$
is a numerical factor of order one that depends on the shape of the
dot. We write the applied field $\vec{B}$ as $\vec{B}=B\left(\sin\Theta\cos\Phi,\sin\Theta\sin\Phi,\cos\Theta\right).$
$\Theta=0$ is the z-axis of the lab frame. Note that, as before,
the relaxation rate decreases as the excited state splitting increases
since the spin mixing which allows the electric field to relax the
qubit is via the excited orbital state. 

The second mechanism is physically distinct in that it does not involve
orbitally excited states; instead the noise is converted to random
time-dependent effective magnetic field on the spin. It is sufficient
to consider a Rashba Hamiltonian, 
\[
H=-\frac{1}{2}g\mu_{B}\vec{B}\cdot\vec{\sigma}-e\alpha_{1}^{0}E_{z}\left(t\right)\left(\sigma_{x}\frac{\partial}{\partial y}-\sigma_{y}\frac{\partial}{\partial x}\right),
\]
which leads to

\[
\frac{1}{T_{1}}=\eta^{\prime}\frac{e^{2}}{\hbar^{2}L^{2}}\left(\alpha_{0}^{1}\right)^{2}~S_{E_{z}}\left(g\mu_{B}B/\hbar\right),
\]
where again $\eta^{\prime}$ is a geometry-dependent constant of order
unity. Here, the rate increases with smaller dot sizes ($L^{2}=\frac{2\hbar^{2}}{m_{t}\Delta}$).
Reasonable values for the parameters are $\alpha_{0}^{1}=10^{-5}~$nm$^{2}$
\cite{ART-Prada-spin-orbitSiGe} , and $L=50$ nm$;$ evaluating this
formula leads to
\[
\frac{1}{T_{1}}\sim0.1s^{-1}\times S_{E_{z}}\left(g\mu_{B}B/\hbar\right),
\]
if $S_{E_{z}}\left(g\mu_{B}B\right)$ is measured in V$^{2}$-s/m$^{2}.$
Zimmerman \textit{et al}. determined the strength of electrical noise
in a Si SET\ structure by measuring fluctuations in the peak separations
of Coulomb blockade oscillations, but so far this type of measurement
has been peformed only at frequencies much less than 1 GHz, which
makes it difficult to estimate the noise magnitude at typical qubit
operating frequencies in real structures. But the only $B$-dependence
in $T_{1}$ comes from $S_{E_{z}}\left(g\mu_{B}B/\hbar\right),$ which
is likely to vary extremely slowly with $B$ for any mechanism that
one can think of. This means that defect-dominated electrical noise
can be easily distinguished from other relaxation mechanisms by the
fact that it is $B$-independent.

Magnetic noise from quantum and thermal current fluctuations in metallic
portions of the circuit will produce a fluctuating magnetic field
at the qubit that can relax the spin. No spin-orbit coupling is required
for this mechanism to operate. This effect has recently been calculated
by Langsjoen \textit{et al}. \cite{ART-Langsjoen-Joynt}. These authors
found values of $T_{1}$ of order seconds for typical quantum dot
architectures. The field and temperature dependence is given by $1/T_{1}\sim B\coth\left(\mu_{B}B/2k_{B}T\right),$
which reflects the photon density of states and the Bose function.
The field and temperature dependences are again distinctive.

\subsection{Spin relaxation due to nuclei}

Hyperfine coupling of the electron spin to nuclei can give a very
small admixture of the opposite spin state into a predominantly up
or down state. \ This mechanism would give a $T_{1}$ that depends
relatively weakly on field. \ However, theoretical estimates give
a small magnitude for this effect \cite{ART-Wilson/Feher-ESR3-1961,ART-Khaetskii-ZeemanFlipInDots-2001,THESIS-DeSousa}.
\ This conclusion would of course be strengthened in isotopically
purified Si$^{28}$. \ Furthermore, this mechanism is not specific
to dots and should occur also in donor spin relaxation, where $T_{1}\approx0.25\times10^{4}$
s for Si:P at $B=0.32$ T and $T=1.25K$ \cite{ART-FeherGere-1959,ART-Wilson/Feher-ESR3-1961}.
It does not seem to have been observed. \ Hence this mechanism is
probably negligible at the fields and temperatures under consideration
here.

\subsection{Two-phonon processes}

At higher temperatures there is an activated two-phonon contribution
from SOC mixing in silicon quantum dots. In the Si:P system, these
Orbach processes dominate for T $>$ 2K \cite{ART-Tryshkin/Lyon-ESRofPSi-2003}.
We can use the methods proposed by Castner \cite{ART-Castner-OrbachSpinLatticeinSi-1967}
to estimate the Rashba + Orbach spin relaxation path. We arrive at
\[
\frac{1}{T_{1}^{Orbach}}\approx M_{SO}^{2}\Gamma_{1\rightarrow g}n\left(E_{1g}\right)
\]
 where we estimate for a circular dot that $M_{SO}\approx\left\langle 1\uparrow\right\vert x\left\vert g\downarrow\right\rangle /\left\langle 1\right\vert x\left\vert g\right\rangle \approx2(\alpha+\beta)\sqrt{m^{\ast}/E_{1g}}$
as the spin-mixing of the two states and $\Gamma$ is the orbital
relaxation time of the first excited state. \ For a circular dot
with a parabolic potential 
\begin{equation}
\frac{1}{T_{1}^{Orbach}}\approx\Upsilon_{xy}\frac{8\left(\alpha^{2}+\beta^{2}\right)\left\vert \Delta\right\vert ^{3}}{\hbar^{4}\pi\rho_{Si}}\exp\left[-\left\vert \Delta\right\vert /kT\right].\label{eq:Orbach}
\end{equation}
 Thus the temperature dependence of $T_{1}$ at higher temperatures
can give an accurate measure of $\Delta,$ a technique already used
to find energy splittings of donor states. \ This can provide a check
on transport spectroscopy determinations of this quantity.

\section{Valley relaxation}

Electrons in lateral silicon quantum dots typically reside in the
two degenerate conduction band minima along the $z$ direction. This
doubles the number of levels in the dot relative to the $\Gamma$-point-centered,
direct band-gap III-V quantum dots as was described in Section \ref{sec:Silicon-Quantum-Dot}.
Here we wish to consider the relaxation times of these excited valley
states as we have done for low-lying orbital and spin states above.
Castner was the first to calculate the relaxation across different
valley states from the $2p$ to $1s$ levels in donors \cite{ART-Castner-RamanSpinLatticeinSi-1963}.
This has been repeated in Ref. \cite{ART-Smelyanskiy-2003} for Li
donors and for P and Li in strained silicon in Ref. \cite{ART-Soykal-PRL-2012}.
A similar calculation can be done for lateral quantum dots where the
interface in $z$ is assumed perfectly flat and smooth, and thus the
valleys can be considered good quantum numbers in the usual Kohn-Luttinger
approximation (and the problem is separable in the three dimensions).
We first consider this ideal (or ``1D'') case (which may be relevant
in some experimental situations) and then comment on the more usual
case of significant valley-orbital wave function mixing due to imperfect
interfaces.

\begin{figure}[ptb]
\begin{centering}
\begin{tabular}{cc}
\includegraphics[scale=0.45]{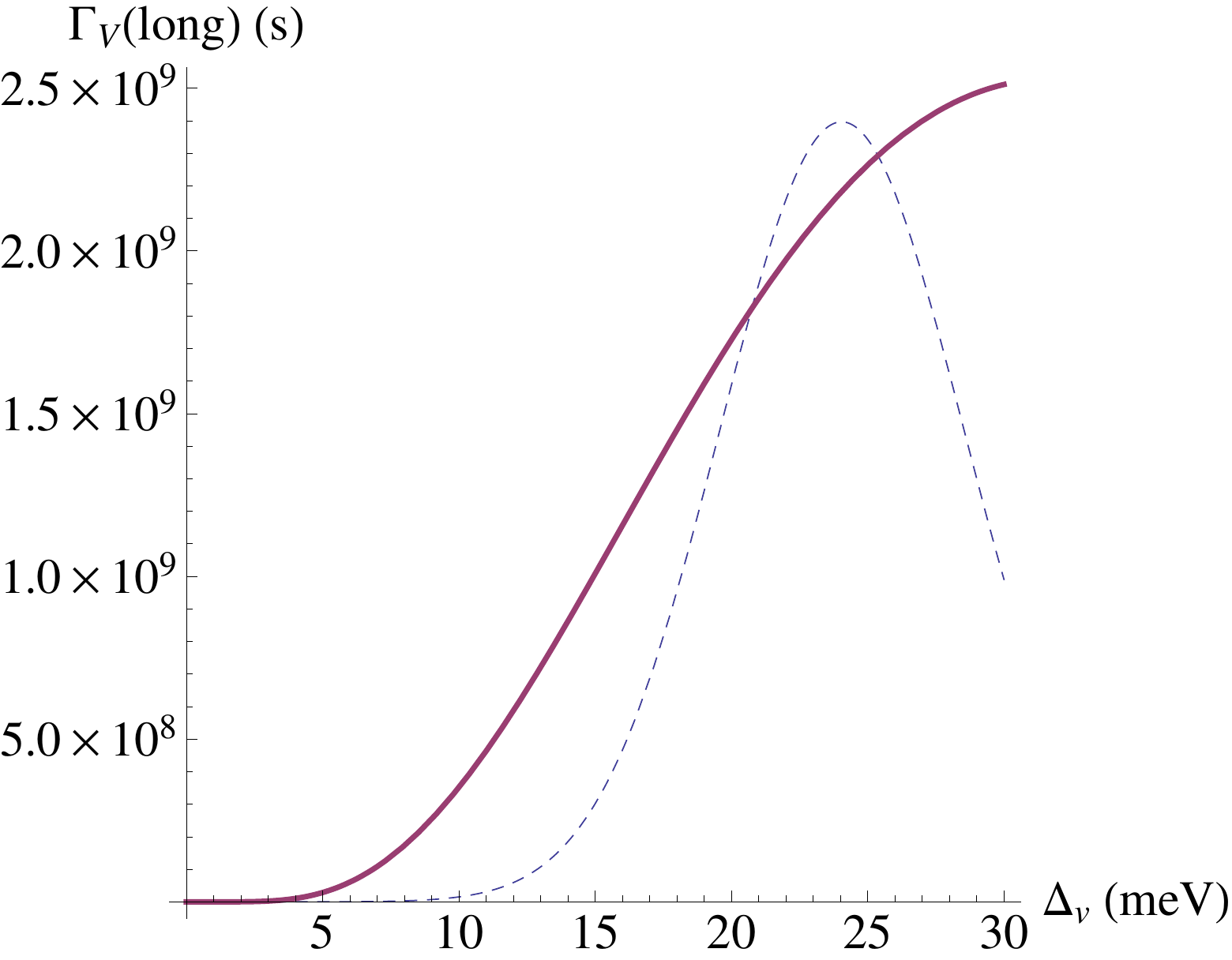}  & \includegraphics[scale=0.48]{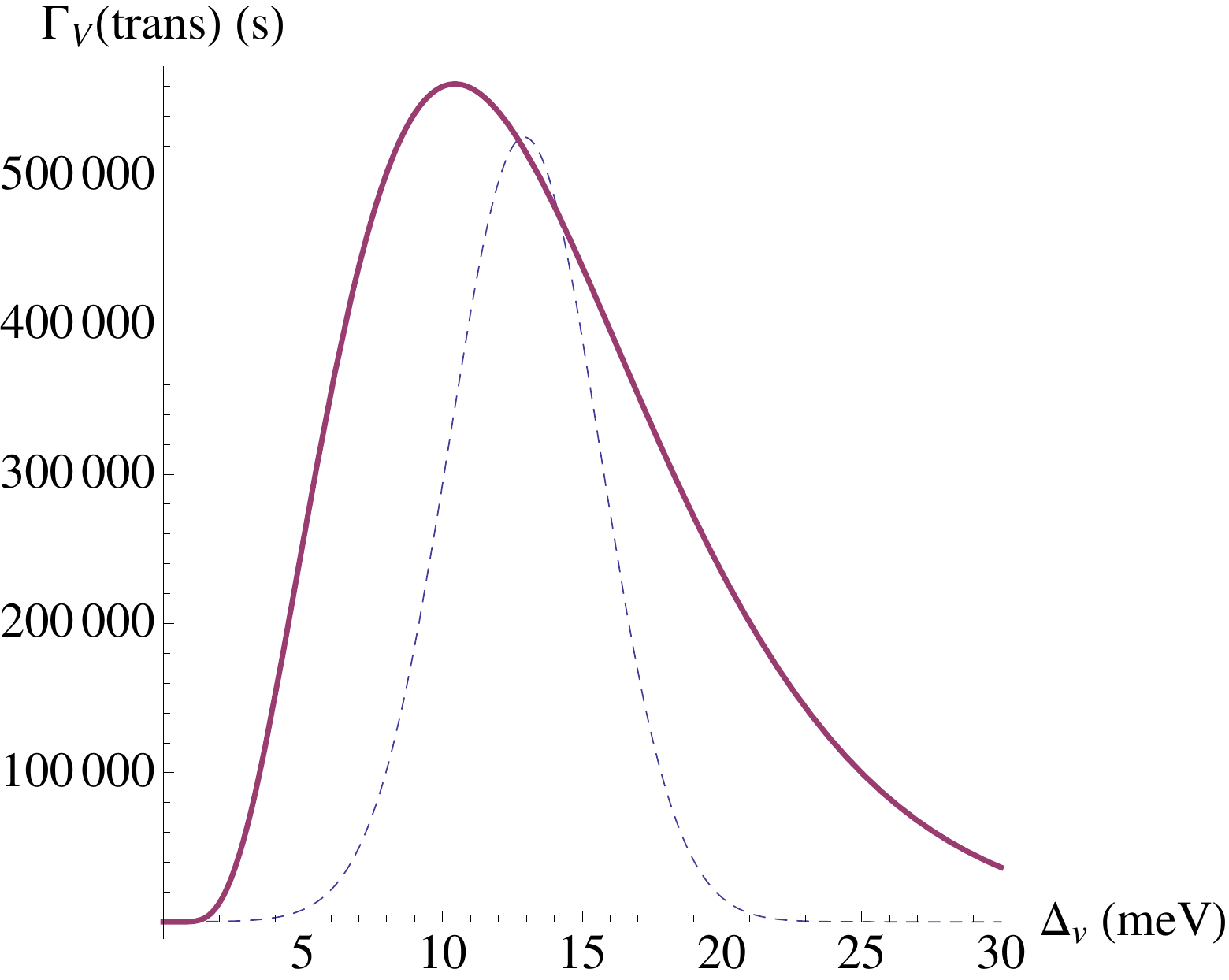}\tabularnewline
\includegraphics[scale=0.45]{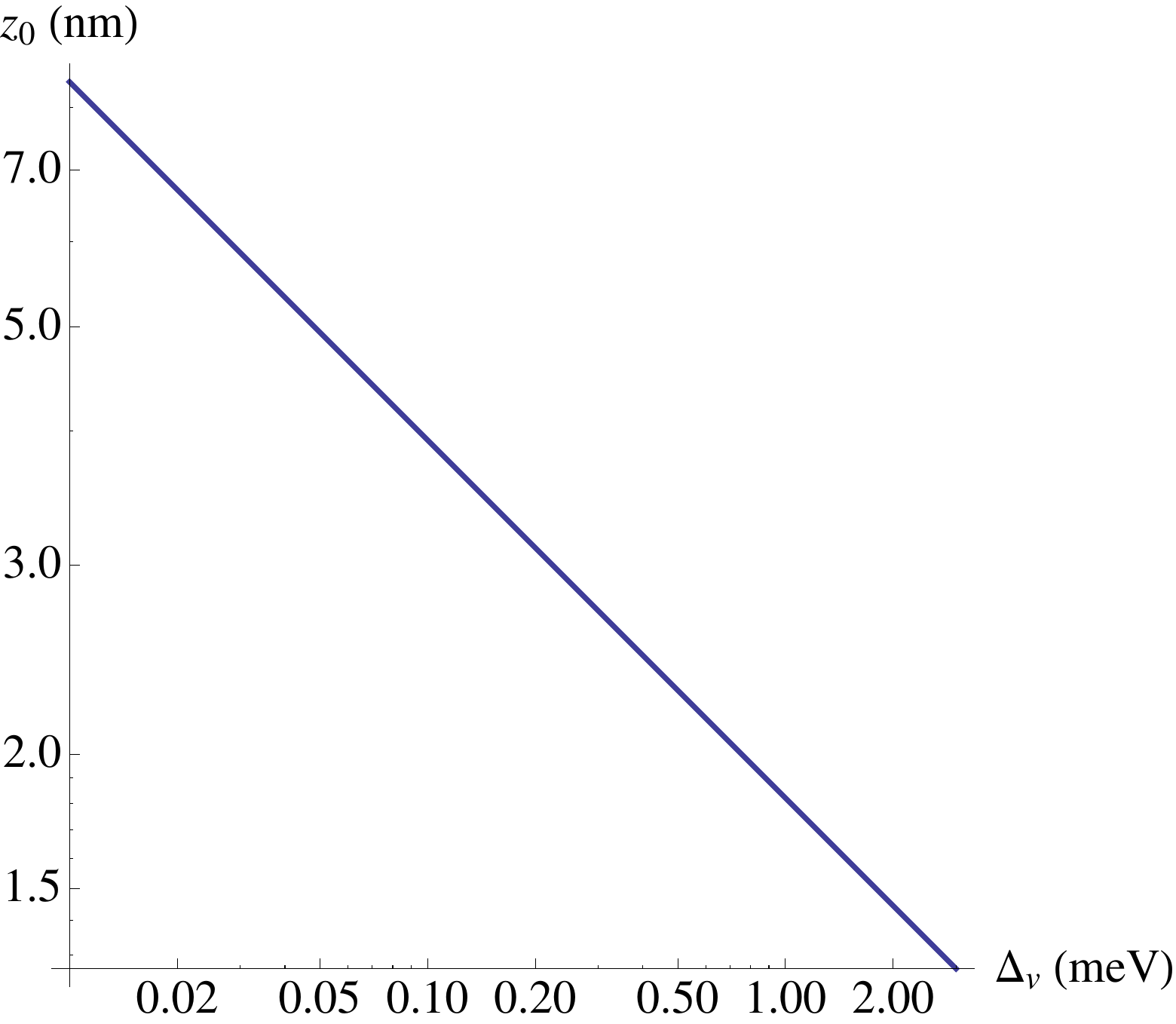}  & \includegraphics[scale=0.45]{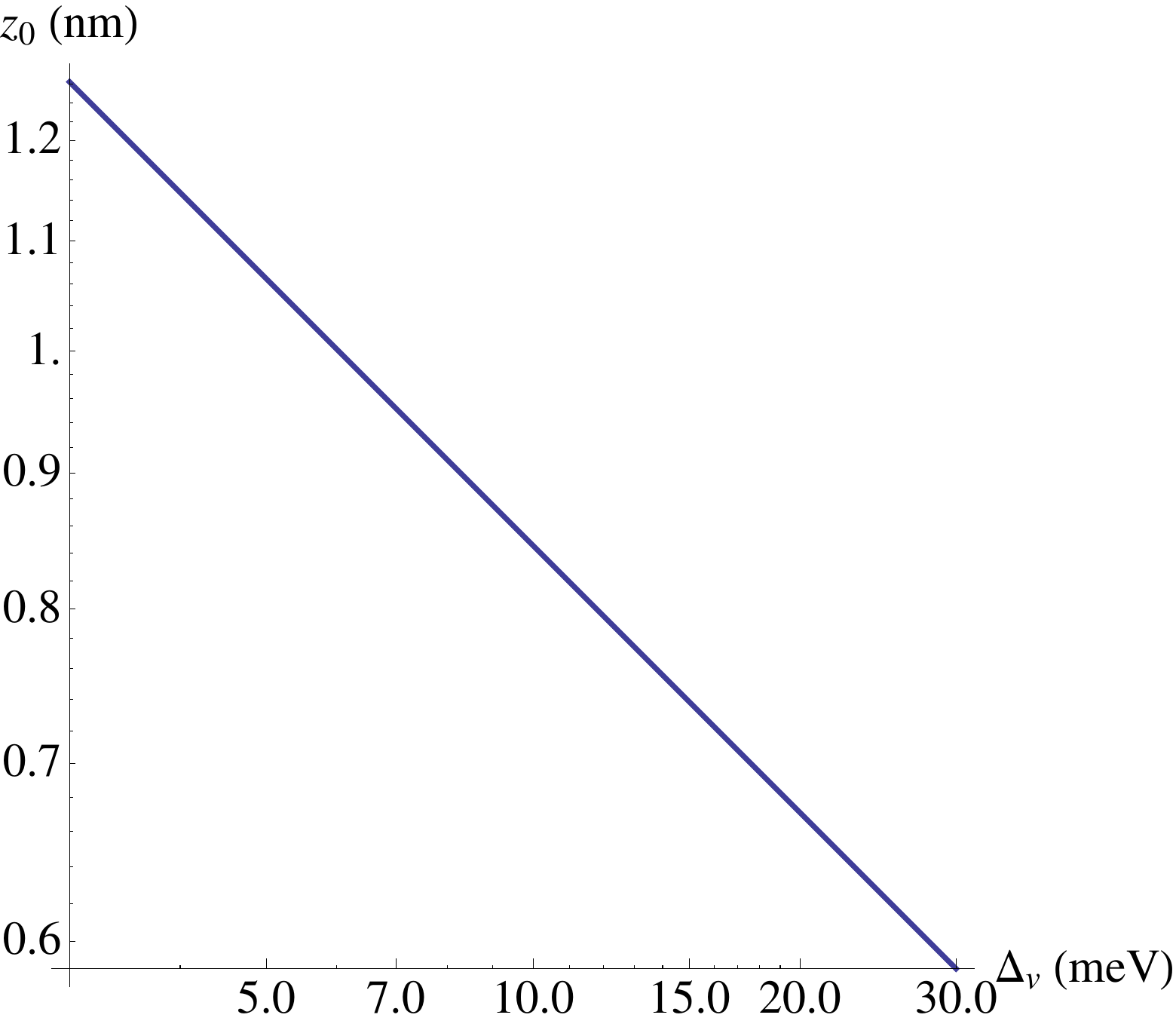}\tabularnewline
\end{tabular}
\par\end{centering}

\caption{\selectlanguage{american}%
Longitudinal and transverse phonon valley relaxation rates for ideal
interfaces versus valley splitting, $\Delta_{v}$. The dashed lines
are with the height of the wave function, $z_{0}$, held constant
(note the peaks at 23 meV and 11 meV respectively) and the solid lines
are the relaxation rates with the changing extent of the wave function
in $z$ (due to the electric field) included. The vertical dimension
of (ideal) triangular quantum well wave function as a function of
valley splitting (theory from Ref. \foreignlanguage{english}{\cite{ART-friesen-valley-orbit}}).
Note that the tighter the confinement in $z$, the faster the relaxation
(due to broadening of the wave function in momentum space.) These
numbers are for $u_{k1}=1$ and should be renormalized down by $u_{k1}^{2}$\foreignlanguage{english}{
(see Appendix \ref{sub:Valley-relaxation-Appendix}). \label{fig:valley relaxation}}\selectlanguage{english}%
}
\end{figure}

\begin{figure}[ptb]
\begin{centering}
\begin{tabular}{c}
\includegraphics[scale=0.45]{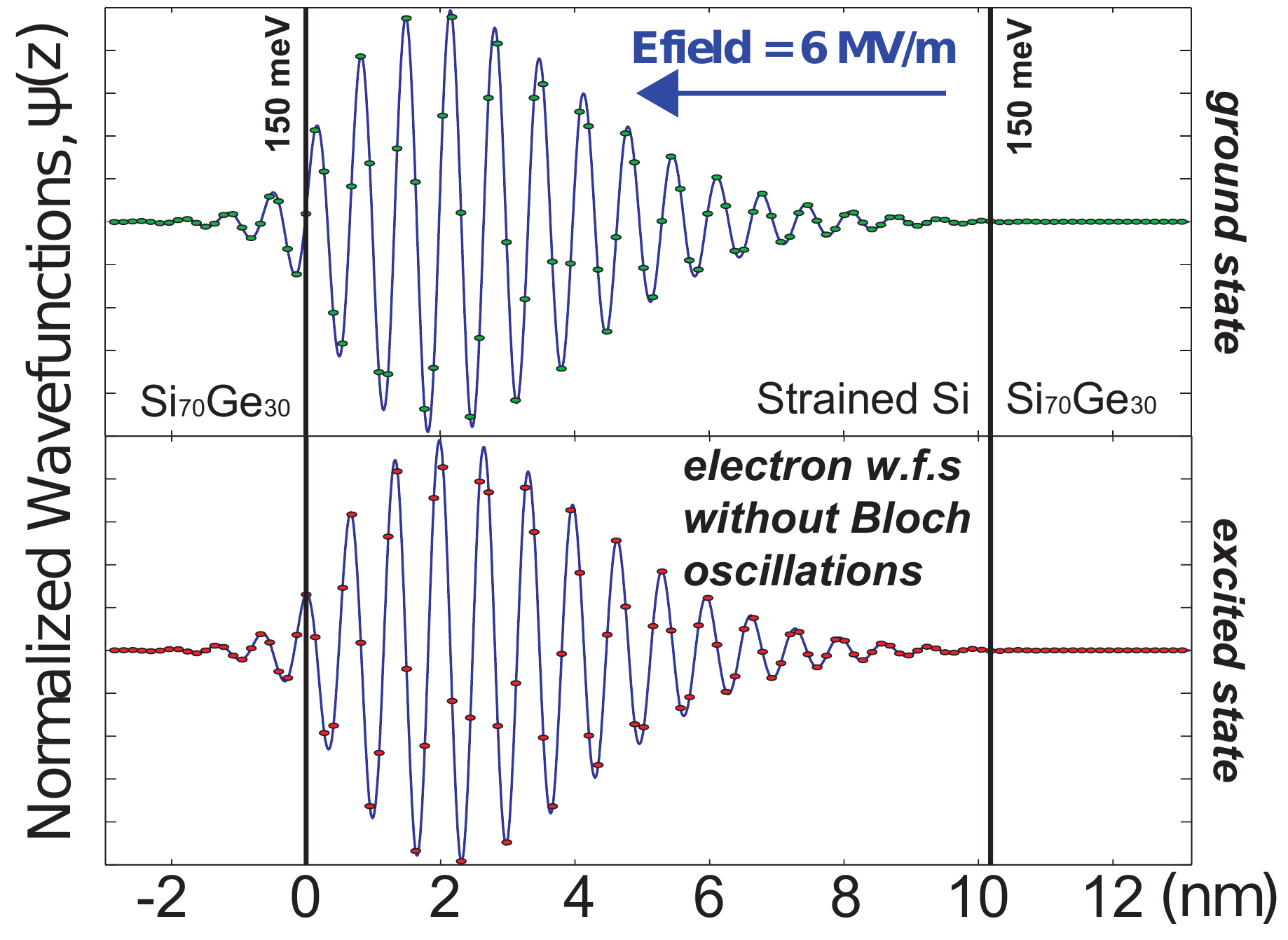}\tabularnewline
\includegraphics[scale=0.3]{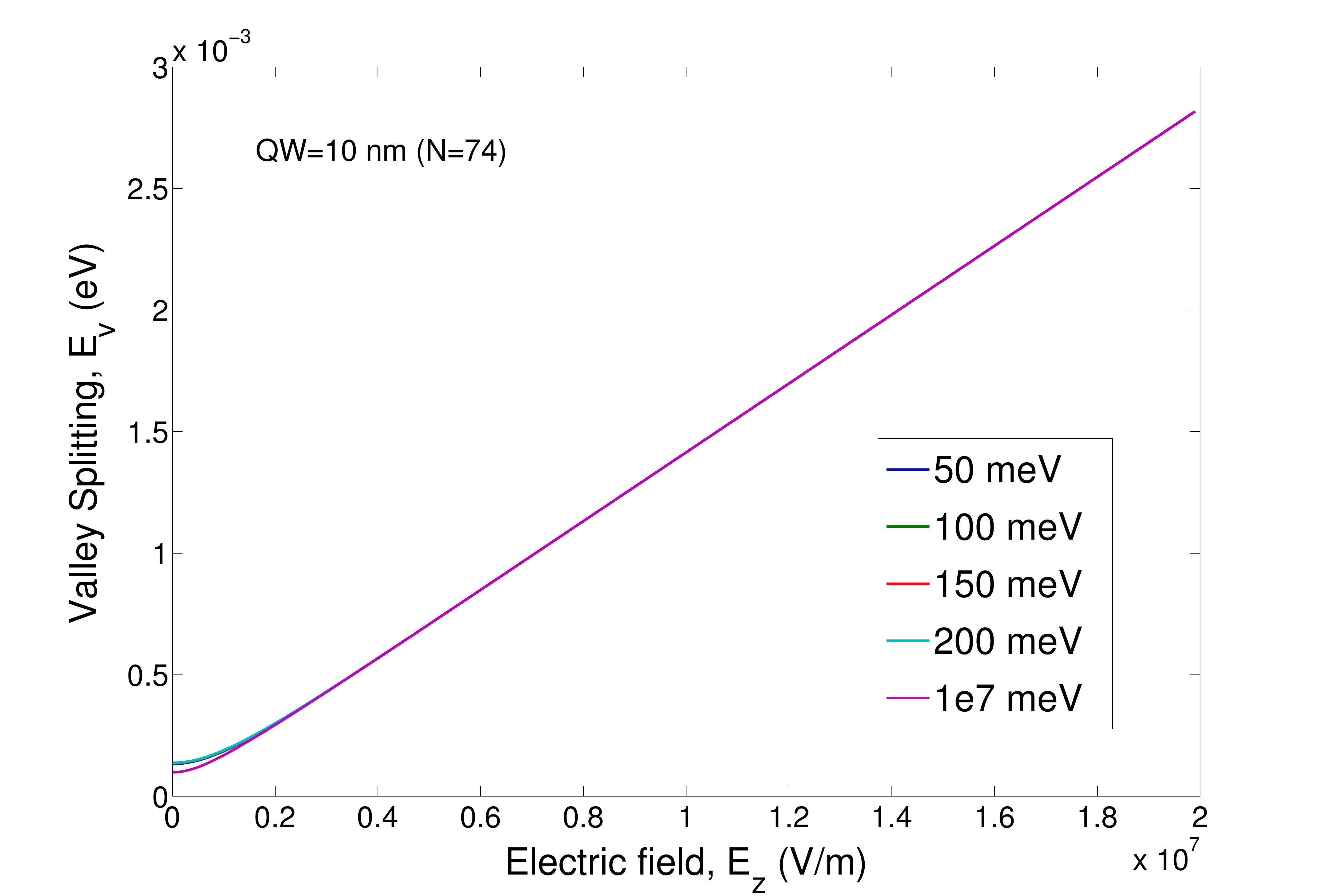}\tabularnewline
\end{tabular}
\par\end{centering}

\caption{\selectlanguage{american}%
Results from 1D tight-binding model for wave functions for ground
and excited valley states and valley splitting for a 10 nm quantum
well. The $z$-component of the electron dot wave function is the
output of a 2-band tight-binding calculation (points) which has been
interpolated (line) for a typical SiGe heterostructure with a quantum
well of 10 nm, barriers of 150 meV, and a large growth direction electric
field due to space-charge separation from the donor layer of $6\times10^{6}$
V/m. Valley splitting in realistic silicon quantum dots will likely
be reduced versus the 1D results presented here due to interface roughness/steps,
etc.\foreignlanguage{english}{ \label{fig:Results-from-1D}}\selectlanguage{english}%
}
\end{figure}
\begin{figure}[ptb]
\begin{centering}
\includegraphics[scale=0.37]{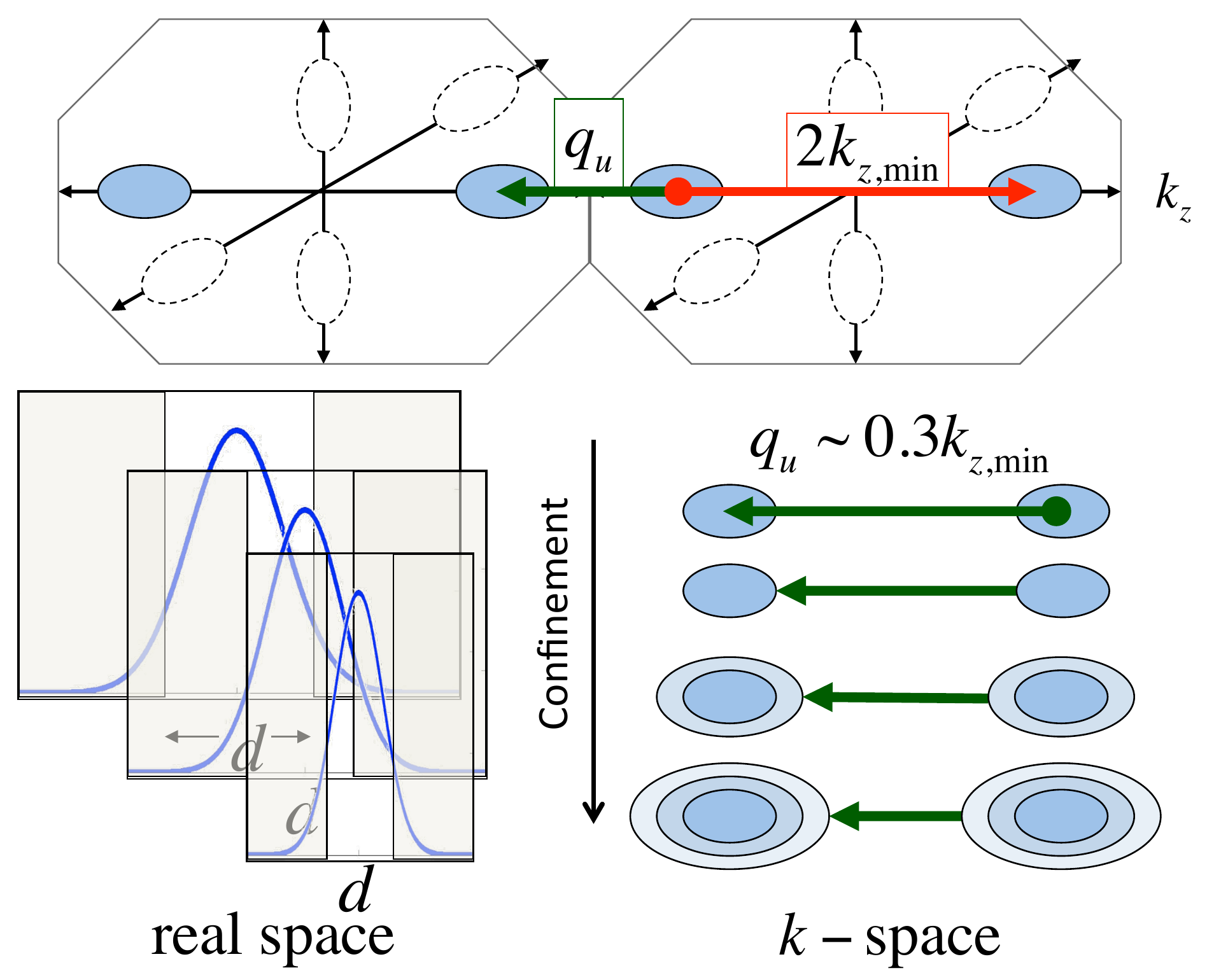}
\par\end{centering}

\caption{\selectlanguage{american}%
Illustration of Umklapp phonon process which enables valley relaxation
in ideal, silicon quantum dots (see text and Appendix \foreignlanguage{english}{\ref{sub:Valley-relaxation-Appendix}).
\label{fig:Umklapp_phonon}}\selectlanguage{english}%
}
\end{figure}
\begin{figure}[ptb]
\begin{centering}
\includegraphics[scale=0.45]{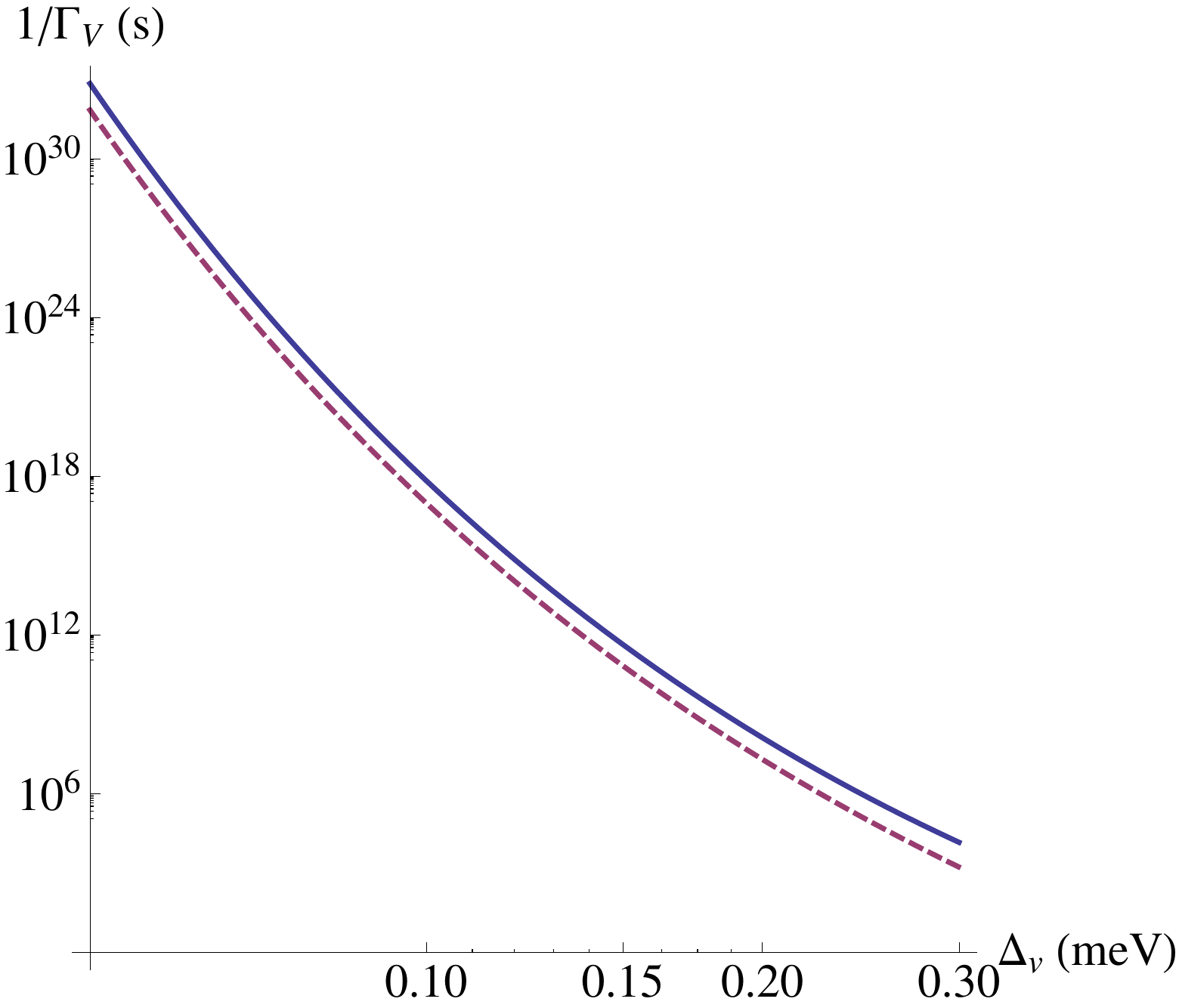}\includegraphics[scale=0.45]{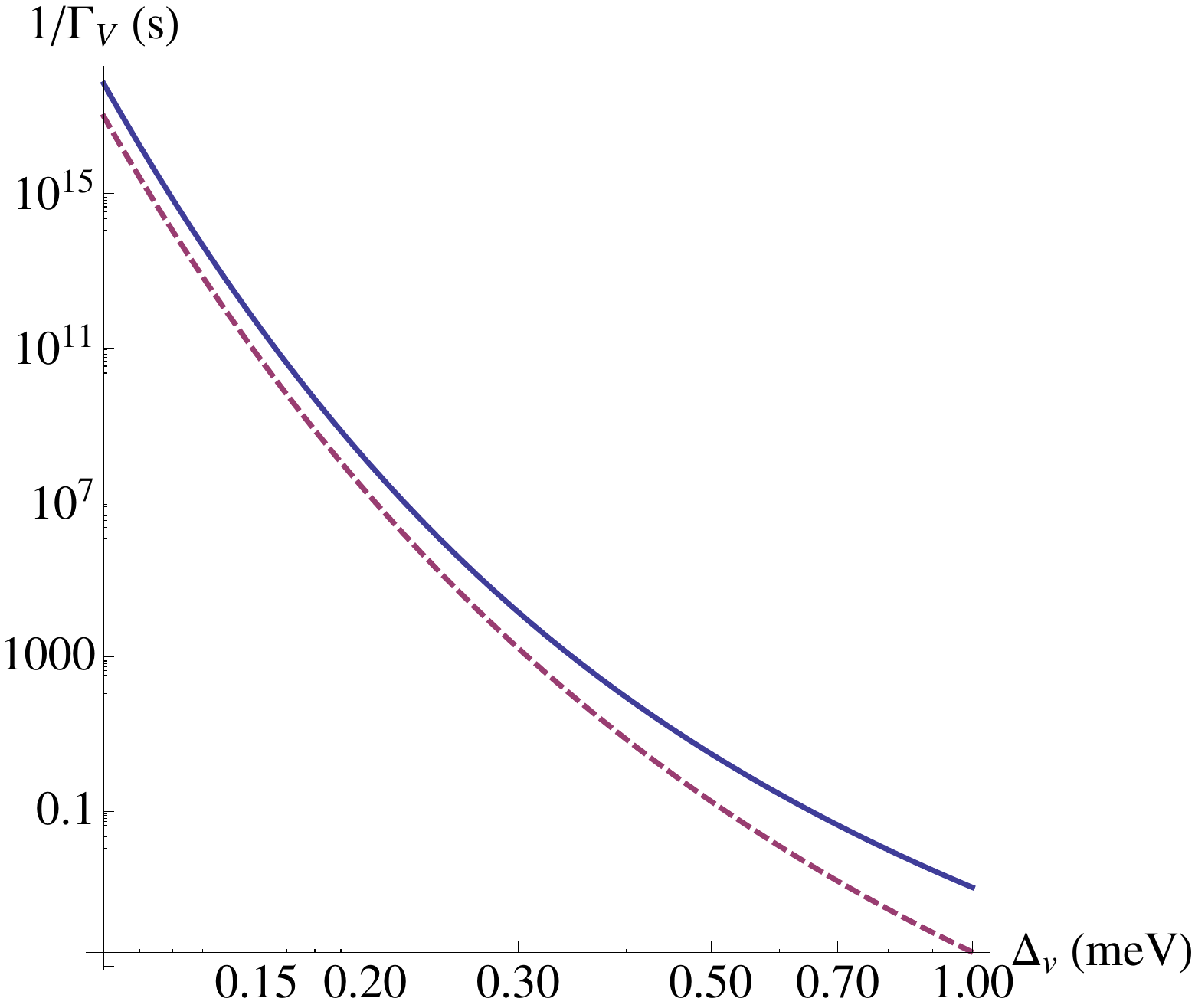}
\par\end{centering}

\begin{centering}
\includegraphics[scale=0.45]{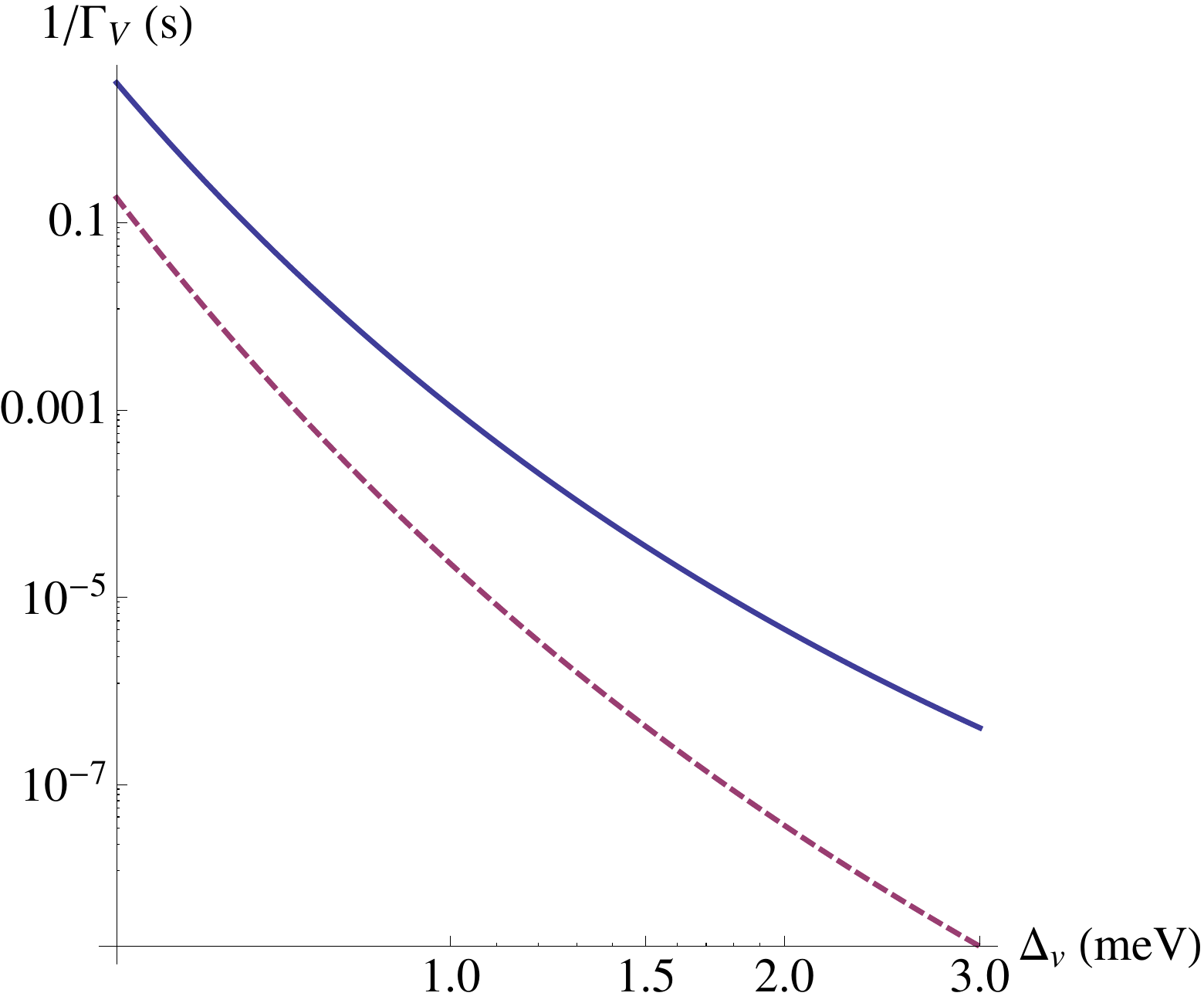}\includegraphics[scale=0.45]{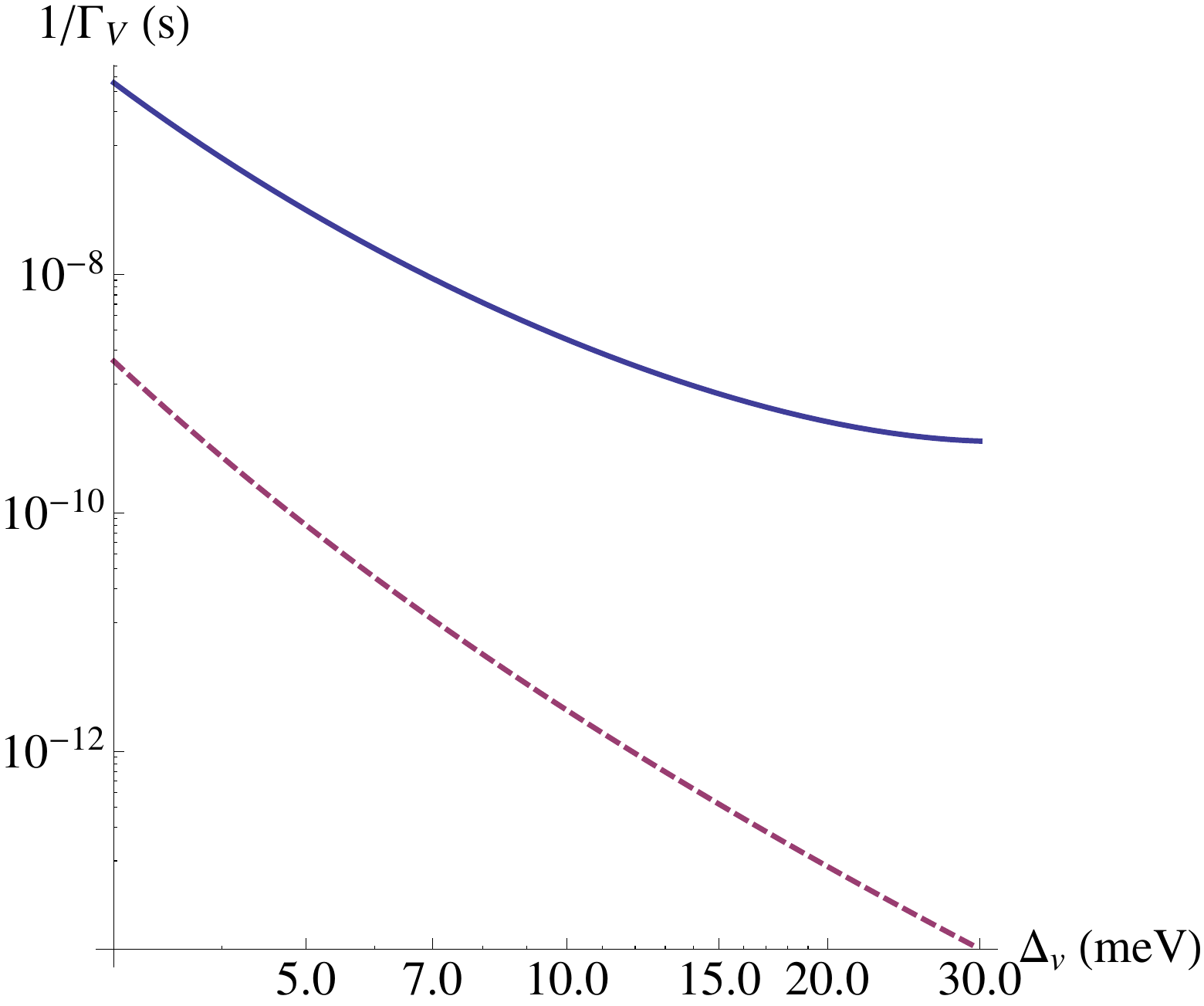} 
\par\end{centering}

\caption{\selectlanguage{american}%
Total valley relaxation time (due to both transverse and longitudinal
phonons) for an ideal-interface, quantum dot excited valley state
in silicon versus valley splitting, $\Delta_{v}$.\foreignlanguage{english}{
}The exact results including all multipole contributions are given
by the blue solid lines while electric dipole approximation (see text)
results are given by the purple dashed lines.\foreignlanguage{english}{
Typical valley splittings in silicon quantum dots are less than 1
meV.\label{fig:Total-valley-relaxation}}\selectlanguage{english}%
}
\end{figure}

\subsection{Ideal (1D) interfaces}

We are concerned with relaxation across the same orbital and spin
states but between valley states ($v=+/-$) in a silicon quantum dot,
particularly the relaxation of the lowest excited valley state with
no change in spin or orbital number (type 3 in Figure \ref{fig:The-circled-numbers}).
Our approach to valley relaxation follows the same procedure as exact
orbital relaxation (Section \ref{sec:Phonon-Bottleneck-Effect} and
Appendix \ref{sub:Orbital-relaxation-Appendix}), where in this case
we replace the matrix element with the inter-valley matrix element:
\[
M_{+-}=\left\langle ns+\right\vert H_{ep}\left\vert ns-\right\rangle .
\]
Assuming no valley-orbit mixing with higher states (separable wave
functions in ($x$, $y$) and $z$), the wave function for an electron
in a lateral silicon quantum dot reads 
\begin{align*}
\psi_{m}^{(v)}(\mathbf{r})=F_{x,y}(x,y) & F_{z}(z)\left[\alpha_{z}^{(v)}u_{z}(\mathbf{r})e^{ik_{m}z}+\alpha_{-z}^{(v)}u_{-z}(\mathbf{r})e^{-ik_{m}z}\right]\\
 & =F_{x,y}(x,y)F_{z}(z)\left[\alpha_{z}^{(v)}\sum_{G_{z}}C_{G_{z}}\exp\left(i(G_{z}+k_{m})z\right)+\alpha_{-z}^{(v)}\sum_{K_{-z}}C_{K_{-z}}\exp\left(i(K_{-z}-k_{m})z\right)\right],
\end{align*}
where $k_{m}$ is the location of the minima along the $z$-axis,
$\alpha^{(\pm)}(z,-z)=1/\sqrt{2}\{1,\pm1\}$ (though these may be
complex in the general case), and we have expanded the Bloch function
in reciprocal lattice vectors, $\mathbf{G}$ or $\mathbf{K}$, $u(\mathbf{r})=\sum_{\mathbf{G}}C_{\mathbf{G}}\exp[i\mathbf{G}\cdot\mathbf{r}].$
The first five terms of the Bloch expansion contribute 90\% of the
wave function amplitude (values from a recent study are listed in
Table 1 of Ref. \cite{ART-Saraiva-Bloch}). Here, the envelope functions
of the two states are the same, the spin states are the same, but
the Kohn-Luttinger oscillations are out of phase (see Figure \ref{fig:Results-from-1D}).
We assume that the wave function consists of Gaussians in all three
dimensions. Following our exact orbital relaxation calculation, the
valley relaxation rate of a parabolic, circular quantum dot in a {[}001{]}-strained
silicon quantum well is 
\begin{align}
\Gamma_{+-} & =\left(C_{k_{0}}^{1}\right)^{2}\frac{\exp\left(-z_{0}(\Delta_{v})q_{u}^{2}/4\right)}{4\pi\rho_{Si}\hbar}\left[\frac{\Delta_{v}^{3}}{\hbar^{3}v_{l}^{5}}\exp\left(\frac{-x_{0}^{2}}{4}\frac{\Delta_{v}^{2}}{\hbar^{2}v_{l}^{2}}\right)\left(\Xi_{d}^{2}P_{l}^{0}+2\Xi_{d}\Xi_{u}P_{l}^{2}+\Xi_{u}^{2}P_{l}^{4}\right)\right.\nonumber \\
 & +\left.\frac{\Delta_{v}^{3}}{\hbar^{3}v_{t}^{5}}\exp\left(\frac{-x_{0}^{2}}{4}\frac{\Delta_{v}^{2}}{\hbar^{2}v_{t}^{2}}\right)\Xi_{u}^{2}\left(P_{t}^{2}-P_{t}^{4}\right)\right]\label{eq:ValleyRelaxExact}
\end{align}
 where 
\[
P_{s}^{n}=\int_{-1}^{1}x^{n}\exp\left[A_{s}x^{2}\right]\sinh^{2}\left[B_{s}x\right]dx
\]
 and 
\begin{align*}
A_{s} & =\frac{1}{4}q_{\Delta s}^{2}\left[x_{0}^{2}-z_{0}^{2}\right],\\
B_{s} & =\frac{1}{8}z_{0}^{2}2q_{\Delta s}q_{u},
\end{align*}
where $\Delta_{v}$ is the valley splitting, $z_{0}$ is the extent
of the wave function (assumed gaussian) in $\hat{z}$ and $q_{u}$
is the phonon wave length of the emitted Umklapp phonon, $q_{u}=0.3k_{max}$.
The details of this calculation are given in Appendix \ref{sub:Valley-relaxation-Appendix}.

Let us compare Eq. \ref{eq:ValleyRelaxExact} to pure orbital relaxation,
Eq. \ref{eq:exact_orbital}. At first glance, the valley relaxation
rate has a $\Delta^{3}$ dependence as opposed to a $\Delta^{4}$
in the orbital case (assuming parabolic dot potentials for both and
matrix elements given due to gaussian wave functions). To understand
this remember that for valley relaxation this transition occurs within
the lowest manifold (both initial and final states have the same $s$-like
envelope function) such that the matrix elements $M\sim1$. In the
orbital case, we must calculate matrix elements from $2p$-like to
$1s$-like states, such that $M\varpropto x_{0}^{2}\varpropto\Delta$.
The valley relaxation expression also includes prominently a $\exp(-z_{0}^{2}q_{u}^{2}/4)$
prefactor absent in the exact orbital case (Eq. \ref{eq:exact_orbital}).
This prefactor predicts that the phonon relaxation rate will be peaked
at the Umklapp phonon energy (assuming $z_{0}$ is constant with $\Delta_{v}$,
which it isn't). Equation \ref{eq:ValleyRelaxExact} also shows the
importance of the $z_{0}$ extent of the wave function; decreasing
$z_{0}$ increases the relaxation rate. These effects are related,
in that Umklapp phonons at $q_{u}=\Delta_{v}/\hbar v_{l,t}$ which
connect valleys in neighboring Brillouin zones are the most efficient
relaxation channel (see Appendix \ref{sub:Valley-relaxation-Appendix}
for more details). Figure \ref{fig:valley relaxation} explicitly
shows the valley relaxation rate in the two cases of fixed $z_{0}$
wave function height and wave function height that changes accurately
with electric field and valley splitting. It turns out that the Bloch
coefficients to the nearest valley at $0.3k_{max}$ are most efficient
and phonons are then emitted in the $z$ direction. As $z_{0}$ gets
compressed, not only does the valley splitting increase due to interface
scattering (approaching the ``critical\textquotedblright{}\ Umklapp
phonon energies at 13.4 meV (longitudinal) and 23.2 meV (transverse)),
but the wave function gets broadened in momentum space, allowing lower
energy phonons to connect the two opposite valleys (see Figure \ref{fig:Umklapp_phonon}).
So Umklapp valley relaxation is possible even at valley splittings
smaller than $q_{u}$. This effect causes the relaxation rate to continually
increase as the valley splitting approaches $q_{u}$, while the other
``typical\textquotedblright{}exponentials kick in at higher splittings
to cause a bottleneck effect as in the orbital case. The line width
should be weakly dependent on the size of the dot in the lateral dimensions
(as is the valley splitting) and much more so dependent on changes
in $z$ extent of the wave function. 

The valley splitting varies roughly linearly with E-field in a perfect
quantum well where the electron only sees one side of the quantum
well. Figure \ref{fig:valley relaxation} shows the valley relaxation
for an ideal interface as a function of valley splitting with a $z_{0}$
that changes correctly with $\Delta_{v}.$ We account for the change
of wave function size in $\hat{z}$ as a function of valley splitting.
We take the ideal theoretical maximum valley splitting as (given in
J) 
\[
\Delta_{v}=\frac{2v_{v}eE}{\Delta E_{c}}\approx2.3\times10^{-29}E
\]
where $E$ (V/m) is the electric field in the $z$-direction, $\Delta E_{c}$
is the conduction band offset, $v_{v}=7.2\times10^{-11}\Delta E_{c}$
(in eV m with $\Delta E_{c}$ in eV) \cite{ART-friesen-valley-orbit}.
Thus, the valley splitting $\Delta_{v}$ depends on the $E$-field,
which also determines the extent of the wave function in $z$. $z_{0}$
now is a function of the E-field in $z$ (which varies by device and
can often be changed somewhat in a single device). For this we define
the wave function in $z$ as (assuming a triangular potential): 
\[
\Psi_{z}=1.4261\sqrt{\kappa}~Ai(\kappa z-2.3381)
\]
 where $\kappa=1/z_{0}=\sqrt[3]{2m^{\ast}eE/\hbar^{2}}$ (in meters).
Now we replace $E$ with $E(\Delta_{v})=\Delta_{v}/(2e\times7.2\times10^{-11})$
(in V/m). $Ai$ is the Airy function. So the extent of the wave function
in, $z_{0}(\Delta_{v})$, changes with the valley splitting as: 
\[
z_{0}(\Delta)=\left(\frac{\hbar^{2}(7.2\times10^{-11})}{m^{\ast}\Delta_{v}}\right)^{1/3}.
\]

For completeness, we may also look for an expression for the valley
relaxation in the electric-dipole approximation. A reasonable approximation
is to set $q_{\Delta s}^{2}=0$ in Equation \ref{eq:ValleyRelaxExact}
which, to leading order for a parabolic potential in all three dimensions,
gives 
\begin{equation}
\Gamma_{v}^{ED}\approx\left|C_{k}\right|^{2}\exp\left(-\frac{1}{4}q_{u}^{2}z_{0}\left(\Delta_{v}\right)^{2}\right)\frac{\Xi_{u}^{2}\Delta_{v}^{3}}{30\pi v_{t}^{5}\rho_{Si}\hbar^{4}}.\label{eq:ValleyRelaxApproximate}
\end{equation}
 The exact and approximate valley relaxation times are compared in
Figure \ref{fig:Total-valley-relaxation}.

\subsection{Comments on imperfect interfaces (tilt, roughness, and alloy composition)}

Our expression, Eq. \ref{eq:ValleyRelaxExact}, for valley relaxation
is for perfectly smooth interfaces. As we discussed above, imperfect
interfaces will cause mixing between the orbital and valley states.
In realistic devices, miscut, alloy variability, surface roughness,
etc. will be present. This distorts or mixes the orbital and valley
states such that valley is no longer a good quantum number \cite{ART-friesen-valley-orbit}
(e.g., the envelope functions within the $s$-manifold states can
now be different and/or non-$s$-like). In this case, one must realize
that the wave functions of orbital states will be different from the
Gaussian wave functions assumed for s-like and p-like dot states used
in the orbital relaxation section. In reality they will be not be
separable, and there will be sample dependence. However, all these
energy relaxation calculations are proportional to $M^{2},$ where
$M\ $ is the dipole matrix element. In contrast, the energy gap dependence
is much higher: $\Delta^{5}$ ($\Delta^{4}$ for parabolic dots) and
the field dependence is $B^{7}.$ $\Delta$ is much easier to determine
than $M$ and is much more important, which means that our predictions
are still useful. A likely exception is our ideal calculation for
possible long-lived valley states. In the non-ideal case, there is
likely to be valley-orbit mixing, providing an avenue for relaxation
via $G=0$ phonons as in the orbital relaxation case. While a full
theory of this mixing is possible (utilizing appropriate wave functions,
e.g., following \cite{ART-friesen-valley-orbit,ART-Chutia-Valley}),
the matrix elements depend on the exact specification of the interface
for the dot being measured (which is difficult to ascertain) and are
not considered here. Because the orbital relaxation is fast, the long-lived
predictions for valley excited states will be wrong in this case.
Although the degree of the this mixing/distortion depends on the specific
device in question, we can crudely write that 
\[
\Gamma_{v}^{imperfect}=f_{mix}^{2}\Gamma_{10}\left(\Delta_{v}\right)+\left(1-f_{mix}\right)^{2}\Gamma_{v}^{ideal}
\]
where $f_{mix}$ is the notional fraction of orbital wave function
mixed in with the valley wave function and $\Gamma_{10}\left(\Delta_{v}\right)$
is the orbital relaxation rate across the measured valley splitting
of the measured state. In some cases $f_{mix}$ could be calculated
\cite{ART-friesen-valley-orbit} but generally it will depend on microscopic
details of the silicon quantum dot.

\section{Comparison with experiments and Conclusions}

Some spin relaxation experiments have been performed on electrons
in silicon dot devices (see Figure \ref{fig:Spin-relaxation-time-1}).
Our theory for spin relaxation in the electric-dipole approximation
should be valid only well before any degeneracy between Zeeman and
orbital splitting is reached, the regime of a silicon spin qubit.
In Ref \cite{ART-Hayes-T1Si}, $T_{1}$ was measured in a lateral,
depletion mode Si/SiGe quantum dot for three different magnetic fields
between 1 and 2 T. Comparing to Eq. \ref{eq:anisotropy}, the data
appears to follow a $B^{-7}$ rate but with relatively few points
it is difficult to make firm conclusions. These data points fall on
the $0.4$ meV orbital splitting line (obtained using the value of
SOC from Ref. \cite{ART-Wilamowsky/Rossler-2002}), which is roughly
consistent with the lithographic size of the dot but differs from
the $\Delta=2$ meV value used for the theory model in that paper
(though it is unclear if the experimental orbital splitting was measured)
\cite{ART-Hayes-T1Si}. $T_{1}$ has also been measured in a metal-oxide-semiconductor
system with the dot near the Si/SiO$_{2}$ interface \cite{ART-House-Jiang-T1};
they report measuring an orbital splitting in this device of roughly
0.4 meV. There are five data points that fit a $T_{1}\sim$ $B^{-7}$
law reasonably well; however, this behavior is observed only for $B>3$
T. Here theory predicts a roughly order of magnitude shorter spin
relaxation time for a dot with $\Delta=0.4$ meV (the data $>$ 3
T behave as if the orbital splitting is 1 meV). For $B<3$ T, $T_{1}$
is roughly independent of $B$, with $T_{1}=40$ ms (possibly limited
by charge noise or some other mechanism). Finally, a recent measurement
of $T_{1}$ in a laterally gated Si/SiGe dot in a doped device yielded
$T_{1}=2.3$ s at a field of $B=1.85$ $T.$ None of the data shows
evidence of hot spot (fast spin relaxation cusp) behavior.

On the whole, these experiments give good evidence that the SOC-mediated
spin-phonon interaction is the dominant channel at high fields: both
the B dependance and, importantly, the overall magnitude are consistent
with theory. The value of the Rashba coefficient that is used in Figure
\ref{fig:Spin-relaxation-time-1} may not be appropriate for a MOS
structure (although inversion layer spin relaxation measurements show
relaxation times \cite{ART-Shankar-Lyon-SiO2} within an order of
magnitude of those found in SiGe QWs) and can vary from device to
device based on material, interface roughness, electric field at the
interface, etc. \cite{ART-Nestoklon-Ivchenko,ART-Prada-spin-orbitSiGe}.
A much smaller SOC constant might explain the order of magnitude difference
between theory and experiment for the various systems. It would be
very useful in the future to attempt to characterize the SOC strength
in these wafers by other means, for example via 2DEG spin relaxation
\cite{ART-Glazov-2DEGspin}, although the SOC may vary at a microscopic
level. It is also very important to check the dependance of $T_{1}$
on field direction. This has not yet been done experimentally. Looking
beyond the electric-dipole approximation, there may be a strong and
sudden increase in the relaxation rate for spin relaxation when the
spin splitting matches the orbital splitting; no such effect is expected
if the 1st excited state is a (pure) valley state. A complication
to this picture may be a lack of mixing with the orbital state if
the SOC is only Dresselhaus-like \cite{ART-Bulaev-Loss} (even in
the ideal case). Lastly, our assumption of Fock-Darwin wave functions
may not be correct (do to dot asymmetry, interface roughness, etc.);
leading to matrix elements between dot states that influence the relaxation
rate up or down.

While these results are encouraging, they do not constitute a complete
vindication of theory. Hence we discuss how to combine the results
of different measurements to fix some universal quantities. Specializing
Eq. \ref{eq:orbital} to the first excited state, neglecting anisotropy
and the dipole moment in the $z$-direction, noting that $v_{t}^{-7}>>v_{\ell}^{-7},$
using Eq. \ref{eq:upsilon}, and taking $k_{B}T<<\Delta,$ we find
an orbital relaxation rate from the first excited state to the ground
state:
\begin{equation}
\Gamma_{12}=\frac{4\left\vert \Delta\right\vert ^{5}\Xi_{u}^{2}}{105~\hbar^{6}\pi\rho_{Si}v_{t}^{7}}\left\vert M_{x}^{\left(12\right)}\right\vert ^{2}.
\end{equation}
 With similar assumptions for the spin relaxation and specializing
to an in-plane field, we have from Eq. \ref{eq:anisotropy} 
\begin{equation}
\frac{1}{T_{1}}=\frac{1}{21}\Xi_{u}^{2}\frac{m_{t}^{2}}{\pi\hbar^{10}\rho_{Si}v_{t}^{7}}\left(g\mu_{B}B\right)^{7}\frac{\xi^{2}}{e^{4}}\beta^{2}.
\end{equation}
 We wish to eliminate the poorly determined quantities $~M_{x}^{\left(12\right)}$
and $\xi,$ both related to the size of the dot, and the electron-phonon
coupling strength and phonon velocity, in favor of measurable numbers
(as far as possible). To do so, we use Eqs. \ref{eq:mxy}, and \ref{eq:xi}:
\begin{equation}
\Gamma_{12}T_{1}=\frac{4}{5}\frac{\Delta^{8}}{E_{Z}^{7}}\frac{\hslash^{2}}{m_{t}\beta^{2}}.
\end{equation}
Here $E_{Z}=g\mu_{B}B$ is the Zeeman splitting. $\Delta$ is the
energy of the first excited state, which can be measured independently,
by transport spectroscopy or looking at Orbach processes at higher
temperatures. \ $\beta$ can also be measured by other means, though
this is not straightforward \cite{ART-Studer-Ensslin}. Thus it would
take a combination of measurements to use the absolute magnitude of
$T_{1}$ as a test of theory. Note that this analysis assumes that
the first excited state is purely orbital in nature and that the fast
orbital relaxation time can be measured. The former may be ameliorated
in the valley case where there is strong valley-orbital mixing, or
if not, $\Gamma_{21}$ can be replaced with the ideal valley relaxation
time. The latter may be a difficult experimental constraint. Failing
this, the $1/T_{1}\sim B^{7},$ $1/T_{1}\sim\left[1+2n_{B}\left(E_{Z}/k_{B}T\right)\right],$
and the field anisotropy given by Eq. \ref{eq:anisotropy} provide
the best tests. We note also that the field anisotropy of the spin
relaxation, Eq. \ref{eq:anisotropy}, can help determine the relative
contributions of the Rashba and Dresselhaus-like SOC contributions.

We have focused on the lowest lying state configurations of a silicon
quantum dot that are most relevant for quantum computing and have
found that, in general, spin-based quantum computing benefits from
orbital and valley states being as high in energy as possible. We
began by considering phonon relaxation of excited orbital states across
the same valley state in a lateral silicon QD. We found that orbital
relaxation could be dramatically faster in biaxially strained silicon
than in the bulk. This, for example, speeds up spin qubit initialization
via optical pumping schemes \cite{ART-Friesen-ReadoutLetter-2003}
as well as possible leakage to excited states via phonon excitation.
The phonon bottleneck effect will eventually decrease the orbital
relaxation rate but only for unrealistically small dots. In contrast,
spin relaxation can easily be seconds (even when the magnetic field
points along the growth direction) and $T_{1}$ increases for small
dots and low magnetic fields. At small magnetic fields, charge noise
could play a dominant role. Valley relaxation can also be long in
ideal dots, especially for small valley splittings. In non-ideal dots,
although not quantitatively considered here, valley relaxation will
likely be comparable to orbital relaxation.

The theory proposed here depends on the correct identification of
excited states as either orbital excited states or valley excited
states. Theoretical considerations for spin and valley relaxation
in cases where the states are not purely orbital, valley, or spin---in
other words they are mixed due to, for example, disorder or surface
roughness in realistic devices---and for regimes beyond small B-field
(where degeneracies come into play) are subjects for future work.
\begin{acknowledgments}
We are grateful for helpful conversations with M. Friesen, S. N. Coppersmith,
M.A. Eriksson, and R. Ruskov. We thank the group of H.W. Jiang for
access to their data. RJ was supported by ARO grant no. W911NF-11-1-0030.
\end{acknowledgments}
%
%
%
%
%
%

%
%
%
%
%
%
\section{Appendix}

\subsection{Orbital relaxation: beyond the electric dipole approximation\label{sub:Orbital-relaxation-Appendix}}

To calculate the matrix element of $H_{ep}$ between orbital states
$m$ and $n$ in Eq. \ref{eq:GoldenRuleOrbital} beyond the electric
dipole approximation, we begin with the full expression (see Eq. \ref{eq:Helectronphonon}):
\begin{align}
\left\langle n\right\vert H_{ep}\left\vert m\right\rangle  & =\int\psi_{m}^{\ast}\left(H_{ep}\right)\psi_{n}dV\nonumber \\
 & =\sum_{\mathbf{q},\lambda}\sum_{i,j}\alpha_{m}^{i}\alpha_{n}^{j}\left[\Xi_{d}\left(\mathbf{e}\left(\mathbf{q},\lambda\right)\cdot\mathbf{q}\right)+\Xi_{u}\left(\mathbf{q}\cdot\mathbf{K}^{\left(i\right)}\right)\left(\mathbf{e}\left(\mathbf{q},\lambda\right)\cdot\mathbf{K}^{\left(i\right)}\right)\right]a_{q}^{\ast}\int F_{i}^{\ast}F_{j}u_{\boldsymbol{k}_{i}}^{\ast}u_{\boldsymbol{k}_{j}}e^{-i\left(-\mathbf{k}_{i}+\mathbf{k}_{j}+\mathbf{q}\right)\cdot\mathbf{r}}dV.\label{eq:FullValleyM}
\end{align}
 We proceed following the derivation by Castner \cite{ART-Castner-OrbachSpinLatticeinSi-1967}.
This derivation will be useful when valley relaxation is considered.
A function which is periodic with the period of the lattice may be
expanded in a Fourier series in the reciprocal lattice vectors $\mathbf{Q}_{\nu}$,
so
\[
u_{k_{i}}^{\ast}(\mathbf{r})u_{k_{j}}(\mathbf{r})=\sum_{\nu}C_{\boldsymbol{k}_{i}-\boldsymbol{k}_{j}}^{\nu}e^{i\mathbf{Q}_{\nu}\cdot\mathbf{r}},
\]
 and the integral in $M_{mn}$ becomes
\[
\sum_{\nu}C_{\boldsymbol{k}_{i}-\boldsymbol{k}_{j}}^{\nu}\int\left\vert F_{i}\right\vert ^{2}e^{-i\left(-\mathbf{k}_{i}+\mathbf{k}_{j}+\mathbf{q}-\mathbf{Q}_{\nu}\right)\cdot\mathbf{r}}d\mathbf{r}.
\]
 The envelope probability can be Fourier transformed, 
\[
\left\vert F_{i}(\mathbf{r})\right\vert ^{2}=\frac{1}{(2\pi)^{3}}\sum_{\mathbf{k}^{\prime\prime}}f\left(\mathbf{k}^{\prime\prime}\right)e^{i\boldsymbol{k}^{\prime\prime}\boldsymbol{\cdot r}}.
\]
 Plugging this into the integral in $M_{mn}$ gives 
\[
\sum_{\nu}C_{\boldsymbol{k}_{i}-\boldsymbol{k}_{j}}^{\nu}\int\frac{1}{(2\pi)^{3}}\sum_{\mathbf{k}^{\prime\prime}}f\left(\mathbf{k}^{\prime\prime}\right)e^{-i\left(-\mathbf{k}_{i}+\mathbf{k}_{j}+\mathbf{q}-\mathbf{Q}_{\nu}-\mathbf{k}^{\prime\prime}\right)\cdot\mathbf{r}}d\mathbf{r}
\]
 which equals 
\[
\sum_{\nu}C_{\boldsymbol{k}_{i}-\boldsymbol{k}_{j}}^{\nu}\frac{1}{(2\pi)^{3}}\sum_{\mathbf{k}^{\prime\prime}}f\left(\mathbf{k}^{\prime\prime}\right)~\delta^{3}\left(\mathbf{k}_{i}-\mathbf{k}_{j}-\mathbf{q}+\mathbf{Q}_{\nu}+\mathbf{k}^{\prime\prime}\right)=\sum_{\nu}C_{\boldsymbol{k}_{i}-\boldsymbol{k}_{j}}^{\nu}f^{ij}(\mathbf{-k}_{i}+\mathbf{k}_{j}+\mathbf{q}-\mathbf{Q}_{\nu}).
\]
 Finally, the matrix element is given by 
\begin{equation}
\left\langle n\right\vert H_{ep}\left\vert m\right\rangle =\sum_{i,j}\alpha_{m}^{i}\alpha_{n}^{j}\left[\Xi_{d}\left(\mathbf{e}_{s}\cdot\mathbf{q}\right)+\Xi_{u}\left(\mathbf{q}\cdot\mathbf{K}^{\left(i\right)}\right)\left(\mathbf{e}_{s}\cdot\mathbf{K}^{\left(i\right)}\right)\right]a_{q}^{\ast}\sum_{\nu}C_{\boldsymbol{k}_{i}-\boldsymbol{k}_{j}}^{\nu}~f^{ij}(\mathbf{-k}_{i}+\mathbf{k}_{j}+\mathbf{q}-\mathbf{Q}_{\nu}).\label{eq:exactorbitalelement}
\end{equation}

We are calculating an intra-valley scattering process (orbital relaxation
with no change in valley state) so $\mathbf{Q}_{\nu}=0$ is the dominant
term, $\mathbf{k}_{i}=\mathbf{k}_{j}$, and $\alpha_{m}=\alpha_{n}$
which gives 
\[
\left\langle n\right\vert H_{ep}\left\vert m\right\rangle =\sum_{i,j}\alpha_{m}^{i}\alpha_{n}^{j}\left[\Xi_{d}\left(\mathbf{e}\left(\mathbf{q},\lambda\right)\cdot\mathbf{q}\right)+\Xi_{u}\left(\mathbf{q}\cdot\mathbf{K}^{\left(i\right)}\right)\left(\mathbf{e}\left(\mathbf{q},\lambda\right)\cdot\mathbf{K}^{\left(i\right)}\right)\right]~C_{\boldsymbol{k}_{i}-\boldsymbol{k}_{j}}^{\nu}~f^{mn}(\mathbf{q}),
\]
 and for the most relevant transition, 
\begin{align*}
\left\langle 2\right\vert H_{ep}\left\vert 1\right\rangle  & =\sum_{i,j}\alpha_{m}^{i}\alpha_{n}^{j}\left[\Xi_{d}q_{l}+\Xi_{u}e_{z}q_{z}\right]~C_{\boldsymbol{k}_{i}-\boldsymbol{k}_{j}}^{0}~f^{12}(\mathbf{q})\\
 & =\left[\Xi_{d}q_{l}+\Xi_{u}e_{z}q_{z}\right]C_{\boldsymbol{k}_{i}-\boldsymbol{k}_{j}}^{0}~f^{12}(\mathbf{q}),
\end{align*}
 where $q_{l}=q_{\Delta}=\Delta/hv_{l}$ for longitudinal phonons
but $q_{\ell}=0$ for transverse phonons ($q_{\Delta}=\Delta/hv_{t}$).
$C_{\boldsymbol{k}_{i}-\boldsymbol{k}_{j}}^{0}$ is the first coefficient
in the Bloch wave expansion (see Table X for the largest contributions).

The envelope function of the ground state QD wave function in the
absence of a magnetic field in the lowest approximation is a \ product
of Gaussians, $F^{(1)}(\mathbf{r})=F(x,y,z)=F(x)F(y)F(z),$ where
$F(x)=\left(2/\pi\right)^{1/4}x_{0}^{-1/2}\exp\left(-x^{2}/x_{0}^{2}\right),$
$F(y)=\left(2/\pi\right)^{1/4}y_{0}^{-1/2}\exp\left(-y^{2}/y_{0}^{2}\right),$
and $F(z)=\left(2/\pi\right)^{1/4}z_{0}^{-1/2}\exp\left(-z^{2}/z_{0}^{2}\right).$
The excited state, if $y_{0}>x_{0}>>z_{0}$, is $F^{(2)}(\mathbf{r})=F(x,y,z)=F(x)F^{(2)}(y)F(z),$
where 
\[
F^{(2)}(y)=\left(2/\sqrt{y_{0}^{3}}\right)\left(2/\pi\right)^{1/4}y\exp\left(-y^{2}/y_{0}^{2}\right).
\]
 Then, the overlap integral is given by
\begin{align*}
f^{(12)}(\mathbf{q}) & =\int F^{(2)}(\mathbf{r})e^{i\mathbf{q}\cdot\mathbf{r}}F^{(1)}(\mathbf{r})d\mathbf{r},\\
f^{(1)}(q_{x})f^{(12)}(q_{y})f^{(1)}(q_{z}) & =\int F^{(1)}(x)^{2}e^{iq_{x}x}dx\int F^{(2)}(y)e^{iq_{y}y}F^{(1)}(y)dy\int F^{(1)}(z)^{2}e^{iq_{z}z}dz\\
 & =\exp\left(-\frac{1}{8}x_{0}^{2}q_{x}^{2}\right)\frac{iy_{0}}{2}q_{y}\exp\left(-\frac{1}{8}y_{0}^{2}q_{y}^{2}\right)\exp\left(-\frac{1}{8}z_{0}^{2}q_{z}^{2}\right).
\end{align*}
 Inserting these into the Golden Rule, we find that
\begin{align}
\Gamma_{21} & =\frac{2\pi}{\hbar}\sum_{\mathbf{q},s}\left\vert M_{21}\right\vert ^{2}\delta\left(\Delta-\hbar\omega_{\mathbf{q},s}\right)\label{eq:GoldenRule}\\
 & =\frac{2\pi}{\hbar}\sum_{s}\frac{V}{(2\pi)^{3}}\int_{0}^{\infty}q^{2}dq\int\sin\theta~d\theta~d\phi~\left\vert M_{21}\right\vert ^{2}\frac{1}{\left\vert -\hbar v_{s}\right\vert }\delta\left(q-\frac{E_{21}}{\hbar v_{s}}\right),
\end{align}
 continuing, 
\begin{align*}
\Gamma_{21} & =\frac{2\pi}{\hbar^{2}}\sum_{s}\frac{V}{(2\pi)^{3}}\int_{0}^{\infty}q^{2}dq\int\sin\theta~d\theta~d\phi\frac{1}{v_{s}}\left\vert \left[\Xi_{d}q_{l}+\Xi_{u}e_{z}q_{z}\right]a_{q}^{\ast}C_{k_{i}}^{0}f^{12}(\mathbf{q})\right\vert ^{2}\delta\left(q-q_{\Delta}\right)\\
 & =\frac{\left(n_{q}+1\right)\left(C_{k_{0}}^{0}\right)^{2}}{2(2\pi)^{2}\rho_{Si}\hbar}\sum_{s}q_{\Delta s}~I_{s}.
\end{align*}
 We are left with calculating the three angular integrals $I$ which
have units kg$^{2}$/s$^{2}$ and are defined as
\begin{align*}
I_{s} & =\int\sin\theta\frac{1}{v_{s}^{2}}\left\vert \left[\Xi_{d}q_{\Delta l}+\Xi_{u}e_{z}q_{\Delta s}\hat{q}_{z}\right]f^{12}(q_{\Delta s}\mathbf{\hat{q}})\right\vert ^{2}d\theta d\phi\\
 & =\int\sin\theta\frac{1}{v_{s}^{2}}\left[\Xi_{d}q_{\Delta l}+\Xi_{u}e_{z}q_{\Delta s}\hat{q}_{z}\right]^{2}\exp\left(-\frac{1}{4}x_{0}^{2}q_{x}^{2}\right)\frac{y_{0}^{2}}{4}q_{y}^{2}\exp\left(-\frac{1}{4}y_{0}^{2}q_{y}^{2}\right)\exp\left(-\frac{1}{4}z_{0}^{2}q_{z}^{2}\right)d\theta d\phi.
\end{align*}
 We can immediately point out that $I_{t_{2}}=0$ because $e_{z}(t_{1})=0$.
Then, 
\begin{align*}
I_{s} & =\int\sin\theta\frac{q_{\Delta s}^{2}y_{0}^{2}}{4v_{s}^{2}}\left[\Xi_{d}q_{\Delta l}+\Xi_{u}e_{zs}q_{\Delta s}\cos\theta\right]^{2}\exp\left(-\frac{1}{4}x_{0}^{2}q_{\Delta s}^{2}\sin^{2}\theta\cos^{2}\phi\right)\\
 & \times\sin^{2}\theta\sin^{2}\phi\exp\left(-\frac{1}{4}y_{0}^{2}q_{\Delta s}^{2}\sin^{2}\theta\sin^{2}\phi\right)\exp\left(-\frac{1}{4}z_{0}^{2}q_{\Delta s}^{2}\cos^{2}\theta\right)d\theta d\phi.
\end{align*}
 If we assume an approximately circular dot in $x$ and $y$, then
using $\cos^{2}\phi+\sin^{2}\phi=1$ we can do the $\phi$ integral
easily,
\begin{align*}
I_{s} & =\pi\exp\left(-\frac{1}{4}x_{0}^{2}q_{\Delta s}^{2}\right)\int\frac{q_{\Delta s}^{2}y_{0}^{2}}{4v_{s}^{2}}\left[\Xi_{d}q_{\Delta l}+\Xi_{u}e_{zs}q_{\Delta s}x\right]^{2}(1-x^{2})\exp\left(\frac{1}{4}(x_{0}^{2}-z_{0}^{2})q_{\Delta s}^{2}x^{2}\right)dx,\\
I_{l} & =\pi\exp\left(-\frac{1}{4}x_{0}^{2}q_{\Delta s}^{2}\right)\frac{q_{\Delta l}^{2}y_{0}^{2}}{4v_{l}^{2}}\left[\Xi_{d}^{2}q_{\Delta l}^{2}(A_{l}^{0}-A_{l}^{2})+2\Xi_{d}\Xi_{u}q_{\Delta l}^{2}(A_{l}^{2}-A_{l}^{4})+\Xi_{u}^{2}q_{\Delta l}^{2}(A_{l}^{4}-A_{l}^{6})\right],\\
I_{t_{2}} & =\pi\exp\left(-\frac{1}{4}x_{0}^{2}q_{\Delta t}^{2}\right)\frac{q_{\Delta t}^{2}y_{0}^{2}}{4v_{t}^{2}}\Xi_{u}^{2}q_{\Delta t}^{2}\left[A_{t}^{2}-2A_{t}^{4}+A_{t}^{6}\right],
\end{align*}
 where 
\[
A_{s}^{n}=\int_{1}^{-1}x^{n}\exp\left(\frac{1}{4}(x_{0}^{2}-z_{0}^{2})q_{\Delta s}^{2}x^{2}\right).
\]
 Finally, the orbital relaxation rate for a parabolic dot (in all
three dimensions) from its first excited state is given by (with $E_{21}=\Delta$,
the common notation)
\begin{align}
\Gamma_{21} & =\frac{\left(n_{q}+1\right)\left(C_{k_{0}}^{0}\right)^{2}}{2(2\pi)^{2}\rho_{Si}\hbar}\frac{\pi y_{0}^{2}}{4}\times\nonumber \\
 & \left\{ \frac{\exp\left(-\frac{1}{4}x_{0}^{2}q_{\Delta l}^{2}\right)}{v_{l}^{2}}\frac{\Delta^{5}}{\hbar^{5}v_{l}^{5}}\left[\Xi_{d}^{2}(A_{l}^{0}-A_{l}^{2})+2\Xi_{d}\Xi_{u}(A_{l}^{2}-A_{l}^{4})+\Xi_{u}^{2}(A_{l}^{4}-A_{l}^{6})\right]\right.\\
 & \left.+\frac{\exp\left(-\frac{1}{4}x_{0}^{2}q_{\Delta t}^{2}\right)}{v_{t}^{2}}\frac{\Delta^{5}}{\hbar^{5}v_{t}^{5}}\Xi_{u}^{2}\left[A_{t}^{2}-2A_{t}^{4}+A_{t}^{6}\right]\right\} .
\end{align}
 This reduces exactly to the expression for orbital relaxation within
the electric dipole approximation, given in Eq. \ref{eq:orbital},
when $q_{\Delta s}^{2}=0$.

\subsection{Valley relaxation (ideal case)\label{sub:Valley-relaxation-Appendix}}

\subsubsection{Valley relaxation in a three-dimensional parabolic quantum dot}

We consider valley relaxation in a lateral silicon quantum dot. We
begin where we left off in our exact consideration of orbital relaxation.
Our expression for the electron-phonon matrix element, Equation \ref{eq:exactorbitalelement},
was 
\[
M_{mn}=\sum_{i,j}\alpha_{m}^{i}\alpha_{n}^{j}\left[\Xi_{d}\left(\mathbf{e}\left(\mathbf{q},\lambda\right)\cdot\mathbf{q}\right)+\Xi_{u}\left(\mathbf{q}\cdot\mathbf{K}^{\left(i\right)}\right)\left(\mathbf{e}\left(\mathbf{q},\lambda\right)\cdot\mathbf{K}^{\left(i\right)}\right)\right]a_{q}^{\ast}\sum_{\nu}C_{k_{i}}^{\nu}f^{ij}(\mathbf{-k}_{i}+\mathbf{k}_{j}+\mathbf{q}-\mathbf{Q}_{\nu}),
\]
 where $f$ is the Fourier transform of the envelope function overlap
integral. Since the valley transition involves a change in crystal
momentum, there are no intravalley terms from this expression and
we must consider high wavenumber phonons that can connect the two
valleys which are separated in the first Brillouin zone by $2k_{min}$.
In this case, $i\neq j$ and $\mathbf{k}_{j}=-\mathbf{k}_{i}$ and
the matrix element becomes 
\[
M_{mn}(i\neq j)=\sum_{i,j}\alpha_{m}^{i}\alpha_{n}^{j}\left[\Xi_{d}q_{l}+\Xi_{u}\left(\mathbf{q}\cdot\mathbf{K}^{\left(i\right)}\right)\left(\mathbf{e}\left(\mathbf{q},\lambda\right)\cdot\mathbf{K}^{\left(i\right)}\right)\right]a_{q}^{\ast}\sum_{\nu}C_{k_{i}}^{\nu}f^{ij}(-2\mathbf{k}_{i}+\mathbf{q}-\mathbf{Q}_{\nu}),
\]
 where $q_{l}=q_{\Delta}$ for longitudinal phonons but zero for transverse
phonons ($q_{\Delta}=\Delta/hv_{s}$). The matrix element will only
be large for values of $\mathbf{q}\approx2\mathbf{k}_{i}-\mathbf{K}_{\nu}$.
The shortest wavenumber phonon to connect the two valleys is the Umklapp
phonon across the Brillouin zone where $\left\vert q_{u}\right\vert =2(q_{max}-q_{min})$,
where $q_{max}=\pi/a_{Si}$. Thus, for the $+z$ and $-z$ valleys,
\begin{equation}
M_{as}=\frac{1}{2}\left[\Xi_{d}q_{l}+\Xi_{u}e_{z}q_{z}\right]a_{q}^{\ast}C_{k_{0}}^{1}\left[f^{(1)}(-q_{u}\hat{z}+\mathbf{q})-f^{(1)}(q_{u}\hat{z}+\mathbf{q})\right].\label{eq:valleymatrixelemen}
\end{equation}
 This is just the result of Castner as a component of his calculation
of Raman spin transitions in donors. The major difference between
the donor and QD calculations (in the ideal case) are due to the different
envelope functions (impurity vs. parabolic). We next require the Fourier
transform of the QD envelope function: 
\begin{align*}
f^{(1)}(\mathbf{q}) & =\int F^{(1)}(\mathbf{r})^{2}e^{i\mathbf{q}\cdot\mathbf{r}}d\mathbf{r}\\
f^{(1)}(q_{x})f^{(1)}(q_{y})f^{(1)}(q_{z}) & =\int F^{(1)}(x)^{2}e^{iq_{x}x}dx\int F^{(1)}(y)^{2}e^{iq_{y}y}dy\int F^{(1)}(z)^{2}e^{iq_{z}z}dz\\
 & =\exp\left(-\frac{1}{8}x_{0}^{2}q_{x}^{2}\right)\exp\left(-\frac{1}{8}y_{0}^{2}q_{y}^{2}\right)\exp\left(-\frac{1}{8}z_{0}^{2}q_{z}^{2}\right).
\end{align*}
 Again we have considered the case where the $z$ dimension of the
wave function can be approximated as a simple Gaussian (which is for
our consideration a good approximation).

Looking at Eq. \ref{eq:valleymatrixelemen}, we see that the $f$-functions
are heavily peaked at $q_{u}$ ($\approx0.3q_{max}=0.3\pi/a=1.74\cdot10^{9}$
m$^{-1}$) in the $z$ direction and at $0$ in the $x$ and $y$
directions. Since phonons of this magnitude are needed to connect
the two valleys, resonant phonons close to this will increase the
matrix element leading to increased relaxation. However, slightly
off-resonant phonons can also cause a transition due to the widths
of the $f$-functions which broaden as $z_{0}$ gets smaller (see
Figure \ref{fig:Umklapp_phonon}). In donors, the valley splitting
tends to be around 11 meV, not far off of this wave vector. In silicon
quantum dots, the theoretically predicted values of the valley splitting
range from 0 to 3 meV depending on the extent of the $z$ wave function.

To calculate the valley transition rate we employ the Golden Rule,
\begin{align}
\Gamma_{as} & =\frac{2\pi}{\hbar}\sum_{\mathbf{q},s}\left\vert M_{as}\right\vert ^{2}\delta\left(\Delta-\hbar\omega_{\mathbf{q},s}\right),\label{eq:GoldenRule}\\
 & =\frac{2\pi}{\hbar}\sum_{s}\frac{V}{(2\pi)^{3}}\int_{0}^{\infty}q^{2}dq\int\sin\theta d\theta d\phi\left\vert M_{as}\right\vert ^{2}\frac{1}{\left\vert -\hbar v_{s}\right\vert }\delta\left(q-\frac{\Delta}{\hbar v_{s}}\right),
\end{align}
 where we have summed over phonons and the emitted phonon has wave
number $q_{\Delta}=\Delta/\hbar v_{s}$. At cryogenic temperature
there are absolutely no large wave number phonons, so we need only
consider spontaneous emission. Incorporating our expression for the
matrix element, we find (with $F(q_{u},\mathbf{q})=\left[f^{(1)}(-q_{u}\hat{z}+\mathbf{q})-f^{(1)}(q_{u}\hat{z}+\mathbf{q})\right]$)
that 
\begin{align*}
\Gamma_{as} & =\frac{2\pi}{\hbar^{2}}\sum_{s}\frac{V}{(2\pi)^{3}}\int_{0}^{\infty}q^{2}dq\int\sin\theta d\theta d\phi\frac{1}{v_{s}}\left\vert \frac{1}{2}\left[\Xi_{d}q_{l}+\Xi_{u}e_{z}q_{z}\right]a_{q}^{\ast}C_{k_{0}}^{1}F(q_{u},\mathbf{q})\right\vert ^{2}\delta\left(q-q_{\Delta}\right)\\
 & =\frac{\left(n_{q}+1\right)\left(C_{k_{0}}^{1}\right)^{2}}{8(2\pi)^{2}\rho_{Si}\hbar}\sum_{s}\int_{0}^{\infty}q\int\sin\theta\frac{1}{v_{s}^{2}}\left\vert \left[\Xi_{d}q_{l}+\Xi_{u}e_{z}q_{z}\right]F(q_{u},\mathbf{q})\right\vert ^{2}\delta\left(q-q_{\Delta}\right)d\theta d\phi dq\\
 & =\frac{\left(n_{q}+1\right)\left(C_{k_{0}}^{1}\right)^{2}}{8(2\pi)^{2}\rho_{Si}\hbar}\sum_{s}q_{\Delta s}I_{s}.
\end{align*}
 Taking the delta function, the rate becomes 
\begin{align*}
\Gamma_{as} & =\frac{\left(n_{q}+1\right)\left(C_{k_{0}}^{1}\right)^{2}}{8(2\pi)^{2}\rho_{Si}\hbar}\sum_{s}q_{\Delta s}\int\sin\theta\frac{1}{v_{s}^{2}}\left\vert \left[\Xi_{d}q_{\Delta l}+\Xi_{u}e_{z}q_{\Delta s}\hat{q}_{z}\right]F(q_{u},q_{\Delta s}\hat{\mathbf{q}})\right\vert ^{2}d\theta d\phi\\
 & =\frac{\left(n_{q}+1\right)\left(C_{k_{0}}^{1}\right)^{2}}{8(2\pi)^{2}\rho_{Si}\hbar}\sum_{s}q_{\Delta s}I_{s}
\end{align*}
 where $\hat{\mathbf{q}}=\mathbf{e}_{l}$. We are again left with
calculating three angular integrals $I$ which have units (kg$^{2}$/s$^{2}$)
and are defined as
\[
I_{s=l,t_{1},t_{2}}=\int_{0}^{2\pi}\int_{0}^{\pi}\sin\theta\frac{1}{v_{s}^{2}}\left\vert \left[\Xi_{d}q_{\Delta l}+\Xi_{u}e_{zs}q_{\Delta s}\hat{q}_{z}\right]\left[f^{(1)}(-q_{u}\hat{z}+q_{\Delta s}\hat{\mathbf{q}})-f^{(1)}(q_{u}\hat{z}+q_{\Delta s}\hat{\mathbf{q}})\right]\right\vert ^{2}d\theta d\phi.
\]
 $I_{t_{2}}=0$. If we assume that the dot is circular\emph{,} then
the $f$-functions simplify,
\begin{align*}
f^{(1)}(\pm q_{u}\hat{z}+q_{\Delta}\hat{\mathbf{q}}) & =\exp\left(-\frac{1}{8}x_{0}^{2}\left(q_{\Delta}\hat{q}_{x}\right)^{2}-\frac{1}{8}x_{0}^{2}\left(q_{\Delta}\hat{q}_{y}\right)^{2}\right)\exp\left(-\frac{1}{8}z_{0}^{2}\left(q_{\Delta}\hat{q}_{z}\pm q_{u}\right)^{2}\right)\\
 & =\exp\left(-\frac{1}{8}x_{0}^{2}q_{\Delta}^{2}\sin^{2}\theta\right)\exp\left(-\frac{1}{8}z_{0}^{2}\left(q_{\Delta}\cos\theta\pm q_{u}\right)^{2}\right).
\end{align*}

Replacing $\hat{q}$ with its components we can further simplify $I$:
\begin{align*}
I_{s=l,t_{1},t_{2}} & =\int_{0}^{2\pi}\int_{0}^{\pi}\sin\theta\frac{1}{v_{s}^{2}}\left\vert \left[\Xi_{d}q_{\Delta l}+\Xi_{u}e_{z}q_{\Delta s}\cos\theta\right]\exp\left(-\frac{1}{8}x_{0}^{2}q_{\Delta}^{2}\sin^{2}\theta\right)\left\{ {}\right\} \right\vert ^{2}d\theta d\phi\\
 & =\int_{0}^{2\pi}\int_{0}^{\pi}\sin\theta\frac{1}{v_{s}^{2}}\left[\Xi_{d}q_{\Delta l}+\Xi_{u}q_{\Delta s}e_{z}\cos\theta\right]^{2}\exp\left(-\frac{1}{4}x_{0}^{2}q_{\Delta}^{2}\sin^{2}\theta\right)\left\{ {}\right\} ^{2}d\theta d\phi
\end{align*}
 where 
\begin{align*}
\left\{ {}\right\} ^{2} & =\left\{ \exp\left(-\frac{1}{8}z_{0}^{2}\left(q_{\Delta l}\cos\theta-q_{u}\right)^{2}\right)-\exp\left(-\frac{1}{8}z_{0}^{2}\left(q_{\Delta l}\cos\theta+q_{u}\right)^{2}\right)\right\} ^{2}\\
 & =\exp\left(-\frac{1}{4}z_{0}^{2}q_{u}^{2}\right)\exp\left(-\frac{1}{4}z_{0}^{2}\left[q_{\Delta}^{2}\cos^{2}\theta\right]\right)4\sinh^{2}\left(\frac{1}{8}z_{0}^{2}2q_{\Delta}q_{u}\cos\theta\right).
\end{align*}
\[
\]
 So, taking the trivial $\phi$ integral (no $e_{zs}$ depends on
$\phi$), 
\begin{align*}
I_{s=l,t_{1},t_{2}} & =8\pi\exp\left(-\frac{1}{4}z_{0}^{2}q_{u}^{2}\right)\int_{0}^{\pi}\sin\theta d\theta\frac{1}{v_{s}^{2}}\left[\Xi_{d}q_{\Delta l}+\Xi_{u}q_{\Delta s}e_{z}\cos\theta\right]^{2}\\
 & \times\exp\left(-\frac{1}{4}x_{0}^{2}q_{\Delta l}^{2}\sin^{2}\theta\right)\exp\left(-\frac{1}{4}z_{0}^{2}\left[q_{\Delta}^{2}\cos^{2}\theta\right]\right)\sinh^{2}\left(\frac{1}{8}z_{0}^{2}2q_{\Delta}q_{u}\cos\theta\right).
\end{align*}
 Now, we explicitly consider the integrals for $s=l$ and $s=t_{2}$.
Evaluating the longitudinal and transverse ($t_{2}$) integrals separately:
\begin{align*}
I_{l} & =8\pi\exp\left(-\frac{1}{4}z_{0}^{2}q_{u}^{2}\right)\exp\left(-\frac{1}{4}x_{0}^{2}q_{\Delta l}^{2}\right)\frac{q_{\Delta l}^{2}}{v_{l}^{2}}\\
 & \times\int_{1}^{-1}dx\left[\Xi_{d}^{2}+2\Xi_{d}\Xi_{u}x^{2}+\Xi_{u}^{2}x^{4}\right]\exp\left(\frac{1}{4}q_{\Delta l}^{2}\left[x_{0}^{2}-z_{0}^{2}\right]x^{2}\right)\sinh^{2}\left(\frac{1}{8}z_{0}^{2}2q_{\Delta l}q_{u}x\right)
\end{align*}
 and, similarly, 
\begin{align*}
I_{t_{2}} & =8\pi\exp\left(-\frac{1}{4}z_{0}^{2}q_{u}^{2}\right)\exp\left(-\frac{1}{4}x_{0}^{2}q_{\Delta t}^{2}\right)\frac{q_{\Delta t}^{2}}{v_{t}^{2}}\\
 & \times\int_{1}^{-1}dx\Xi_{u}^{2}\left[x^{2}-x^{4}\right]\exp\left(\frac{1}{4}q_{\Delta t}^{2}\left[x_{0}^{2}-z_{0}^{2}\right]x^{2}\right)\sinh^{2}\left(\frac{1}{8}z_{0}^{2}2q_{\Delta t}q_{u}x\right).
\end{align*}
 The integrals have no analytical solution so we define a numerically
tractable integral function
\[
P_{s}^{n}=\int_{-1}^{1}x^{n}\exp\left[Ax^{2}\right]\sinh^{2}\left[Bx\right]dx
\]
 where 
\begin{align*}
A & =\frac{1}{4}q_{\Delta s}^{2}\left[x_{0}^{2}-z_{0}^{2}\right],\\
B & =\frac{1}{8}z_{0}^{2}2q_{\Delta s}q_{u},
\end{align*}
 and rewrite our solutions: 
\[
I_{l}=8\pi\exp\left(-\frac{1}{4}z_{0}^{2}q_{u}^{2}\right)\exp\left(-\frac{1}{4}x_{0}^{2}q_{\Delta l}^{2}\right)\frac{q_{\Delta l}^{2}}{v_{l}^{2}}\left[\Xi_{d}^{2}P_{l}^{0}+2\Xi_{d}\Xi_{u}P_{l}^{2}+\Xi_{u}^{2}P_{l}^{4}\right]
\]
 and 
\[
I_{t_{2}}=8\pi\exp\left(-\frac{1}{4}z_{0}^{2}q_{u}^{2}\right)\exp\left(-\frac{1}{4}x_{0}^{2}q_{\Delta t}^{2}\right)\frac{q_{\Delta t}^{2}}{v_{t}^{2}}\Xi_{u}^{2}\left[x^{2}P_{t}^{2}-x^{4}P_{t}^{4}\right].
\]
 Finally, the valley relaxation rate of a parabolic, circular quantum
dot in a {[}001{]}-strained silicon quantum well is ($n_{q}=0$ and
$C=1$) 
\begin{align*}
\Gamma_{as} & =\left(n_{q}+1\right)\left(C_{k_{0}}^{1}\right)^{2}\frac{\exp\left(\frac{-z_{0}^{2}q_{u}^{2}}{4}\right)}{4\pi\rho_{Si}\hbar}\left[\exp\left(\frac{-x_{0}^{2}q_{\Delta l}^{2}}{4}\right)\frac{q_{\Delta l}^{3}}{v_{l}^{2}}\left(\Xi_{d}^{2}P_{l}^{0}+2\Xi_{d}\Xi_{u}P_{l}^{2}+\Xi_{u}^{2}P_{l}^{4}\right)\right.\\
 & +\left.\exp\left(\frac{-x_{0}^{2}q_{\Delta t}^{2}}{4}\right)\frac{q_{\Delta t}^{3}}{v_{t}^{2}}\Xi_{u}^{2}\left(P_{t}^{2}-P_{t}^{4}\right)\right]
\end{align*}

\bibliographystyle{apsrev4-1}


%

\end{document}